\RequirePackage[running]{lineno}

\documentclass[aps,prd,onecolumn,nofootinbib,superscriptaddress,letterpaper,amsmath,amssymb]{revtex4}
\pdfoutput=1

\usepackage{ulem}
\usepackage{graphicx}   % needed for figures
\usepackage{dcolumn}    % needed for some tables
\usepackage{bbm}        % for math
\usepackage{amsmath}
\usepackage{amsfonts}
\usepackage{amssymb}   % for math
\usepackage{epstopdf}
\usepackage{slashed}
\usepackage{hyperref}
\usepackage{multirow}
\usepackage{color}
\usepackage{float}
\usepackage{subcaption}

\hypersetup{colorlinks=true, linkcolor=black}

\newcommand{\xspec}{{\sc xspec}}
\newcommand{\surrMR}{ML-Likelihood$_{\text{M,R}}$}
\newcommand{\surrEOS}{ML-Likelihood$_{\text{EOS}}$}
\newcommand{\mrnet}{MR\_Net}

\begin{document}

\title{ Deducing Neutron Star Equation of State from Telescope Spectra\\ with Machine-learning-derived Likelihoods}

\author{Delaney Farrell}
\affiliation{Department of Physics, San Diego State University, San Diego, CA 92115, United States}
\author{Pierre Baldi}
\author{Jordan Ott}
\affiliation{Department of Computer Science, University of California Irvine, Irvine, California 92697, USA}
\author{Aishik Ghosh}
\affiliation{Department of Physics and Astronomy, University of California Irvine, Irvine, California 92697, USA}
\affiliation{Physics Division, Lawrence Berkeley National Laboratory, Berkeley, CA 94720, USA}
\author{Andrew W. Steiner}
\affiliation{Department of Physics and Astronomy, University of Tennessee, Knoxville, Tennessee 37996, USA}
\affiliation{Physics Division, Oak Ridge National Laboratory, Oak Ridge, Tennessee 37831, USA}
\author{Atharva Kavitkar}
\affiliation{Department of Computer Science, TU Kaiserslautern, Germany }
\author{Lee Lindblom}
\affiliation{Center for Astrophysics and Space Sciences,
University of California at San Diego, San Diego, CA 92093, United States}
\author{Daniel Whiteson}
\affiliation{Department of Physics and Astronomy, University of California Irvine, Irvine, California 92697, USA}
\author{Fridolin Weber}
\affiliation{Department of Physics, San Diego State University, San Diego, CA 92115, United States}
\affiliation{Center for Astrophysics and Space Sciences,
University of California at San Diego, San Diego, CA 92093, United States}

\begin{abstract}
The interiors of neutron stars reach densities and temperatures beyond the limits of terrestrial experiments, providing vital laboratories for probing nuclear physics. While the star's interior is not directly observable, its pressure and density determine the star's macroscopic structure which affects the spectra observed in telescopes. The relationship between the observations and the internal state is complex and partially intractable,  presenting difficulties for inference.  Previous work has focused on the regression from stellar spectra of parameters describing the internal state. We demonstrate a calculation of the full likelihood of the internal state parameters given observations, accomplished by replacing intractable elements with machine learning models trained on samples of simulated stars. Our machine-learning-derived likelihood allows us to perform  {\it maximum a posteriori} estimation of the parameters of interest, as well as full scans. We demonstrate the technique by inferring stellar mass and radius from an individual stellar spectrum, as well as equation of state parameters from a set of spectra. Our results are more precise than pure regression models, reducing the width of the parameter residuals by 11.8\% in the most realistic scenario. The neural networks will be released as a tool for fast simulation of neutron star properties and observed spectra.
\end{abstract}

\maketitle
\tableofcontents

\vspace{.25in}

\section{Introduction}

Neutron stars are valuable astrophysical laboratories for studying matter under extreme conditions. With masses generally between 1 to 2 $M_\odot$ and radii between 10 and 15 km, the inner regions of these neutron-rich stars can reach density regimes well beyond those accessible in terrestrial laboratories. Matter at such high densities can potentially experience transitions to stable but unusual states of matter such as exotic baryons made of hyperons and $\Delta$ isobars~\cite{Tolos:2020aln,li2018competition, spinella2019hyperonic,Malfatti:2020,Sedrakian:2023PPNP};  deconfined up, down, and strange quarks~\cite{Alcock:1986,orsaria2014quark};  color superconducting phases~\cite{Alford:2001,Alford:2007xm,zdunik2013maximum}; or Bose-Einstein condensates made of negatively charged pions or $K^-$ mesons~\cite{baym1973pion,KAPLAN,Glendenning:1999,ellis1995kaon,Ramos:2001}.  A better understanding of the internal composition of these stars would shed light on many areas of current interest, including various astrophysical phenomena such as core-collapse supernovae \cite{steiner2013core} and binary star mergers \cite{hotokezaka2011binary}, nuclear laboratory physics, QCD and relativistic gravity, as well as the early Universe. A long-standing issue in experimental and theoretical astrophysics is the determination of the equation of state (EOS) of matter within the cores of neutron stars, which describes the underlying relationship between pressure $P$ and energy density $\epsilon$. Theoretical EOS models make various assumptions regarding the state of matter within neutron stars' interiors, resulting in widely varying relationships between pressure and density. 

The interiors of neutron stars are not available for direct observation. However, observational data from these stars, such as electromagnetic emission, is growing rapidly from telescopes like the \textit{Chandra} X-ray Observatory, the Neutron star Interior Composition Explorer (NICER), and the Five-hundred-meter Aperture Spherical radio Telescope (FAST). The star's emitted spectrum is influenced by macroscopic stellar properties such as mass and radius, which are determined by the star's underlying EOS. The mass and radius, paired with other stellar parameters, control the emitted radiation, which when convoluted with the telescope response, determines the observed spectra. Therefore, in principle, the EOS can be inferred from observations of stellar spectra. In practice, crucial elements of the likelihood are not tractable, preventing straightforward inference. For example, while the mass-radius relation can be calculated from the EOS using the relativistic stellar structure equations (SSEs), the equations cannot be trivially inverted\cite{Rutledge,Heinke06,Lattimer01, PhysRevD.82.103011,Steiner10te, Lindblom:2013xkra,issp}. The task is further complicated by the small number of neutron star observations and the relatively large uncertainty of the individual measurements. Extraction of the maximal information content, with robust uncertainties, is therefore of paramount importance.

Recently, machine learning tools have demonstrated breakthroughs in data analysis in the natural sciences~\cite{baldi2021deep}, increasing the power of valuable data~\cite{baldi2014searching} while naturally propagating uncertainties~\cite{Ghosh:2021roe}. Applications in particle physics have used larger and deeper neural networks powered by advances in hardware processing to tackle more complex and higher-dimensional tasks~\cite{Baldi:2016fzo,Guest:2016iqz}.  In astrophysical contexts such as regression of neutron star parameters where the likelihood is not tractable and analytical inversion is not feasible, simulation-based inference techniques~\cite{Cranmer_2020} have been explored. For example, recent work from Fujimoto {\it et al}~\cite{Fujimoto:2019hxv,Fujimoto:2021zas} demonstrated regression of neutron star matter EOS from mass-radius pairs, where networks are trained on samples of simulated stars and uncertainties are drawn from an ad-hoc Gaussian model.  Morawski~\cite{Morawski:2020izm} also performed regression from mass-radius pairs to EOS assuming Gaussian uncertainty and attempted to reduce EOS parameterization dependence. Soma {\it et al}~\cite{soma2022neural} regressed both EOS and mass-radius information through a set of connected modes: density values are used to determine pressure values using a deep neural network, and the regressed pressure values are then used to generate mass-radius pairs using a generative deep learning model. Ferreira~\cite{Ferreira:2019bny} instead regressed EOS from radius and tidal deformation information using support vector machines (SVMs) and deep neural networks (DNNs). Our recent work~\cite{delaney-eos} demonstrated regression of EOS directly from simulated X-ray spectra without collapsing to intermediate states of observable properties like mass and radius, and where uncertainty due to 
unknown values of other stellar parameters is fully propagated. This was the first demonstration that neural networks have the capacity to directly analyze high-dimensional astrophysical data such as X-ray spectra. 

\begin{figure}[hbt!]
    \centering
\includegraphics[width=1.0\textwidth]{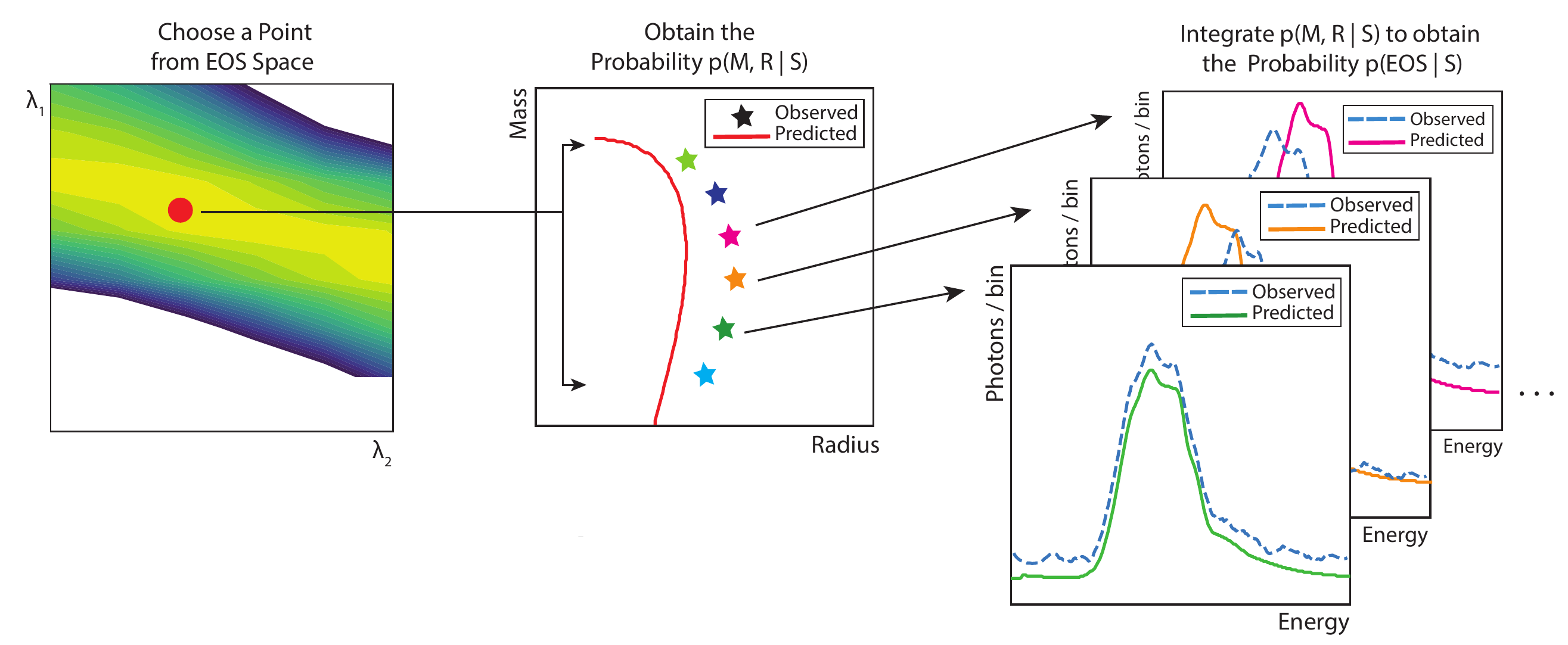}
    \caption{ Schematic diagram depicting the evaluation of the likelihood of producing a set of observed stellar spectra by comparing it to the predicted spectra along the mass-radius curve determined by EOS parameters $\lambda_1$, $\lambda_2$. Each value of the EOS parameters determines a curve in the mass-radius plane. Integrating along the curve, the probability of observing each star is evaluated as in Fig.~\ref{fig:schematic_mr}.}
    \label{fig:schematic_eos}
\end{figure}

Machine learning is, however, capable of more than regression. 
In this paper, we introduce a novel approach to the inference of EOS parameters from neutron star X-ray spectra, in which the likelihood is made tractable by replacing the intractable elements with neural networks trained on samples of simulated stars. Rather than directly learning the entire likelihood~\cite{nle} in one step, we leverage our knowledge of the problem by replacing only the crucial missing pieces, which focuses the learning task and allows the interpretation of network outputs as physically meaningful quantities. The resulting {\it machine-learning-derived likelihood} of observing a set of stellar spectra given EOS parameters allows for estimation of the EOS parameters via Bayesian {\it maximum a posteriori} and use of standard error estimation techniques. The derivation of this machine-learning-derived likelihood is shown schematically in Figure~\ref{fig:schematic_eos}.

This paper is outlined as follows. In Section~\ref{sec:bg}, we describe the physics of neutron star interiors. Section~\ref{sec:ml} gives an introduction to the relevant machine learning techniques. We describe the generation of our samples of simulated stars in Section~\ref{sec:data}, and the strategy for machine-learning-derived likelihoods in Section~\ref{sec:mllh}. Sections~\ref{sec:mr} and~\ref{sec:eos} demonstrate the likelihood calculation and extraction of parameter estimates for mass-radius information and EOS parameters, respectively. Sections~\ref{sec:disc} and~\ref{sec:conc} present a discussion of the results and future directions, respectively.

\section{Background}
\label{sec:bg}

Observation of neutron stars has long served as a way to place constraints on our knowledge of superdense matter due to the theoretical connection between features deduced from observation (such as gravitational mass and radius) and the underlying EOS. The theoretical connection comes from the relativistic stellar structure equations, derived from Einstein's field equations, where the star's internal structure is used to derive the stellar properties of a compact object. 

For spherically symmetric, non-rotating stars comprised of isotropic material in hydrostatic equilibrium, the stellar structure is compactly summarized in the Tolman-Oppenheimer-Volkoff (TOV) equation (in geometerized units $G = c = 1$): 
\begin{equation}
\label{eq:tov}
    \frac{dP}{dr} = -\frac{(\epsilon+P)(m+4\pi r^3 P)}{r^2\left(1-\frac{2m}{r}\right)},
\end{equation}
where $m$ is the gravitational mass enclosed within a sphere of radius $r$, $P$ is the pressure, and $\epsilon$ is the energy density. This equation creates a one-to-one map from the EOS to a mass-radius relation \cite{Lindblom:2013xkra}, where the inverse mapping would provide constraints on the EOS from mass and radius values. While inverting the relativistic SSEs has been demonstrated to be numerically intractable, mass and radius values determined by observation provide valuable constraints on EOS; observations of high-mass pulsars (such as PSR J0952-0607, where $M = 2.35 \pm 0.17 \, M_{\odot}$ \cite{romani2022psr}) have ruled out previously popular EOS models which failed to predict masses above $2 M_{\odot}$. 

Observation of neutron star emission has long served as a way to constrain mass and radius for neutron stars (eg. \cite{hebeler2013equation, Steiner18ct}), which in turn places constraints on the nuclear EOS. Some of the most precise constraints thus far come from observing thermal emission from quiescent low-mass X-ray binaries (qLMXBs). These binary systems are identified in well-studied globular clusters where distances, ages, and reddening are well-constrained \cite{Heinke_2003}. Additionally, their low magnetic fields result in minimal effects on the radiation transport or temperature distribution on the star's surface \cite{2016ApJ...831..184B, campana1998neutron, potekhin2014atmospheres}, making them particularly desirable for placing strong constraints on neutron star structure.

Thermal emission from qLMXBs is observed with powerful telescopes such as  NASA’s \textit{Chandra} X-ray Observatory, where high-resolution imaging and spectroscopy have provided powerful insight into neutron star properties such as cooling \cite{cooling, wijnands2017cooling}, mass and radius limits \cite{2016ApJ...831..184B}, and binary mergers of exotic stars \cite{mag}. \textit{Chandra}'s telescope contains a system of four pairs of mirrors that focus incoming X-ray photons to the Advanced CCD Imaging Spectrometer (ACIS), which measures the energy of each incoming X-ray. The observed spectrum then determines neutron star bulk properties like mass and radius through spectral fitting, where both the spectrum and corresponding instrument response are fit to an appropriate atmosphere model, where the surface composition is known or can be determined by the X-ray spectrum. An extensive list of such models exists in \xspec~\cite{xspec}, an X-ray spectral fitting package distributed and maintained by the aegis of the GSFC High Energy Astrophysics Science Archival Research Center (HEASARC). \xspec is a state-of-the-art tool to analyze X-ray data from \textit{Chandra} and other spectrometers like \textit{NICER}, \textit{Nustar}, and \textit{XMM-Newton} - making it a valuable resource for inference of neutron star properties.

\section{Machine Learning}
\label{sec:ml}

Machine learning is a branch of computer science, and the main component of today’s artificial intelligence, where computers learn to compute useful functions from data. In deep learning, a modern re-branding of neural networks and today’s major branch of machine learning, networks of simple neurons connected by adjustable synaptic weights are used to implement flexible classes of functions. In a typical case, the parameters of such models, i.e. the synaptic weights, are adjusted incrementally from the training data via a stochastic gradient descent process in order to minimize some error function. For instance, in the case of a regression problem, neural networks can be trained to minimize the standard least-squared error over the training data. This can be viewed as a generalization of standard linear regression, where the linear model is replaced by a more complex, non-linear, neural network model. The least-squares error can be viewed as the negative log-likelihood of the data under a Gaussian noise model. The neural network learns to predict the parameters of the negative log-likelihood function, i.e. the mean of the Gaussian distribution in the standard regression framework. 

Machine learning, and in particular deep learning, have many applications in science \cite{baldi2021deep} and have greatly benefited from the increase in computing power over the last few decades, especially in the form of GPUs (Graphical Processing Units) which are well suited for the matrix-vector multiplications - the computational workhorse of neural networks. Deep learning methods can flexibly handle data sets, models, and dimensions that range from very small to very large. For instance, deep learning models with millions, or even billions, of adjustable parameters, are not uncommon. These methods can handle and leverage raw data, without the need for pre-processing using heuristic or manual simplifications, which often discard valuable information. 
In physics applications, this often leads to significant performance improvements in terms of accuracy, detection rates, and so forth (e.g. \cite{baldi2014searching,baldi2016jet,fenton2022permutationless}). In the rest of this work, we use deep learning methods for regression to deterministically compute the parameters of complex  functions needed to calculate likelihoods.

\section{Training Data}
\label{sec:data}

Samples of simulated stars are used both to train and evaluate network performance.
The samples used in these studies were originally produced for Ref.~\cite{delaney-eos}; we use the identical samples here to allow direct comparison to previous work. Each simulated star is described by two parameters of interest and three additional physical stellar properties. The parameters of interest are mass $M$ and radius $R$, derived from EOS variations using the stellar structure equations. The other three quantities, deemed nuisance parameters, affect the stellar spectrum but are not related to the EOS: the effective temperature of the star's surface, $T_{\mathrm{eff}}$, the distance to the star, $d$, and the hydrogen column, $N_H$ which parameterizes the reddening of the spectrum by the interstellar medium. All five parameters determine the simulated X-ray spectra, generated using \xspec. Details of each data generation step are outlined below.

\subsection{Generation of Equation of State}

As described in Ref.~\cite{delaney-eos}, the EOS models used to generate many simulated neutron stars are variations of the relativistic, non-linear mean-field model GM1L~\cite{Typel}. The parameterization of GM1L used in these studies accounts only for protons and neutrons; the corresponding saturation properties of this parametrization are shown in Table 1 of Ref.~\cite{delaney-eos}, which follow current constraints from nuclear experiments and astrophysical observations \cite{Oertel}. In order to represent the EOS in a compact fashion and limit parameters of interest, we represent GM1L in a parametric representation based on spectral fits; the process for constructing such fits is outlined in detail in Refs.~\cite{PhysRevD.97.123019} and \cite{PhysRevD.82.103011}. 

As in Ref.~\cite{delaney-eos}, the parametric representation of GM1L is given by the coefficients of order two spectral expansion: $\lambda_1$ and $\lambda_2$. We randomly vary the two original coefficients to create $10^4$ unique EOS variations in order to generate the many models needed for training the networks discussed below. Each EOS variation was then used to generate a mass-radius relation using the TOV equation (\ref{eq:tov}). From each mass-radius relation, 100 $(M,R)$ pairs whose mass falls in the physical range of neutron star masses (1 to 3 $M_\odot$) are selected for training and validation.

\subsection{Modeling Spectra}

Generation of a simulated neutron star X-ray spectrum requires the stellar mass and radius, generated from a particular EOS model as described above, as well as the three nuisance parameters.  Beyond spectral fitting, the \xspec~program~\cite{xspec} can create simulated spectra via the \texttt{fakeit} command. The command requires a specified theoretical model and telescope response matrix. As described in greater detail in Ref.~\cite{delaney-eos}, the theoretical hydrogen atmosphere model \texttt{NSATMOS}~\cite{Heinke_2006} was used, paired with the \textit{Chandra} telescope response specified in Ref.~\cite{Heinke_2006} which describes the instrument response and telescope effective area. 
    
\subsection{Nuisance Parameters}

The \texttt{NSATMOS} model implemented in \xspec~ requires five inputs for each simulated spectrum. The first two, gravitational mass $M$ in units of $M\odot$ and radius $R$ in units of kilometers, are taken from mass-radius relations from the GM1L EOS variations. The remaining three parameters are the effective temperature of the star's surface $T_{\mathrm{eff}}$, the distance to the star, $d$, and the hydrogen column density, $N_H$. These three additional parameters are referred to as nuisance parameters (NPs) in Ref.~\cite{delaney-eos} and are given values drawn from reasonable physical limits.

To determine the physical limits of the distance to the star and the hydrogen column density $N_H$, we turn to Table 1 in Ref.~\cite{Steiner18ct}. The distances typically range between 2 and 10 kpc, and hydrogen columns lie between 0.2 and $5 \times 10^{21}~\mathrm{cm}^{-2}$. From Table 3 in Ref.~\cite{Lattimer14ns}, effective temperatures at the surface typically lie between 50 and 200 eV, or from $6 \times 10^5$ and $2.4 \times 10^6$ K, with temperatures in the core increasing by a few orders of magnitude.

Examples of generated X-ray spectra are shown in Ref.~\cite{delaney-eos}. The three NPs, which we will refer collectively to as $\nu$, directly impact the simulated spectra and the uncertainty in the spectral fitting process. In order to propagate the uncertainty from these parameters and gauge our networks' sensitivity to each parameter, we define three example scenarios of uncertainties: ``true", ``tight", and ``loose". The three scenarios describe the quality of prior information on the NP values for each star.

In the ``true" scenario, the NPs are set to the true value chosen from the physical range used to generate the spectra, such that the NP prior is essentially a delta function. In the ``tight" scenario, the uncertainty is described as a narrow  Gaussian for each NP, with distance having a width of 5\%, hydrogen column having a width of 30\%, and $\log(T_\textrm{eff})$ having a width of 0.1. In the ``loose" scenario, the uncertainties are described by a wider Gaussian, with distance having a width of 20\%, hydrogen column having a width of 50\%, and $\log(T_\textrm{eff})$ having a width of 0.2. Table~\ref{tab:nps} shows the width of the prior information of the three NPs for the three uncertainty scenarios.

\begin{table}[]
    \caption{Description of ``true", ``tight", and ``loose" nuisance parameter (NP) scenarios. Shown are the width of each Gaussian distribution representing the prior knowledge of each NP. For distance and $N_H$, width is relative; for log($T_{\text{eff}}$), it is absolute. See text for details and references.}
    \label{tab:nps}
    \centering
    \begin{tabular}{lcrrr}
             \hline \hline
         Nuis. Param. & True & Tight & Loose \\
                  \hline 
        Distance & exact  & 5\% & 20\% \\
        Hydrogen Column $N_H$ &exact & 30\% & 50\% \\
        log($T_{\text{eff}}) $ &exact    & $\pm$0.1  & $\pm$0.2  \\ 
         \hline \hline
    \end{tabular}
\end{table}

\section{Machine-Learning Derived Likelihood Calculation}
\label{sec:mllh}

\begin{figure}[hbt!]
    \centering
\includegraphics[width=0.75\textwidth]{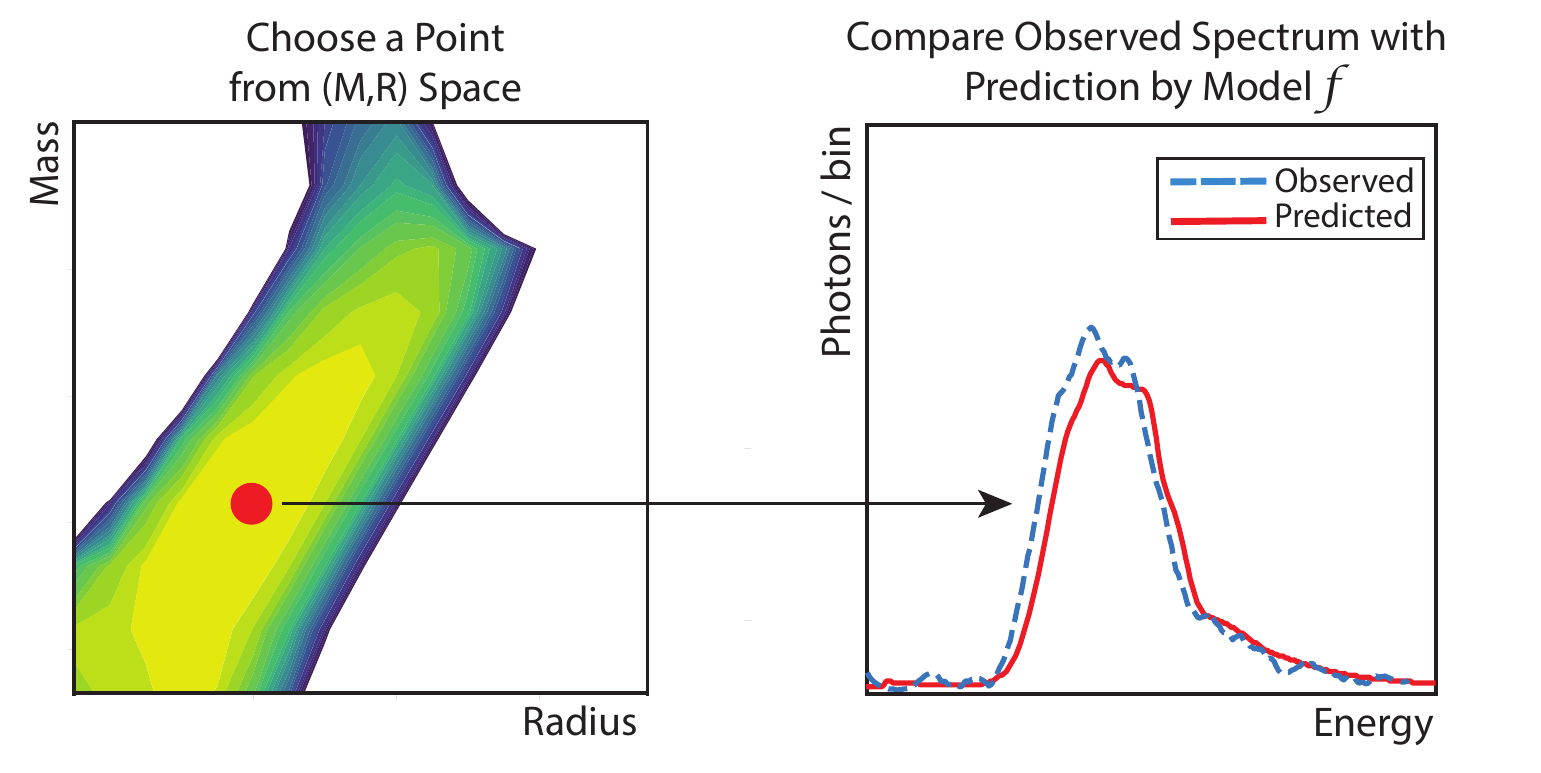}
    \caption{Schematic diagram depicting the evaluation of the likelihood of producing on the observed stellar spectrum by comparing it to the predicted spectrum for assumed values of stellar mass and radius. Varying the assumed mass and radius allows for an evaluation of the likelihood across the $(M,R)$ parameter space.}
    \label{fig:schematic_mr}
\end{figure}

If the likelihood $p(x|\theta)$ of observing some data $x$ for given values of the theoretical model parameters $\theta$ were known, estimating the value of $\theta$ for a given $x$ would be a straightforward task. In a Bayesian approach, one would use {\it maximum a posteriori} (MAP) to find $\hat\theta$, the value of $\theta$ which maximizes the product of the likelihood and the prior $p(\theta)$.

In the case of neutron star EOS estimation, the likelihood $p(S|\lambda_1,\lambda_2)$ of observing a set of stellar spectra $S$ for a given choice of EOS parameters $\lambda_1,\lambda_2$ is not tractable. Instead, simulation-based inference~\cite{Cranmer_2020} techniques often circumvent the need for a closed-form likelihood by approximating $p(S|\lambda_1,\lambda_2)$ using density estimation with simulated samples. But accurately estimating the density in high-dimensional spaces such as those for stellar spectra, with dimensionality of $\approx \mathcal{O}(1000)$, would require prohibitively large samples of simulated events. Instead, we use machine learning to estimate the likelihood. But rather than estimating the entire likelihood in one step, our machine-learning-derived likelihood calculation focuses the statistical power where it is most needed for a given set of observations by combining the available analytic components with neural networks to replace the intractable components. 

For the estimation of parameters such as EOS or mass and radius, many elements of the likelihood calculation are known, such as the Poisson fluctuations within each energy bin of a stellar spectrum. However, there are two elements without closed-form expressions that require machine-learning assistance. We overview those intractable components here briefly and provide more detail below.
 
The first missing piece is a prediction for the expected photon rates observed in a telescope given stellar parameters $M$ and $R$ and nuisance parameters $\nu$. Below, we train a neural network to estimate this quantity and demonstrate its application by doing MAP estimates of $M$ and $R$, in which we integrate over the nuisance parameters $\nu$.
 
The second missing piece is a prediction for the mass-radius values allowed by given EOS parameters $\lambda_1,\lambda_2$.  We train a neural network to output the allowed radius for a given stellar mass when provided with the EOS parameters. This allows integration over the $M-R$ curve defined by the EOS.

Together, these two machine-learned elements allow for a tractable calculation of the likelihood $p(S|\lambda_1,\lambda_2)$ and provide a MAP estimate of $\lambda_1,\lambda_2$ for a given set of spectra $S$. We describe each piece in turn below.
 
\section{Stellar Mass and Radius Inference}
\label{sec:mr}

The mass and radius of a neutron star can be estimated from its spectrum $s$ if one can calculate and maximize $p(s|M,R) p(M,R)$ for a fixed $s$. This requires access to $p(s|M,R)$, which depends on $p(s|M,R,\nu)$, the likelihood to produce this spectrum $s$ given the full set of stellar parameters including the nuisance parameters $\nu$:

\[ p(s|M,R) = \int d\nu\ p(s|M,R,\nu)\ p(\nu) , \] 
If the spectra consists of a set of energy bins of telescope photon counts,

  \[ s = (N_1^\gamma,N_2^\gamma,...,N^\gamma_{n_{\textrm{bins}}}), \]

\noindent
then the joint likelihood $p(s|M,R,\nu)$ over the bins can be written as

  \[p(s|M,R,\nu) = \prod_j^{n_\textrm{bins}}  \textrm{Pois} ( N^\gamma_j, \mu_j (M,R,\nu)), \]

\noindent
where $\mu_j$ is the expected number $\langle N_j^\gamma\rangle$ of photons in bin $j$ of $n$ bins:
\[ \mu(M, R, \nu) = (\langle N_1^\gamma\rangle,\langle N_2^\gamma\rangle,...,\langle N^\gamma_{n_{\textrm{bins}}}\rangle), \]

\noindent
determined by $M,R,\nu$ via complex physics of the stellar emission model as well as the telescope response. In principle, for this specific scenario, this function is contained within \xspec, but is not exposed to the user in a convenient fashion which would allow for a rapid evaluation of many points, as needed for this application. More generally, one may want to learn a function in scenarios where no function is available, or where the training sample contains examples generated from a mixture of models where no single simple function can describe all samples. 

When we do not have access to a simple expression for $\mu(M, R, \nu)$, it is possible to train a neural network to learn a function $f[M, R, \nu] $ 

\[ f[M, R, \nu] \rightarrow (\frac{dN_1^\gamma}{dt},\frac{dN_2^\gamma}{dt},...,\frac{dN_{n_{\textrm{bins}}}^\gamma}{dt}), \]

\noindent
such that $\mu(M, R, \nu)$ can be estimated from $f(M, R, \nu)$ scaled by the observation time $\Delta t$, 

\[ \mu \approx  f[M, R ,\nu] \Delta t. \]

The likelihood  $p(s|M,R,\nu)$ can then be estimated as

 \[  p(s|M,R,\nu) = \prod_j^{n_\textrm{bins}}  \textrm{Pois} ( N^\gamma_j, \mu_j = f[M, R ,\nu]_j \Delta t  ),  \]

\noindent
where the final likelihood can be obtained by marginalizing over the nuisance parameters:
  
  \begin{equation} 
  \label{eq:mrlhood}
  p(s|M,R) = \int d\nu\ p(s|M,R,\nu)\ p(\nu) = \int d\nu \prod_j^{n_\textrm{bins}}  \textrm{Pois} ( N^\gamma_j, \mu_j = f[M, R ,\nu]_j \Delta t  ) p(\nu).
    \end{equation}

This estimate allows for a scan of the $M,R$ plane for the values which maximize $p(s|M,R)p(M,R)$ to find the MAP estimate for $(M,R)$ given a fixed $s$. Schematically, this process is shown in Fig~\ref{fig:schematic_mr}.

%\begin{figure}[hbt!]
%    \centering
%\includegraphics[width=0.75\textwidth]{figs/spectra_to_mr_fwd_schematic_paper.pdf}
%    \caption{Schematic diagram depicting the evaluation of the likelihood of producing on observed stellar spectrum by comparing it to the predicted spectrum for assumed values of stellar mass and radius. Varying the assumed mass and radius allows for an evaluation of the likelihood across the $(M,R)$ parameter space.}
 %   \label{fig:schematic_mr}
%\end{figure}

\subsection{Learning the Model \textit{f} for Stellar Spectra}

The estimated likelihood requires learning a function $f[M, R, \nu]$ which produces the expected stellar spectrum in each bin. This function is modeled by a deep neural network with five inputs ($M$, $R$, and the three components of $\nu$: effective temperature, distance, and $N_H$) and 250 outputs, one for each of the spectral bins.

The network architecture includes two input branches, one to process the mass and radius, and another to process the nuisance parameters. Each branch contains a series of nine layers, with 2048 nodes and leaky ReLU activation, that process its inputs in isolation. Following these initial layers, the output from the branches is combined together, forming a single volume with all information. This grouping is then passed to another successive set of nine layers which produce the generated spectra. The choice of hyperparameters has a significant impact on performance of this network. A complete list of hyperparameters tried is given in \ref{appendix:hypo}; the architecture used for model $f$ was the configuration with the best performance on the validation set.

Generating the spectra is formulated as a supervised learning problem, where the network learns to minimize the error between the true spectra and its predictions for a set of simulated examples generated by \xspec. The weights are updated with gradient descent via backpropagation. The Huber loss function is used, which is a standard loss function for regression robust to outliers, and the Adam optimizer computes gradients and schedules the backward passes. 

\begin{figure}[hbt!]
    \centering
\includegraphics[width=0.45\textwidth]{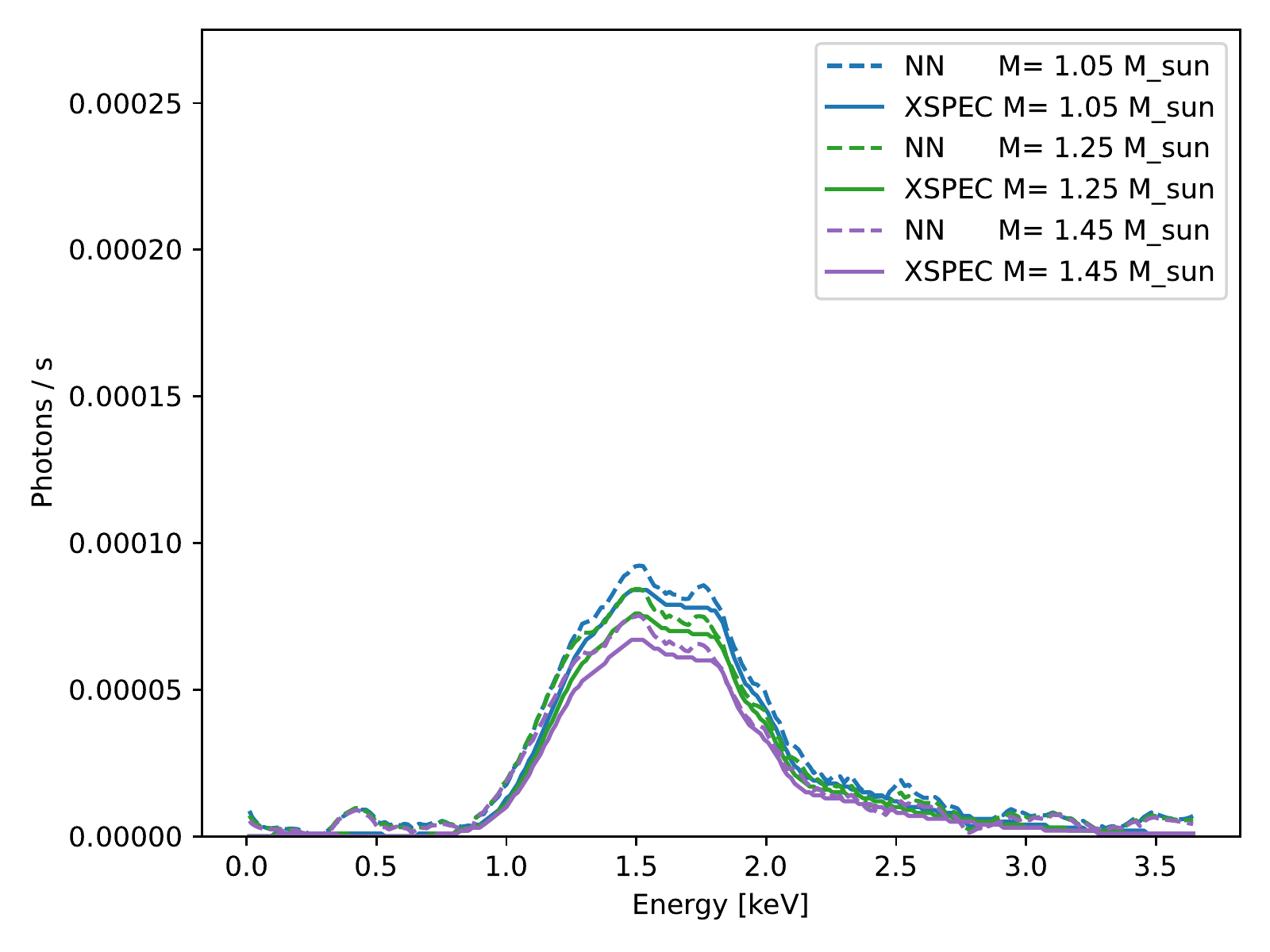}
         \includegraphics[width=0.45\textwidth]{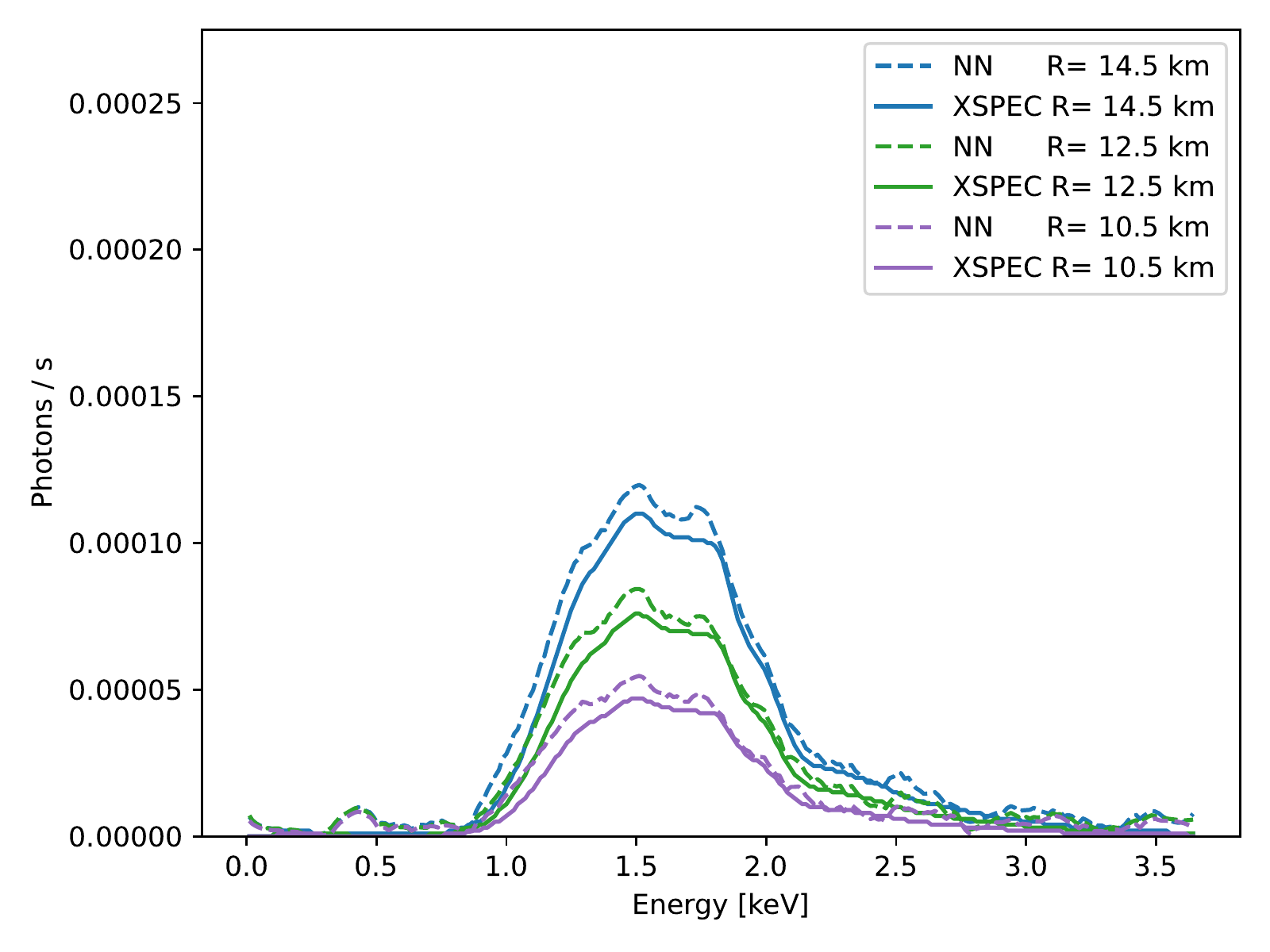}
    \includegraphics[width=0.45\textwidth]{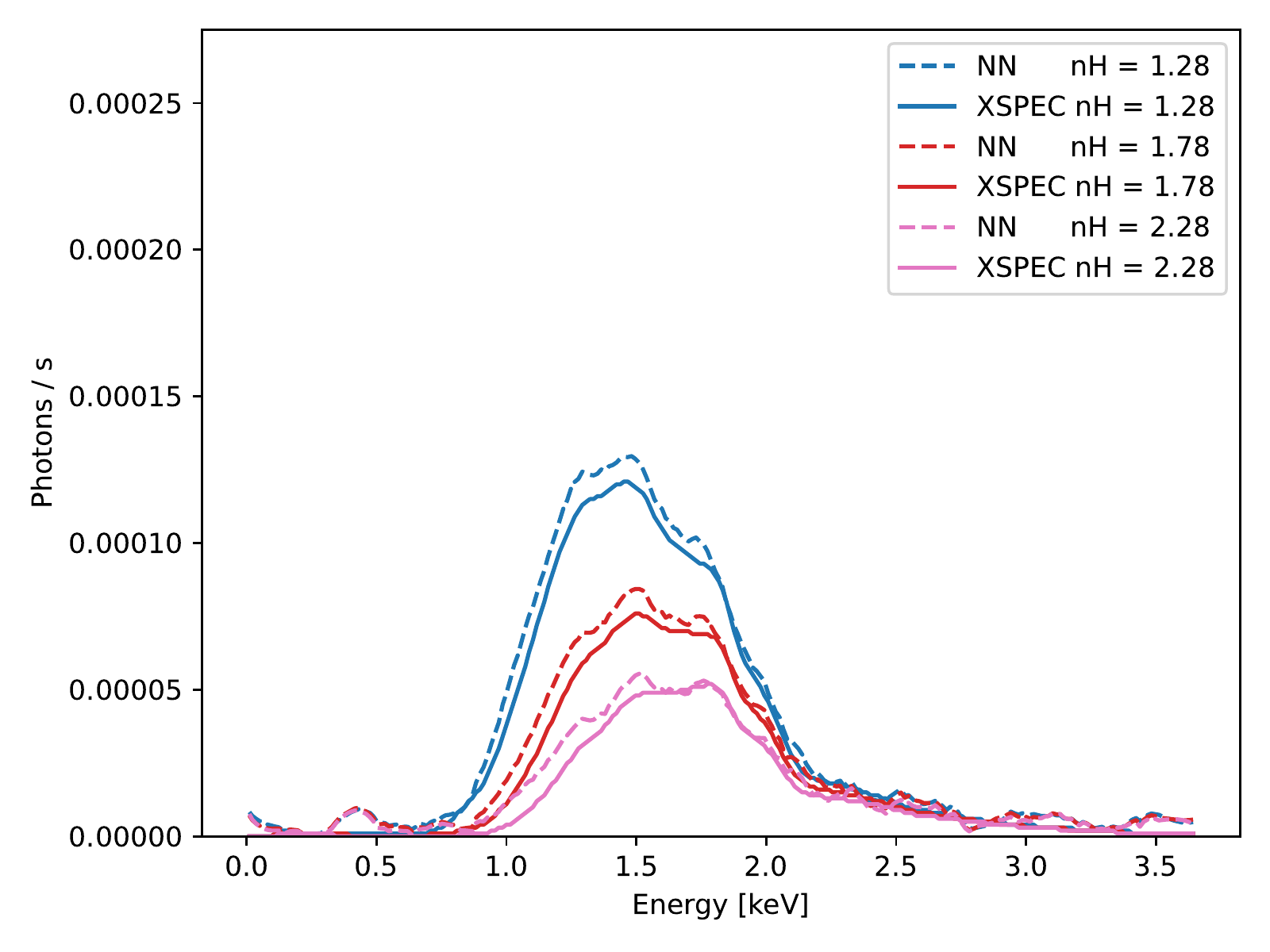}
    \includegraphics[width=0.45\textwidth]{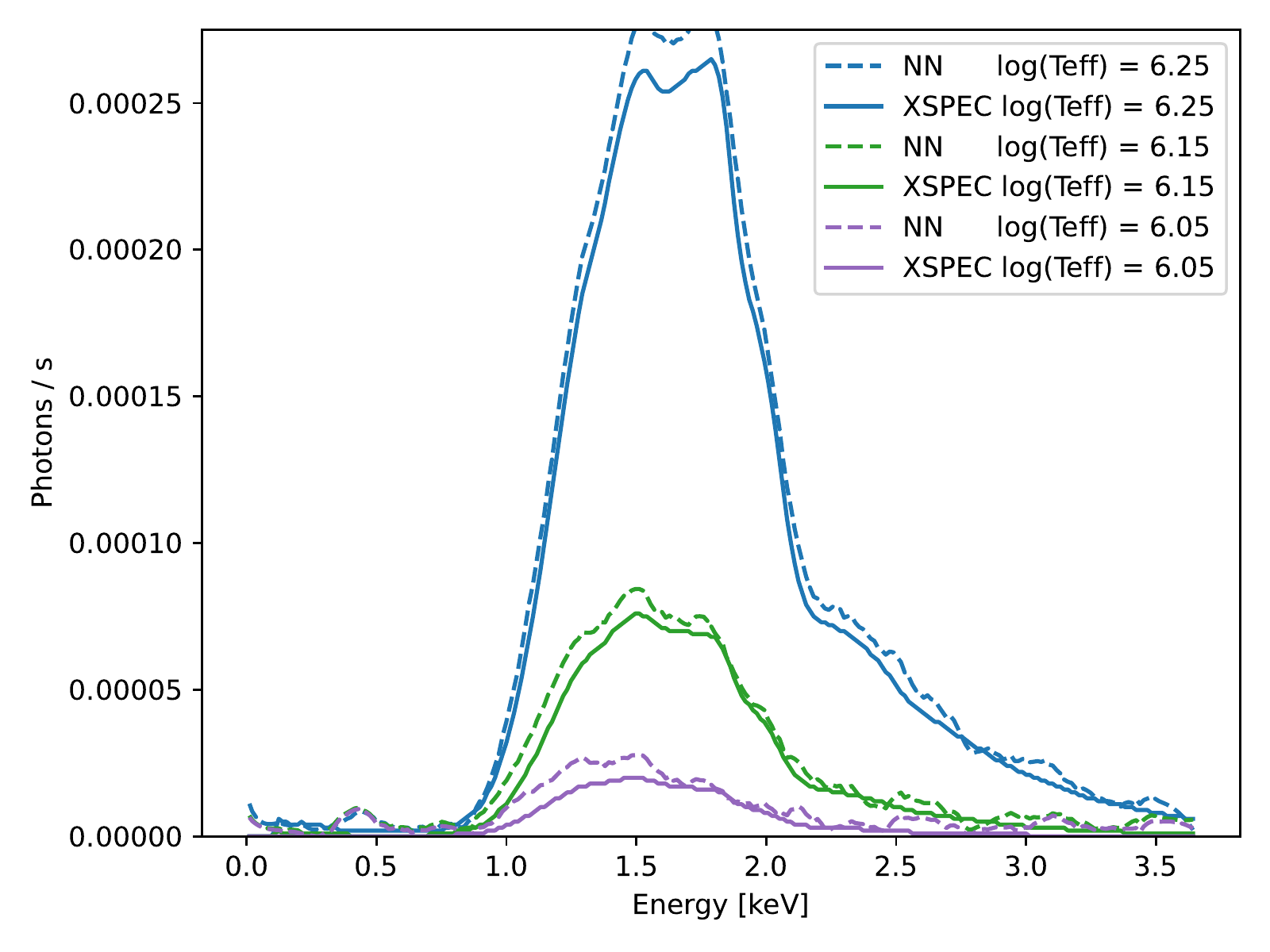}
    \includegraphics[width=0.45\textwidth]{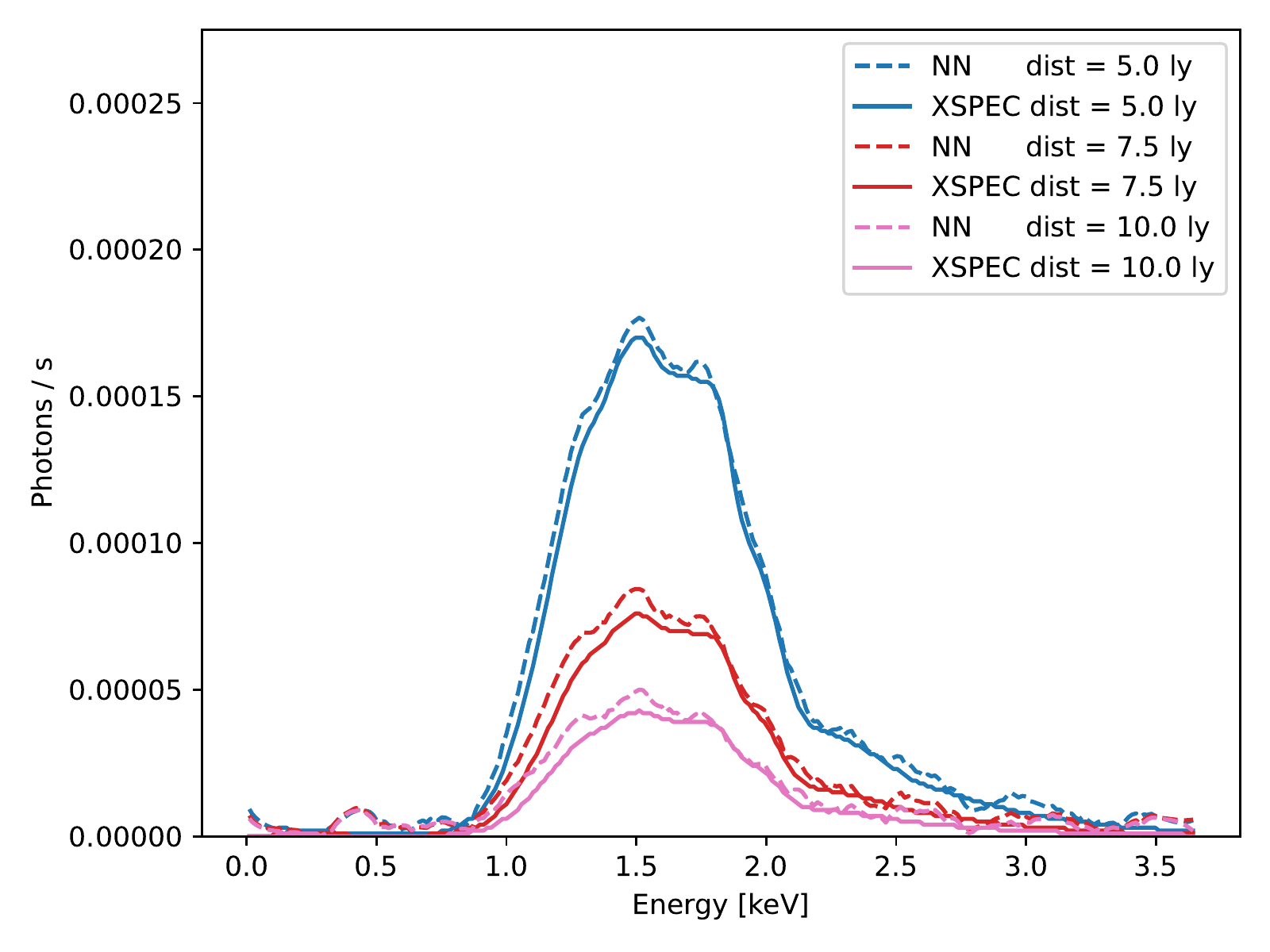}    
    \caption{ Comparison of neutron star X-ray spectra predictions (dashed) from our network $f[M, R, \nu]$ described in the text, as compared to training data generated by \xspec~ (solid).   Each pane shows the expected rate of photons ($\frac{dN_1^\gamma}{dt}$) in Chandra per energy bin, for variations of the parameters of interest (mass $M$, radius $R$) as well as for variations of the nuisance parameters $\nu$ ($n_H$, log($T_{\textrm{eff}}$), distance).}
    \label{fig:genspectra}
    %%%%% Plot generation notes
    % HPC3 /data/homezvol1/daniel/ns_moredata/NeutronStars
    % ns_env make_scan.py 1.25 12.5. produces "spectra.txt"
    % laptop /Users/danielwhiteson/ns/paper_plots/sim
    % plot_scan_comp.py spectra.txt
\end{figure}

Figure \ref{fig:genspectra} shows several examples of generated spectra. Each subplot contains generated results using various values of mass, radius, and nuisance parameters. The predictions are shown with their corresponding version from \xspec. The predictions track the \xspec~ values remarkably well across a range of parameter values. In general, we notice a slight overprediction by the network; below, we estimate the potential bias due to this overprediction and find it to be negligible compared to statistical uncertainties and other systematic uncertainties.

\subsection{Results}

\begin{figure}
    \centering
     \includegraphics[width=0.37\textwidth]{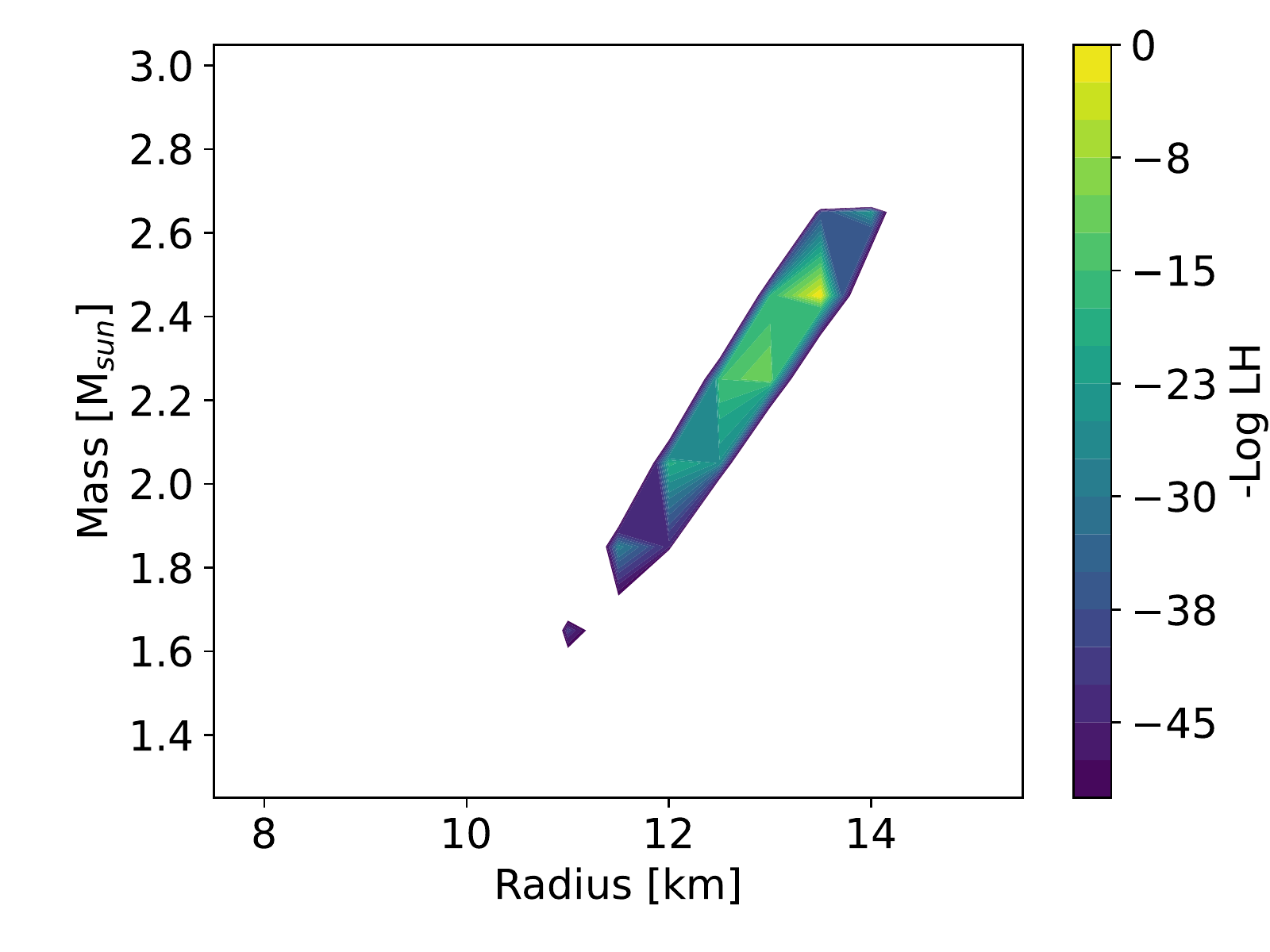}
    \includegraphics[width=0.37\textwidth]{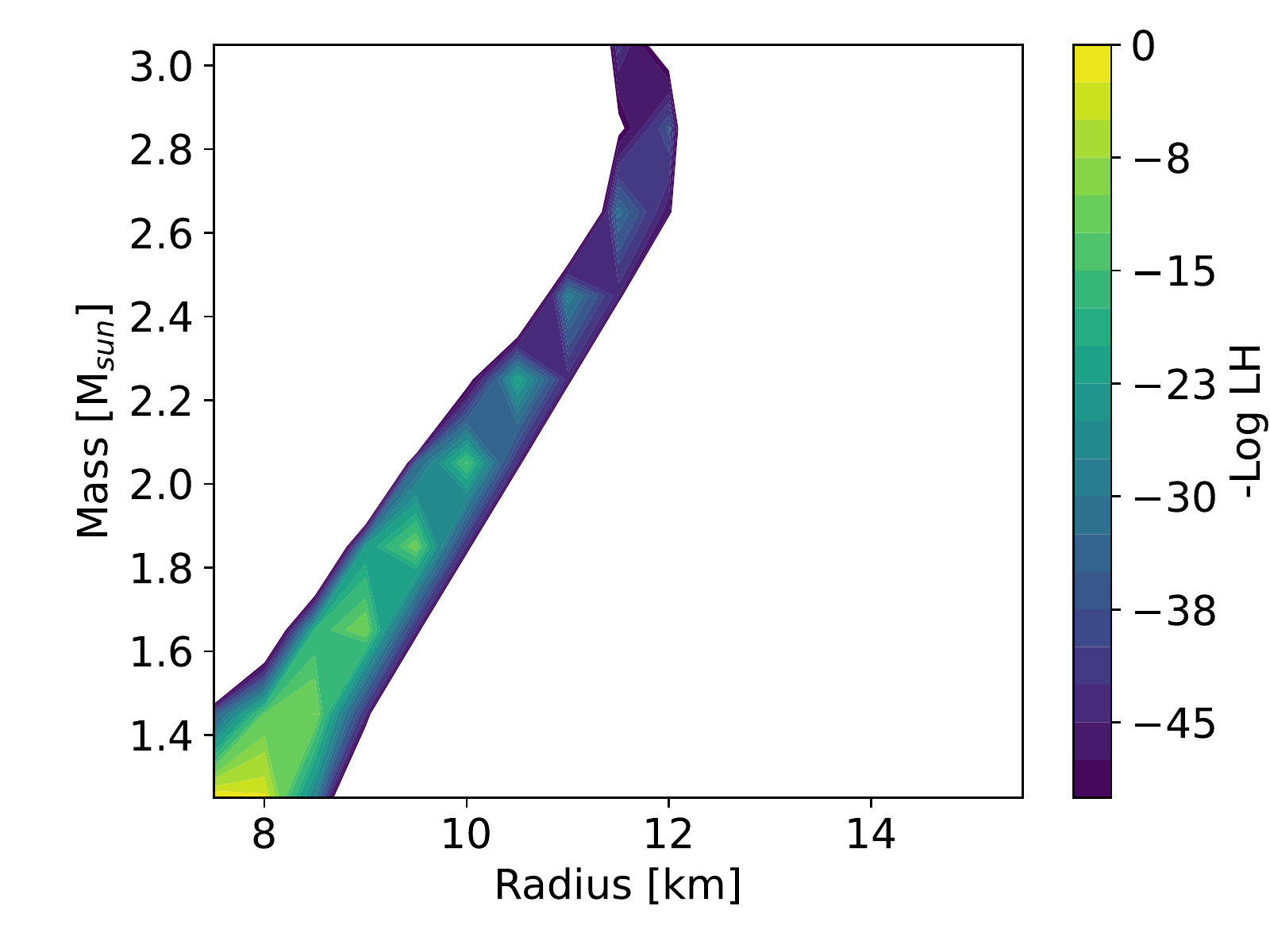}
     \includegraphics[width=0.37\textwidth]{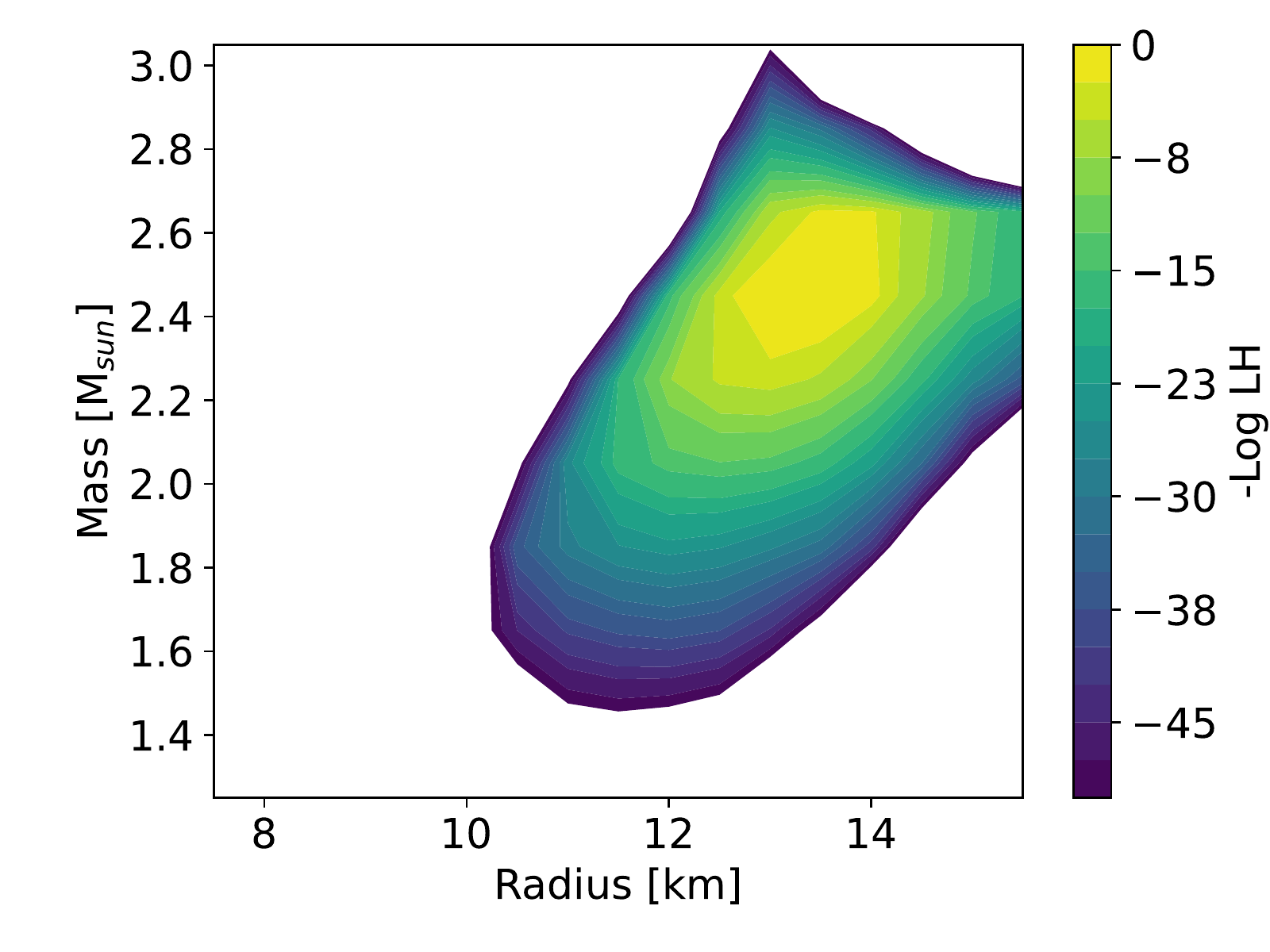}
    \includegraphics[width=0.37\textwidth]{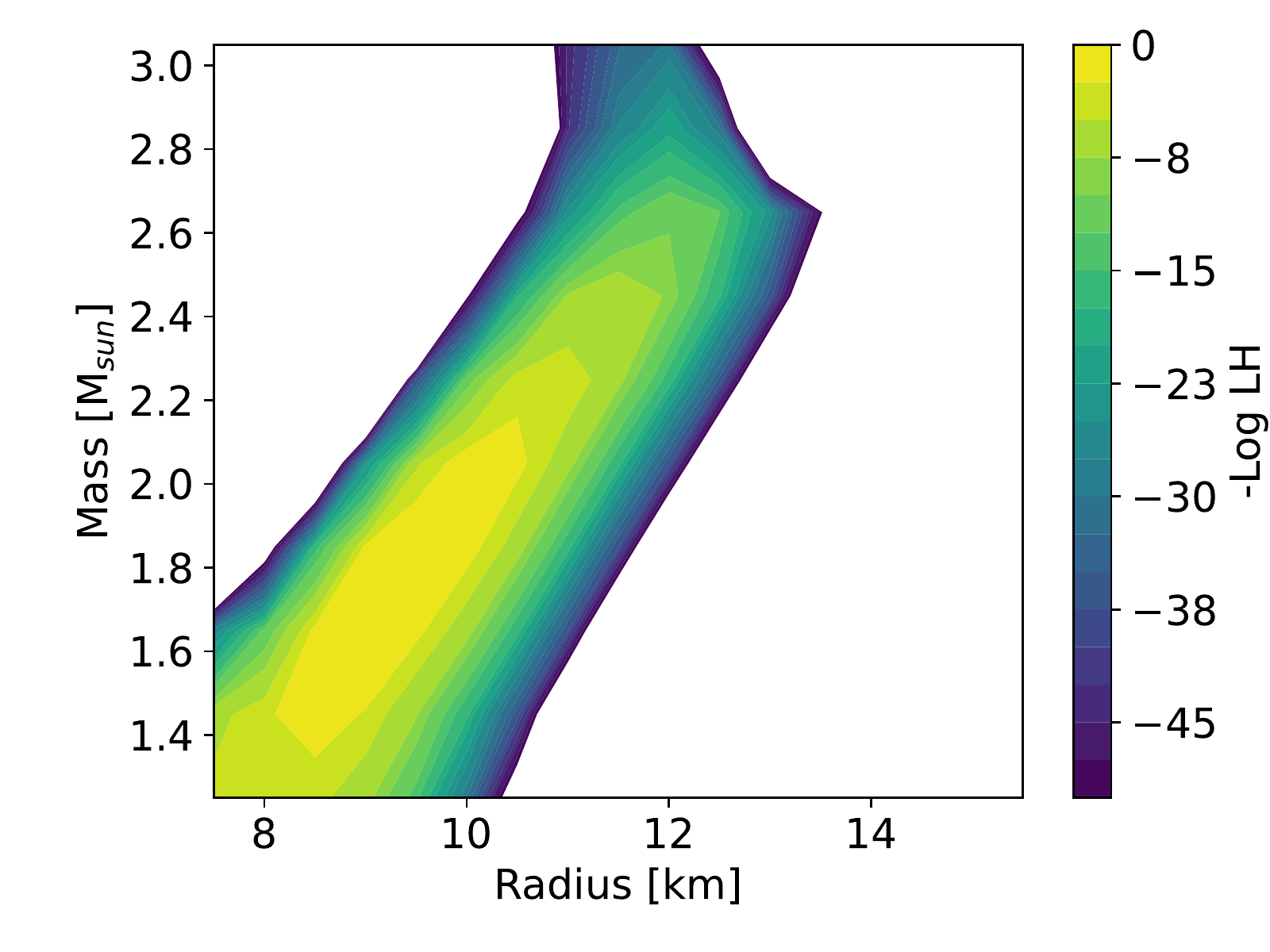}
    \includegraphics[width=0.37\textwidth]{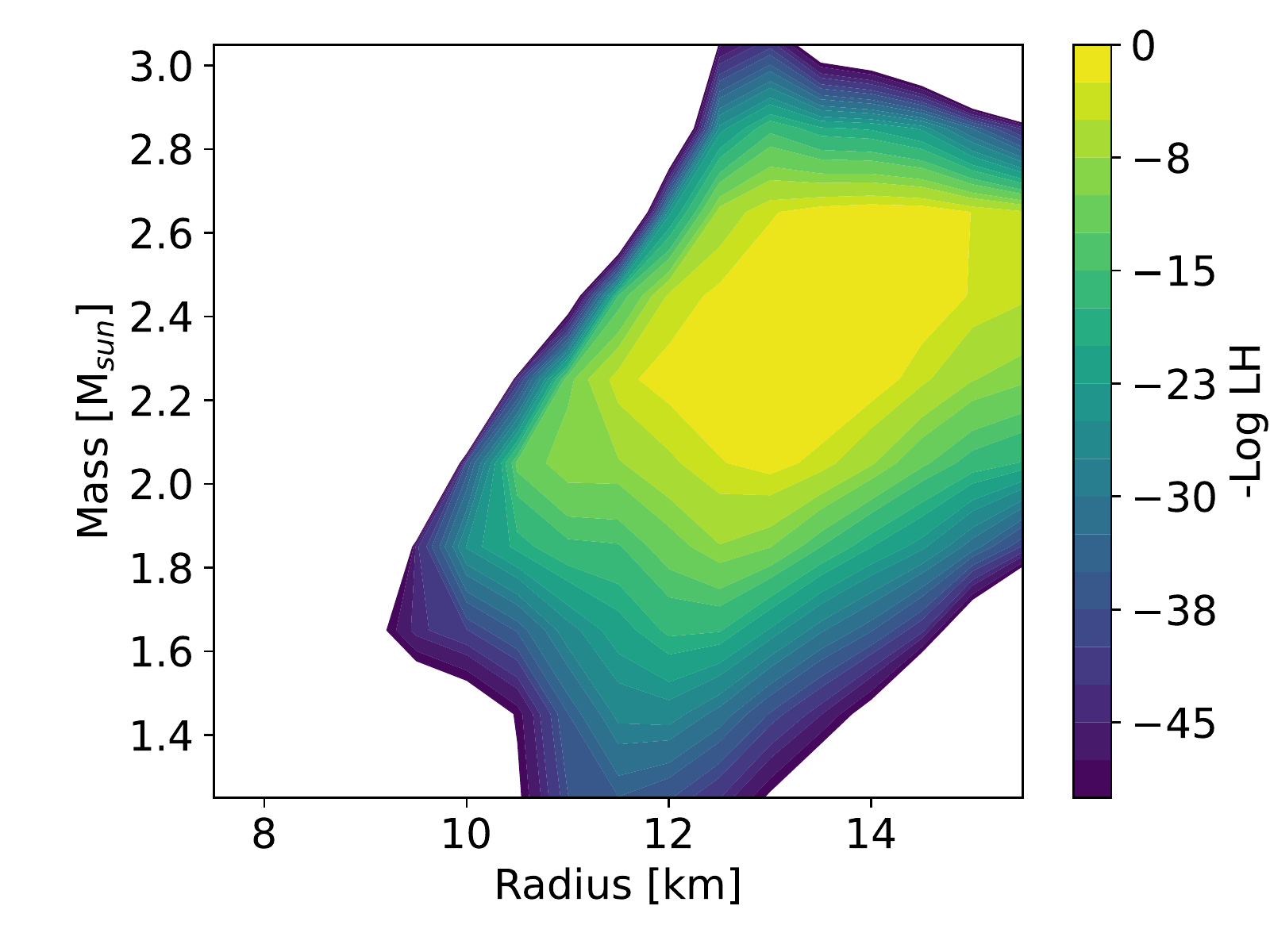}
   \includegraphics[width=0.37\textwidth]{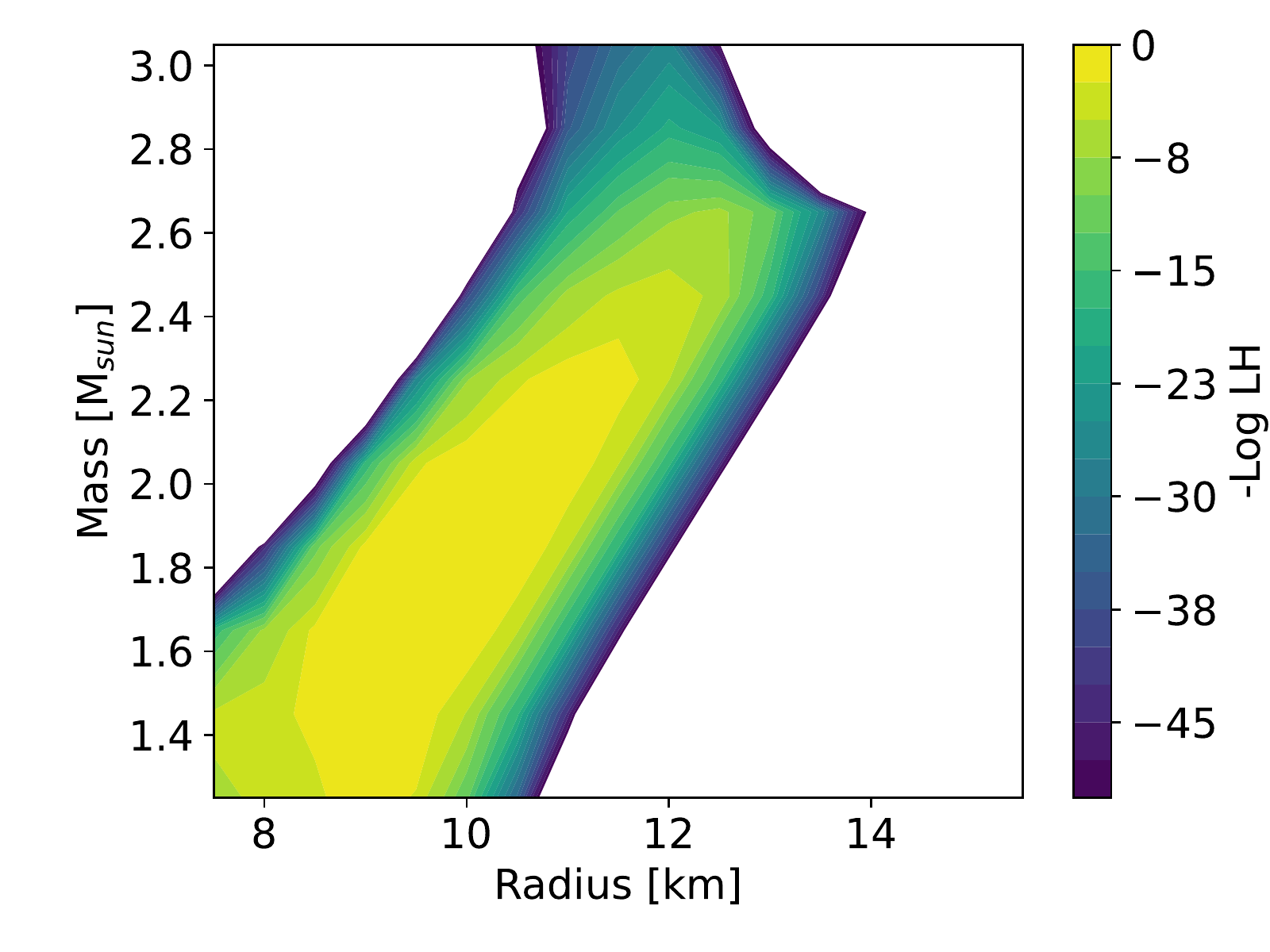}
    \caption{ Scans of the likelihood for two example stellar spectra $s$ (left, right) versus stellar mass and radius. Top demonstrates the ideal nuisance parameter (NP) conditions where the NPs are fixed to their true values. For the same simulated observed spectra, center shows a more realistic ``tight'' scenario , and bottom shows a ``loose'' scenario in which the NPs are not well constrained by priors. In the ``loose'' and ``tight'' scenarios, dependence on the nuisance parameters has been integrated out as described in the text.}
    \label{fig:lhmr_loose}
\end{figure}

The machine-learning-derived likelihood is tractable, allowing for a scan of $p(s|M,R)$ for individual stars. Figure~\ref{fig:lhmr_loose} shows examples of two individual simulated stars under the three nuisance parameter scenarios.

The likelihood can then be used to produce estimated values of neutron star mass and radius for a given spectrum, after marginalizing over the nuisance parameters. The mass-radius plane is scanned using adaptive optimization to find values that maximize the product of the likelihood and the prior.

\begin{figure}[hbt!]
    \centering
    \includegraphics[width=0.45\textwidth]{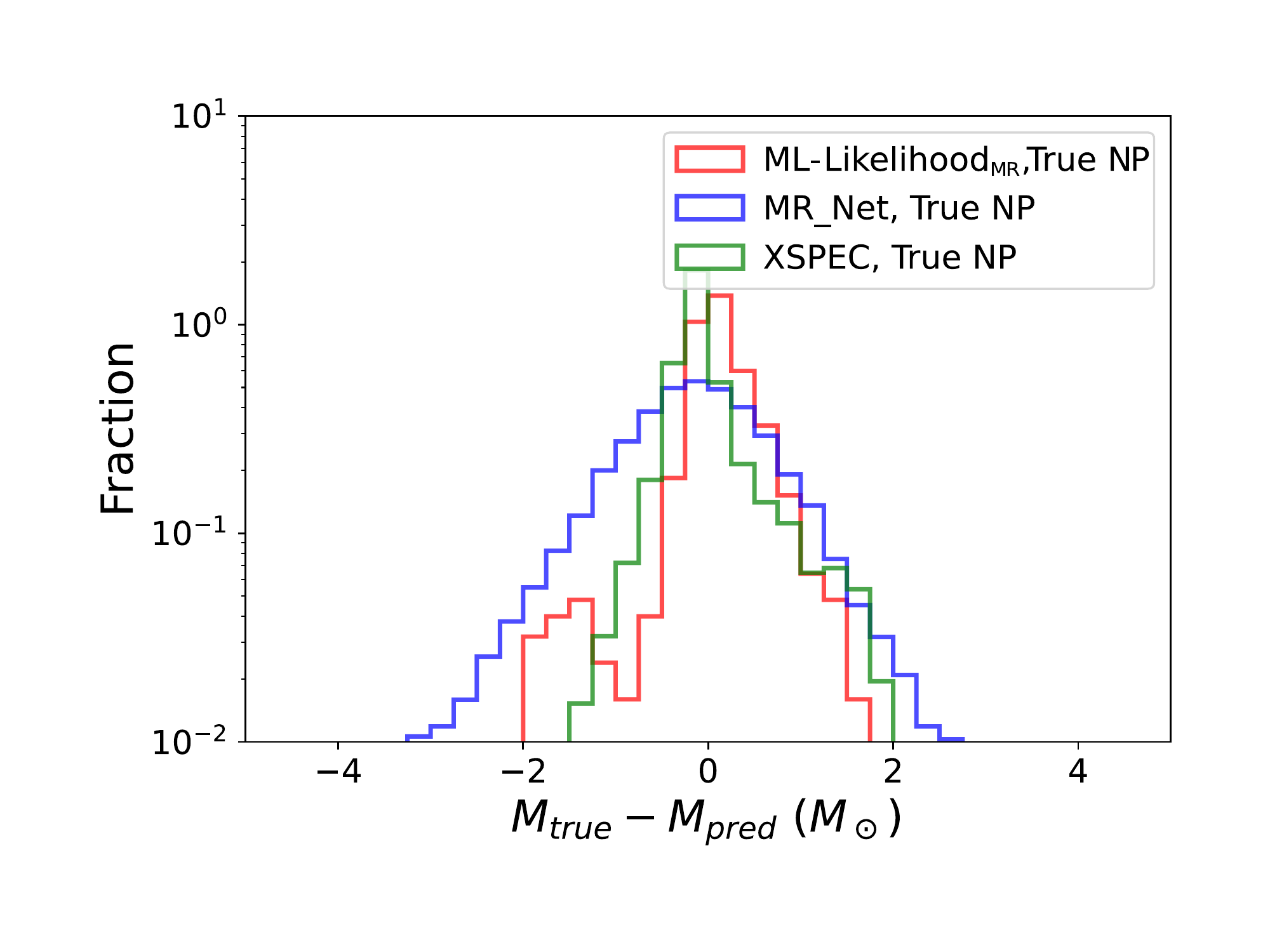}
        \includegraphics[width=0.45\textwidth]{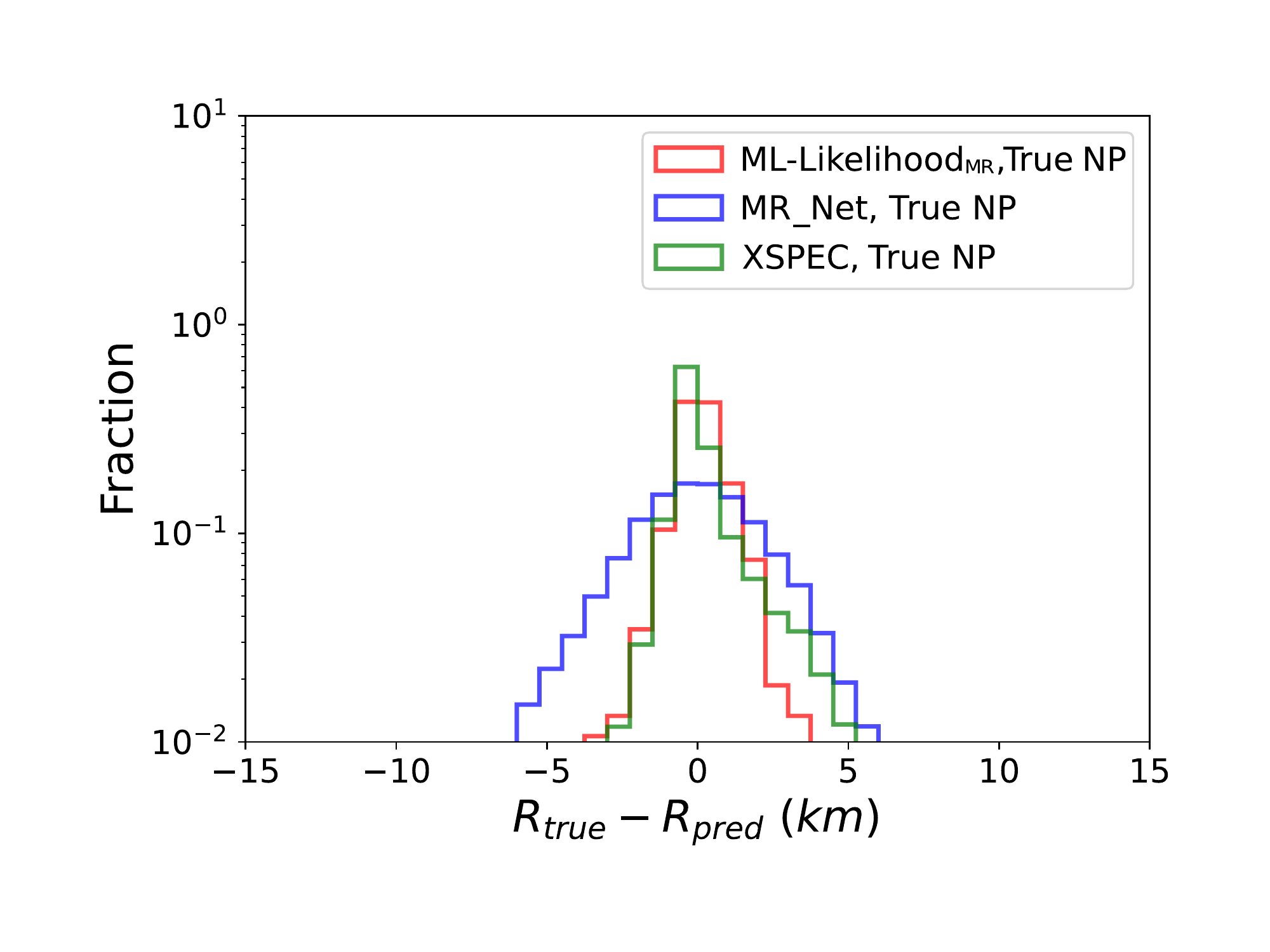}
    \includegraphics[width=0.45\textwidth]{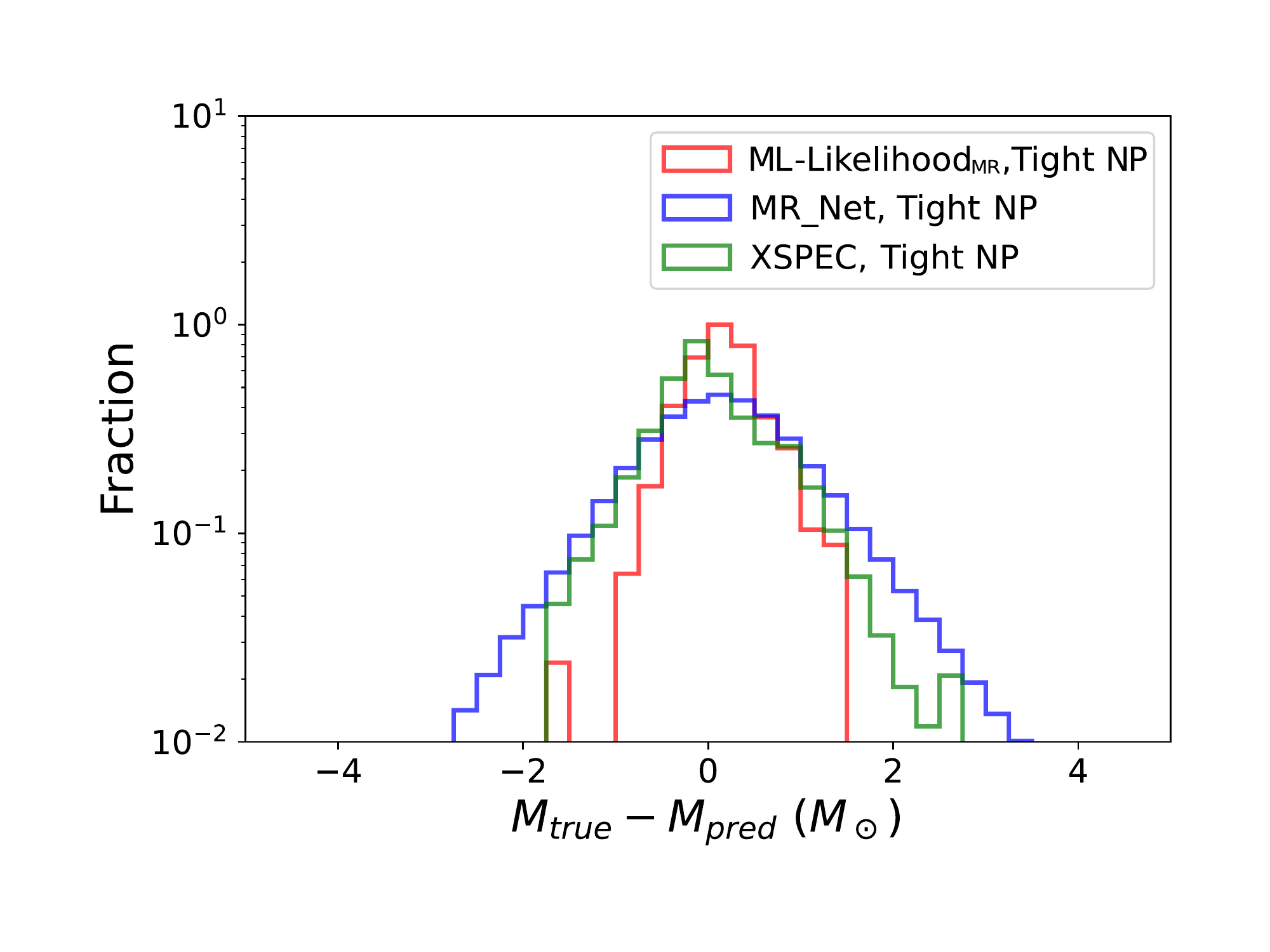}
        \includegraphics[width=0.45\textwidth]{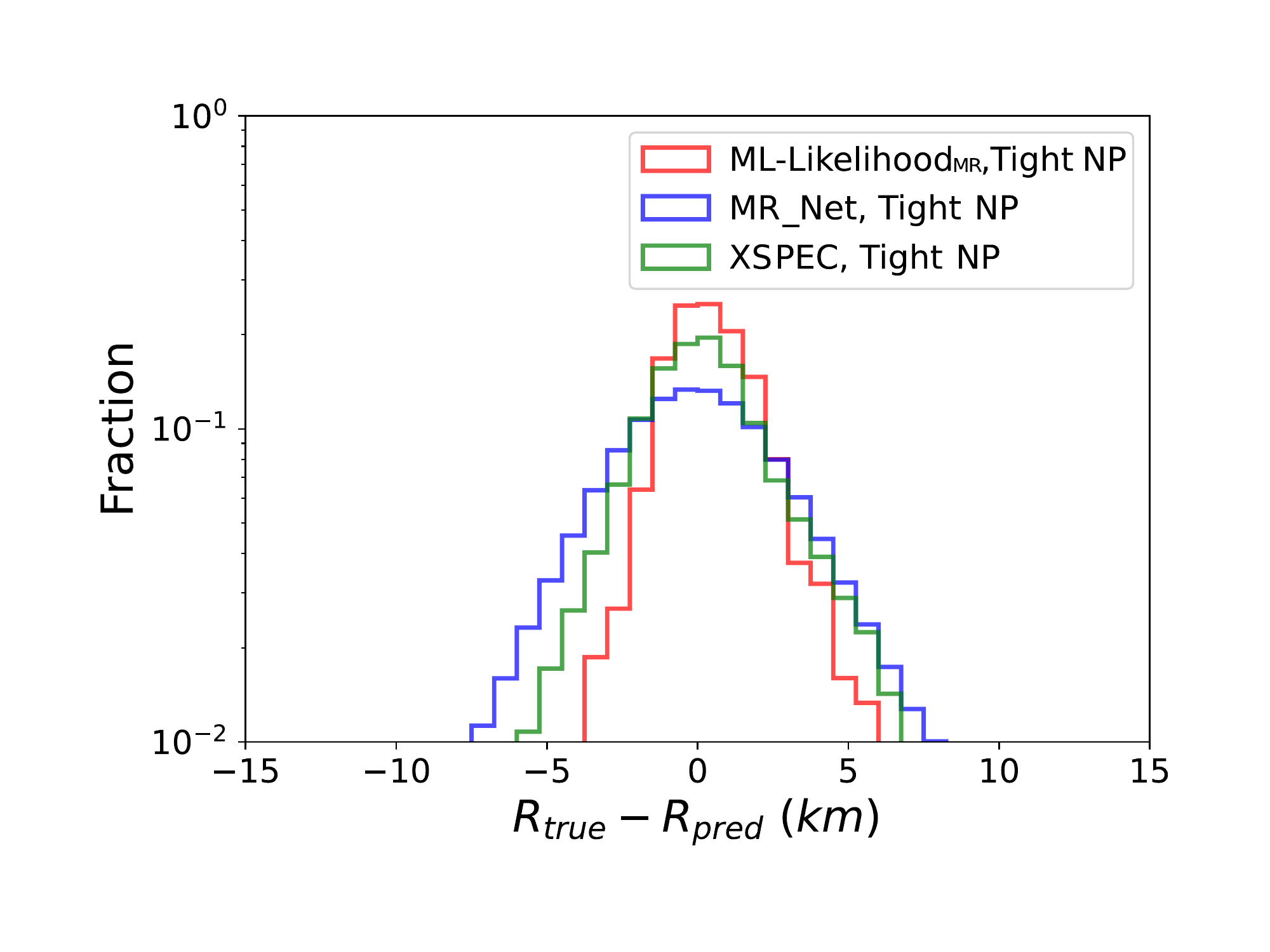}
    \includegraphics[width=0.45\textwidth]{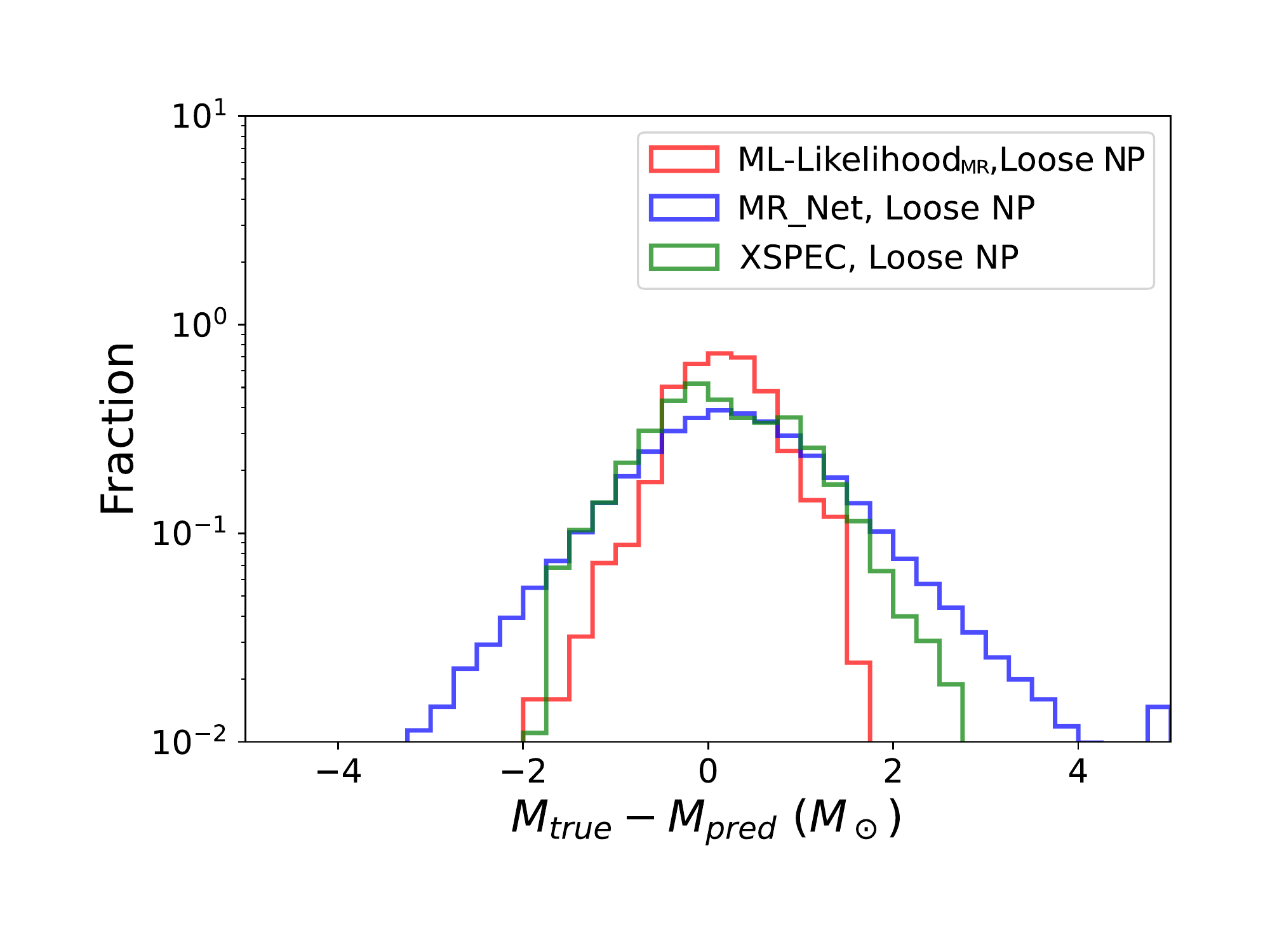}
        \includegraphics[width=0.45\textwidth]{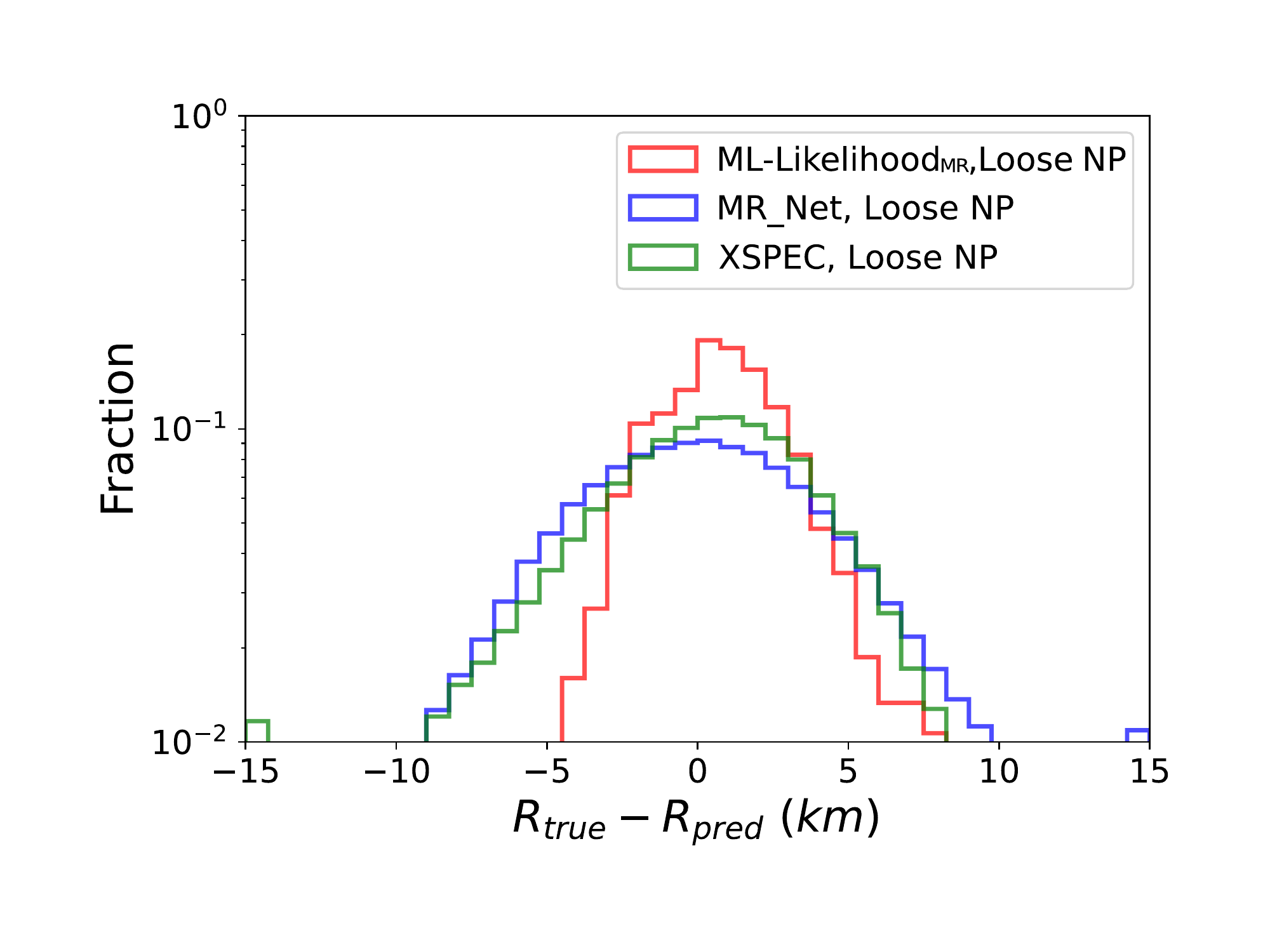}
    \caption{ Performance of our estimation of neutron star mass (left) and radius (right) using an approximate likelihood which incorporates neural networks, \surrMR, in comparison to the performance of a pure regression network, \mrnet~\cite{delaney-eos} and the \xspec~ tool. Shown is the residual, the difference between the true and predicted values, for three scenarios of nuisance parameter uncertainties.    In the ``true" case, the NPs are fixed to their true values; in the ``tight" and ``loose" cases, they are drawn from narrow or wide priors, respectively; see text for details. }
    \label{fig:res_forward_mr}
\end{figure}

We assess the performance of this method to produce likelihood estimates, \surrMR, by calculating the residual between the true values of the mass and radius and the estimates produced by our method.  As benchmarks, we compare the residuals to those generated by \xspec~ itself, which has access to the true likelihood used to generate the simulated samples, as well as a regression-based method, \mrnet, described in Ref.~\cite{delaney-eos}. Figure~\ref{fig:res_forward_mr} and Table~\ref{tab:mr} shows a comparison of the results.

It is striking that \surrMR~ outperforms \mrnet~ dramatically, despite being trained on the same dataset. \surrMR~ benefits from the physics knowledge encoded in the approximate likelihood, which requires ML solutions only for a specific sub-task, rather than having to blindly learn the entire problem, as \mrnet~ must. \surrMR~ performs essentially as well as \xspec, even having smaller residuals when nuisance parameters have the most uncertainty, despite not having access to the explicit likelihood used to generate the data, as \xspec~ does. Note that the performance comparison here highlights a {\it practical} difference between our method and use of \xspec, rather than a {\it principled} difference in the methods. That is, our method of replacing the difficult calculational step with a learned ML model allows for convenient and rapid evaluation of the likelihood over many values of the nuisance parameters, enabling us to marginalize over them, which explains the improved relative performance. While in principle one could perform the same operation with \xspec, its lack of a convenient programmable interface makes this impractical. The practical distinction is crucial, however, as it allows for greater general flexibility, such as interpolation across several models, or in application to the broader EOS inference problem described below, where likelihoods are not just hidden behind inconvenient interfaces, but completely unavailable.

\begin{table}[h]
    \caption{ Performance of our estimation of neutron star mass (left) and radius (right) using an approximate likelihood which incorporates neural networks, \surrMR, in comparison to the performance of a pure regression network, \mrnet~\cite{delaney-eos} and the \xspec~ tool. Shown are the mean ($\mu$) and standard deviation ($\sigma$) of the residual distributions under three scenarios of nuisance parameter uncertainties.    In the ``true" case, the NPs are fixed to their true values; in the ``tight" and ``loose" cases, they are drawn from narrow or wide priors, respectively; see text for details.}
    \label{tab:mr}
    \centering
    \begin{tabular}{lcrrrrr}
             \toprule
         & Nuis. & \multicolumn{2}{c}{Mass} & \ \ & \multicolumn{2}{c}{Radius}  \\
        \cline{3-4} \cline{6-7}\\
         Method & Params & $\mu$ & $\sigma$ && $\mu$ & $\sigma$ \\
                  \hline 
        \surrMR   & True  & 0.11  & 0.47  && 0.28  & 1.43     \\
        \xspec~\   & True  & -0.01 & 0.50  && 0.23  & 1.44     \\
        \mrnet\   & True  & -0.14 & 0.93  && -0.7  & 2.80  \\
         \hline
         \surrMR   & Tight & 0.15  & 0.50  && 0.57  & 1.87     \\
        \xspec~\   & Tight & 0.06  & 0.73  && 0.24  & 2.61     \\
        \mrnet\   & Tight & 0.17  & 1.06  && 0.06  & 3.52  \\
         \hline
         \surrMR   & Loose & 0.15 & 0.58 && 0.91  & 2.42    \\
        \xspec~\   & Loose & 0.18  & 0.86  && -0.15 & 4.32     \\
        \mrnet\   & Loose & 0.28  & 1.29  && 0.14  & 4.93  \\
        
         \hline \hline
         \end{tabular}
\end{table}

\section{Equation of State Inference}
\label{sec:eos}

The ultimate goal is to estimate the EOS parameters ($\lambda_1,\lambda_2$) given a  set of spectra $S = (s_1,s_2...,s_{n_{\textrm{stars}}})$.  In principle, this would be straightforward if one could evaluate  $p(S|\lambda_1,\lambda_2)p(\lambda_1,\lambda_2)$, which would allow for maximization to find an estimate for $\lambda_1,\lambda_2$ for a fixed $S$. 

We begin with the assumption that the EOS parameters have a uniform prior within their physical boundaries of $\lambda_1 \in [4.75,5.25], \lambda_2 \in -[1.85,-2.05]$ (see Fig 3 of Ref.~\cite{delaney-eos}).  The remaining step is evaluating $p(S| \lambda_1,\lambda_2)$.

First we express the probability over the entire set $S$ as the joint probability for each star $s_i$:

\[ p(S | \lambda_1,\lambda_2 ) = \prod^{n_{\textrm{stars}}}_i  p(s_i|\lambda_1,\lambda_2).\]

The obstacle is that we do not know how to evaluate $p(s|\lambda_1,\lambda_2)$, only $p(s|M,R)$, which depends on stellar parameters $M$ and $R$. Linking these expressions is not trivial, as the EOS parameters $\lambda_1,\lambda_2$  do not uniquely determine stellar parameters $M$ and $R$, instead they only determine the $M$-$R$ {\it relation}. That is, each point in ($\lambda_1,\lambda_2$) space specifies a curve in $M$-$R$ space. The solution is to integrate over the $M$-$R$ curve allowed by the EOS parameters $\lambda_1,\lambda_2$. This is most directly accomplished by expressing the integral over the mass-radius plane, constrained by a delta function which traces out the $M$-$R$ curve determined by the EOS parameters $\lambda_1,\lambda_2$:

\[ p(s|\lambda_1,\lambda_2) = \int dM dR\  p(M,R|\lambda_1,\lambda_2) p(s|M,R), \]

%\[ p(S | \lambda_1,\lambda_2,\bar{\nu} ) = \prod^{n_{\textrm{stars}}}_i \int dM_i dR_i\  p(M_i,R_i|\lambda_1,\lambda_2) p(s_i|M_i,R_i,\nu_i)\]

\noindent
where $p(M,R|\lambda_1,\lambda_2)$ describes the allowed $M$-$R$ relation given the EOS specified by $\lambda_1,\lambda_2$, as

\[ p(M,R|\lambda_1,\lambda_2)  = \delta( h_{\lambda}[M] - R )\ \ p (M|\lambda_1,\lambda_2), \]

\noindent
where $h_\lambda[M] \rightarrow R$ is a function that gives the allowed value of $R$ for a value of $M$, determined by the EOS parameters $\lambda_1, \, \lambda_2$.  The function $h_\lambda[M]$ encodes all of the physics which translates the EOS into stellar mass and radius, and is not available analytically or tractable numerically. It is possible, however, to train a neural network to learn this function, as we do below. Assuming $h_\lambda[M]$ is available, we choose to integrate over mass, as each mass is mapped to a unique $R$; the same is not true for scanning in $R$, see Figure~\ref{fig:nn_mr}.  

The delta function reduces the double integral in $M$ and $R$  to a single integral over mass:

\[ p(s | \lambda_1,\lambda_2) = \int dM\ p (M|\lambda_1,\lambda_2)\ p(s|M,R=h_\lambda[M]),   \]

\noindent
where the range of the mass integral is limited to the physical region, from $1.2 M_{\odot} $ to  $1.6-3.25 M_{\odot}$, depending on the radius, see Figure~\ref{fig:nn_mr}. This allows us to write  an expression for the joint probability over the set of stars:

\[ p(S | \lambda_1,\lambda_2 ) = \prod^{n_{\textrm{stars}}}_i \int dM_i\ p (M_i|\lambda_1,\lambda_2)\ p(s_i|M_i,R_i=h_\lambda[M_i]).   \]

We have now expressed the likelihood $p(S | \lambda_1,\lambda_2 )$ in terms of the likelihood $p(s|M,R)$, which we previously learned to calculate. The equation for $p(s|M,R)$ now allows for the expression of a joint likelihood over the stars and the bins:

  \[ L_S(\lambda_1,\lambda_2)= p(S|\lambda_1,\lambda_2) = \prod^{n_{\textrm{stars}}}_i \int dM_i\ p (M|\lambda_1,\lambda_2) \int d\nu\prod^{n_{\textrm{bins}}}_j  \textrm{Pois} ( N^\gamma_{ij}, \mu_{ij}( M_i, R_i=h_\lambda[M_i],\nu ) p(\nu), \]

\noindent
where we can replace each of the $\mu_{ij}$ as we did above with $f[M, R ,\nu] \Delta t$

  \begin{equation} 
  \label{eq:eoslhood}
  L_S(\lambda_1,\lambda_2)= p(S|\lambda_1,\lambda_2) = \prod^{n_{\textrm{stars}}}_i \int dM_i\ p (M|\lambda_1,\lambda_2) \int d\nu \prod^{n_{\textrm{bins}}}_j  \textrm{Pois} ( N^\gamma_{ij}, \mu_{ij}= f[M_i, h_\lambda[M_i],\nu ]_j  \Delta t ) p(\nu).
  \end{equation}

This expression can be evaluated, assuming one can learn a function $h_\lambda[M] \rightarrow R$. The determination of the likelihood $L_S(\lambda_1,\lambda_2)$ is shown schematically in Fig.~\ref{fig:schematic_eos}.

%\begin{figure}[hbt!]
%    \centering
%\includegraphics[width=1.0\textwidth]{figs/spectra_to_eos_fwd_schematic_paper.pdf}
%    \caption{ Schematic diagram depicting the evaluation of the likelihood of producing a set of observed stellar spectra by comparing it to the predicted spectra along the mass-radius curve determined by EOS parameters $\lambda_1$, $\lambda_2$. Each value of the EOS parameters determines a curve in the mass-radius plane. Integrating along the curve, the probability of observing each star is evaluated as in Fig.~\ref{fig:schematic_mr}.}
 %   \label{fig:schematic_eos}
%\end{figure}

\subsection{Learning the Model $h_\lambda[M]$ for Stellar Radius}

\begin{figure}
    \centering
    \includegraphics[width=0.6\textwidth]{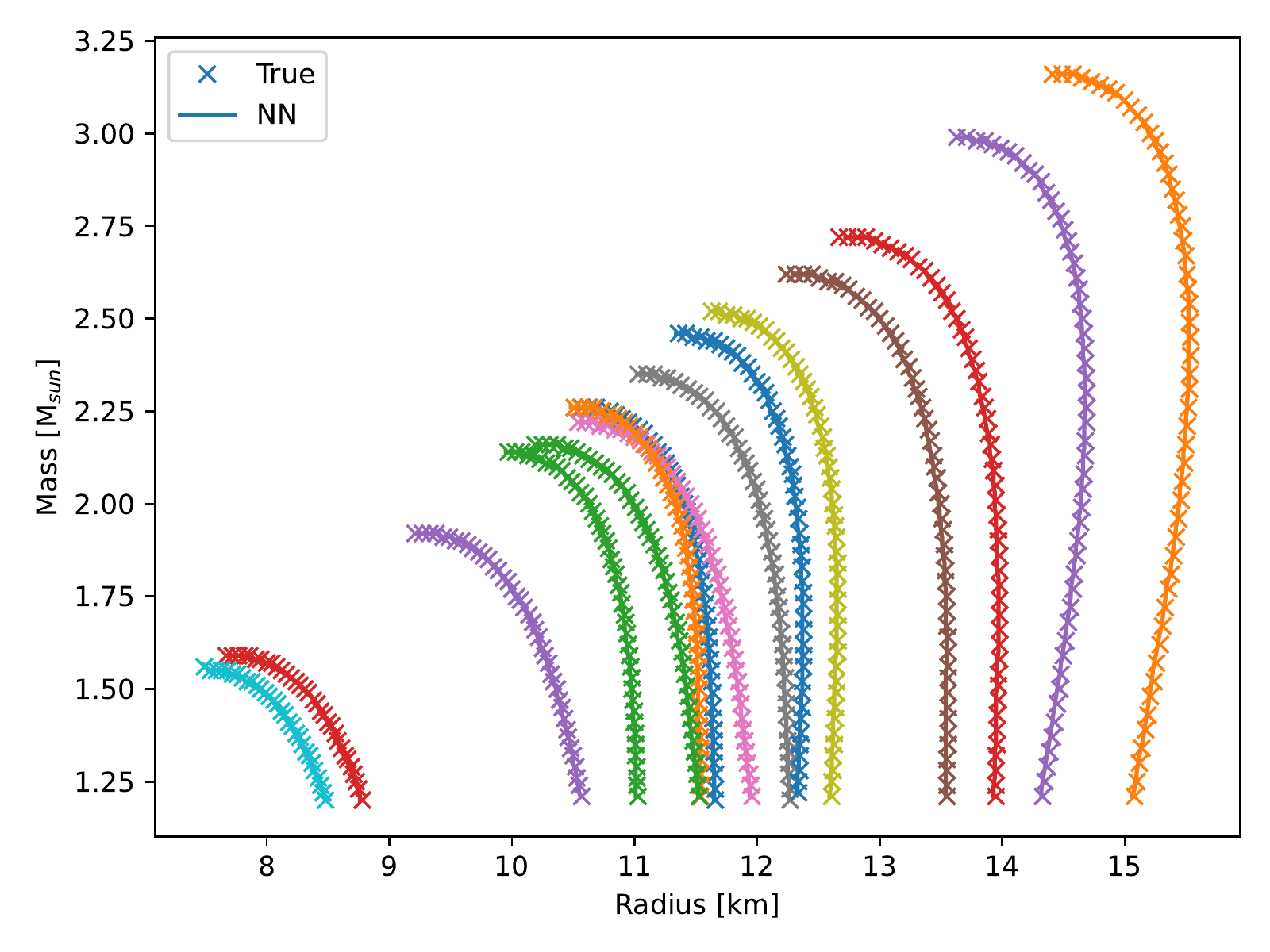}
    \caption{ Relationship between neutron star mass and radius, as determined by equation of state parameters $\lambda_1,\lambda_2$. Each color represents a single choice of EOS parameters, which determine a curve in the mass-radius plane. Individual calculations as described in the text are shown (crosses), as compared with the output of a neural network function $h_\lambda[M]$ (solid line), which estimates the radius corresponding to an input value of $M$ as determined by the EOS parameters.}
    \label{fig:nn_mr}
\end{figure}

The approximate likelihood above requires learning a function $h_\lambda[M]$ which estimates the stellar radius for a given stellar mass as determined by the EOS parameters $\lambda_1,\lambda_2$.  Note that one could equivalently estimate the mass from the radius, but this has the additional complication of degenerate outputs for some radii, see Fig.~\ref{fig:nn_mr}. It is important here to note that model $h_\lambda[M]$ is trained on M-R relations created by equations of state that do not feature a phase transition; parameterizing the M-R relation as a function of $R = h_\lambda[M]$ may miss details stemming from more exotic features, like those caused by a strong, first-order phase transition, in the EOS.

We model $h_\lambda[M]$ with a deep neural network comprising 10 hidden layers with 32 nodes each and ReLU activation. The output layer is a single node with linear activation, which is standard for regression. The number of hidden layers, their widths, and activation functions were optimized for the functionality of $h_\lambda[M]$; the relatively small width of 32 nodes and ReLU activation were found to perform well. The network was trained with Mean Squared Error (MSE) loss and the Adam optimizer. Figure~\ref{fig:nn_mr} demonstrates how the network $h_\lambda[M_i] \rightarrow R$ performs for a few example values of the EOS parameters $\lambda$. Generation of the training data as described above required approximately 24 hours of CPU time for $10^6$ stars; in comparison, evaluation of the network $h_\lambda[M]$ generates $10^6$ stellar radii in 400 ms, a relative speed enhancement of $10^5$.

\subsection{Results}

\begin{figure}
    \centering
     \includegraphics[width=0.35\textwidth]{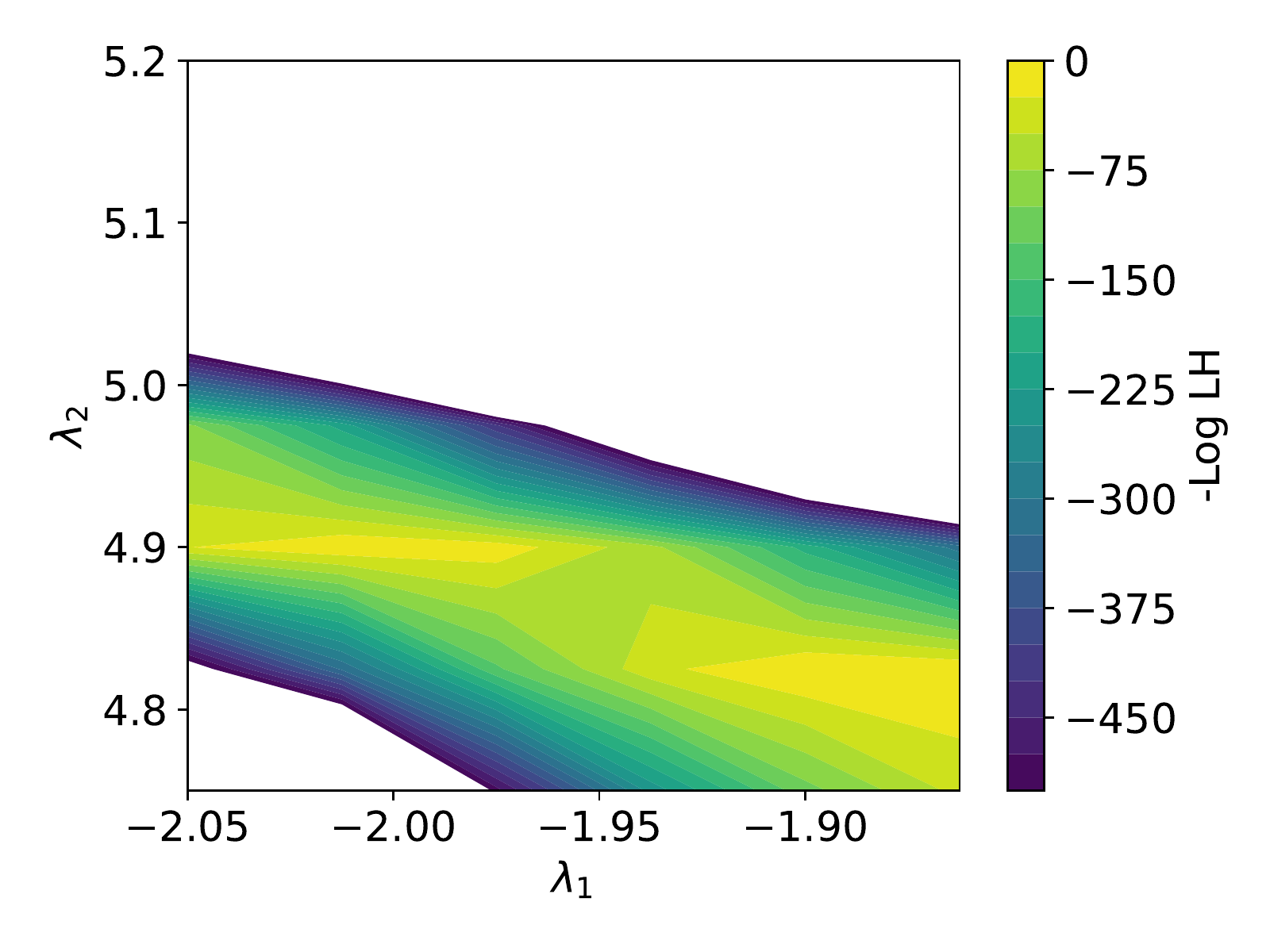}
    \includegraphics[width=0.35\textwidth]{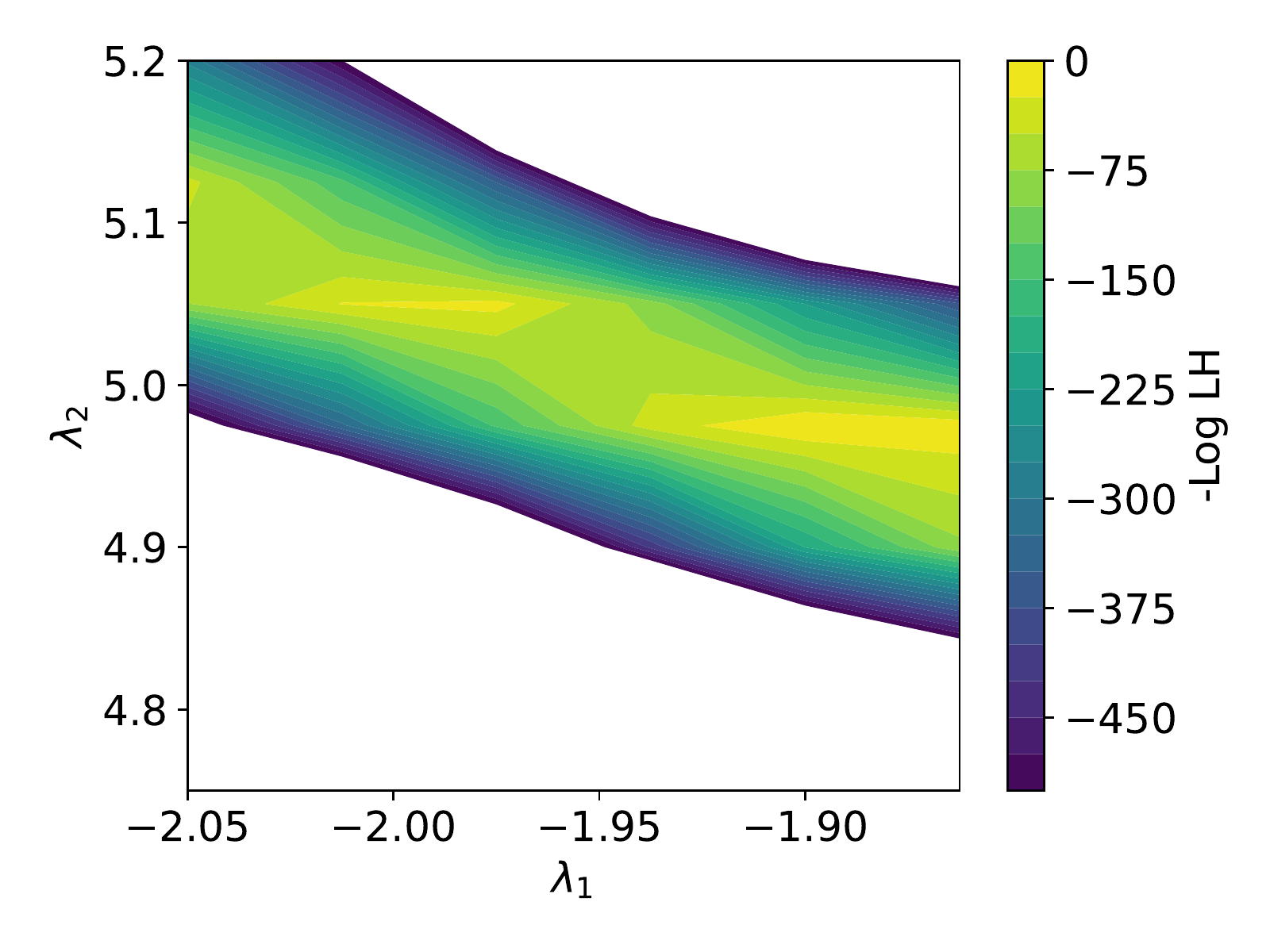}
     \includegraphics[width=0.35\textwidth]{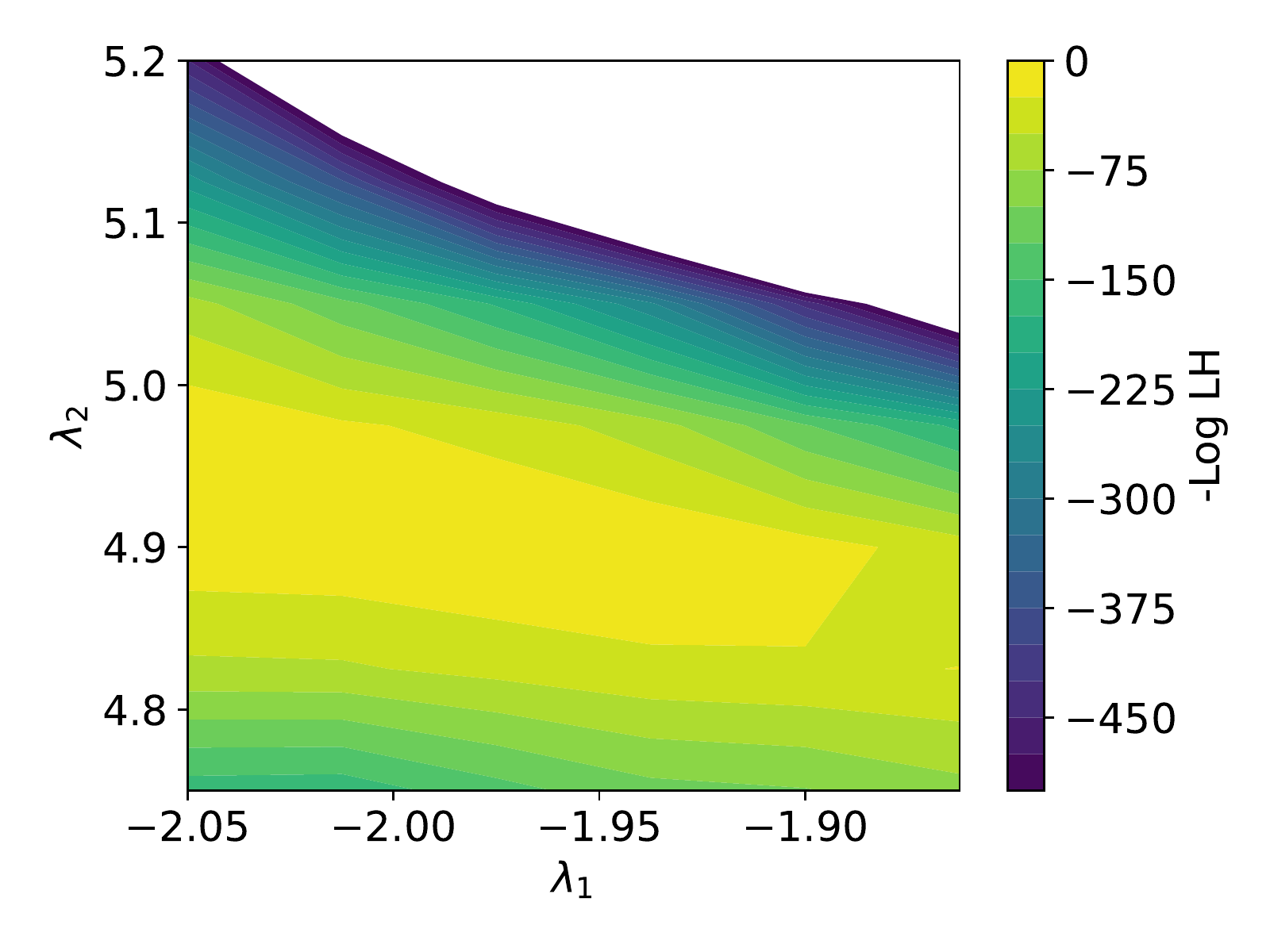}
    \includegraphics[width=0.35\textwidth]{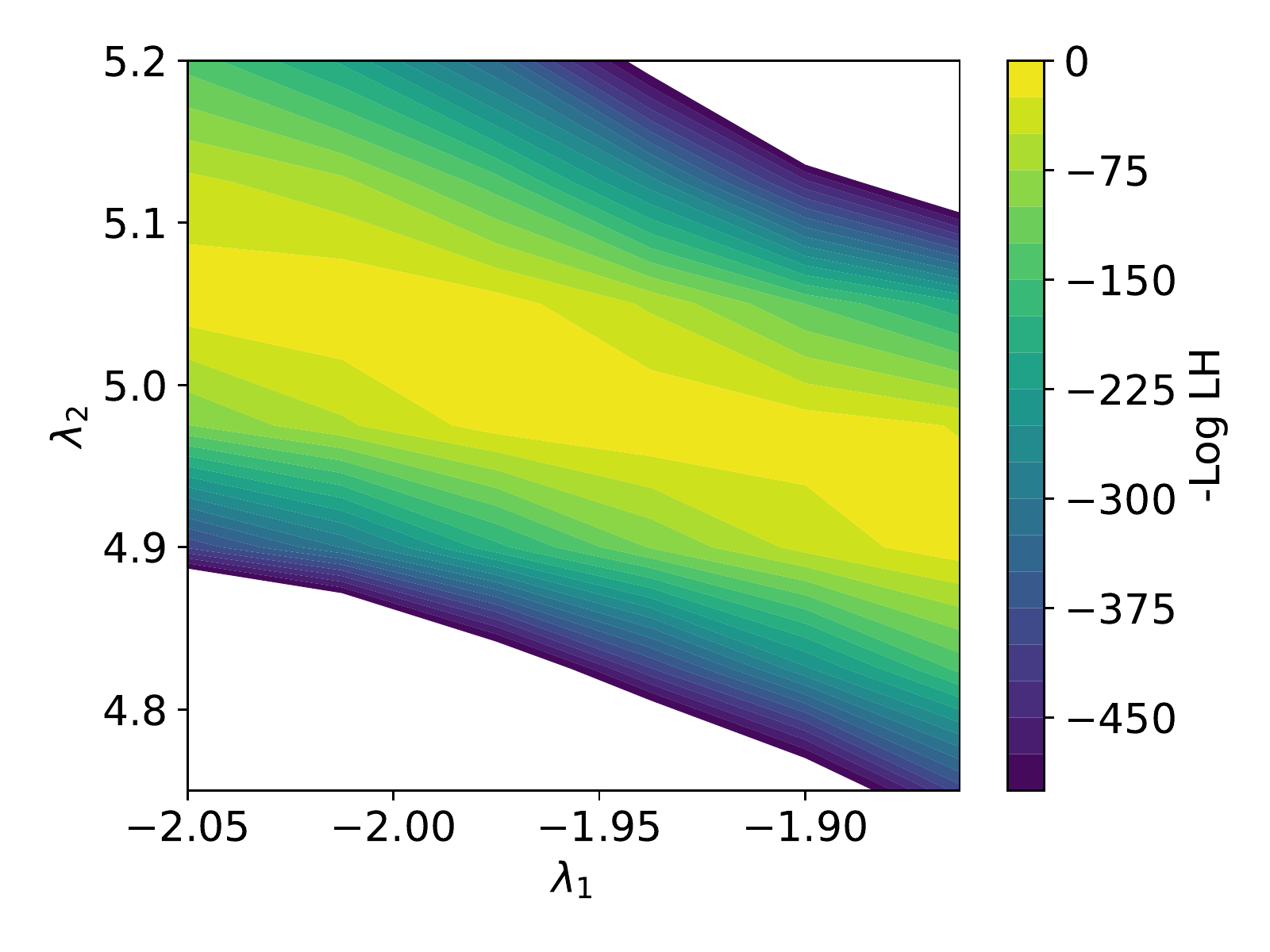}
    \includegraphics[width=0.35\textwidth]{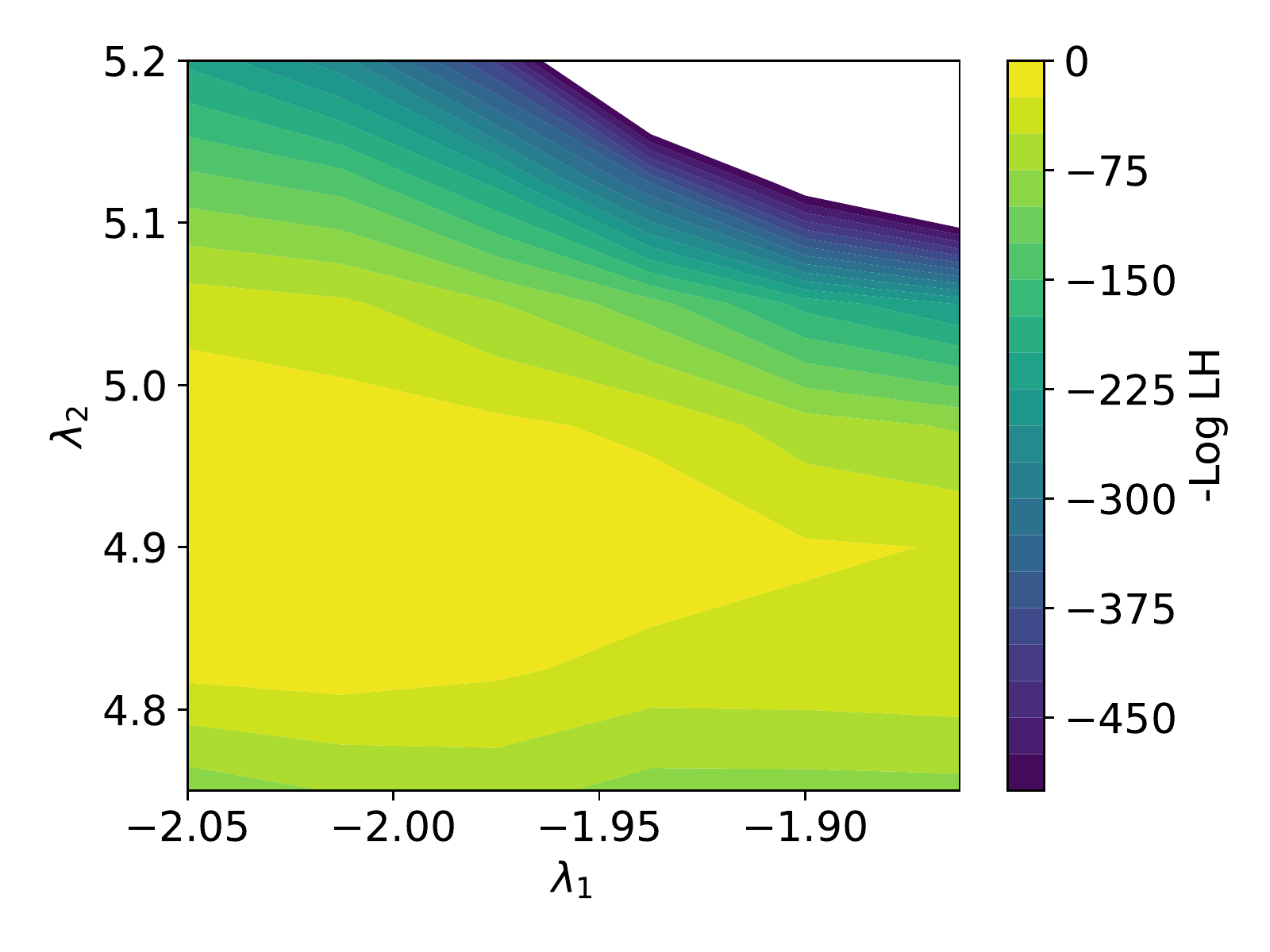}
   \includegraphics[width=0.35\textwidth]{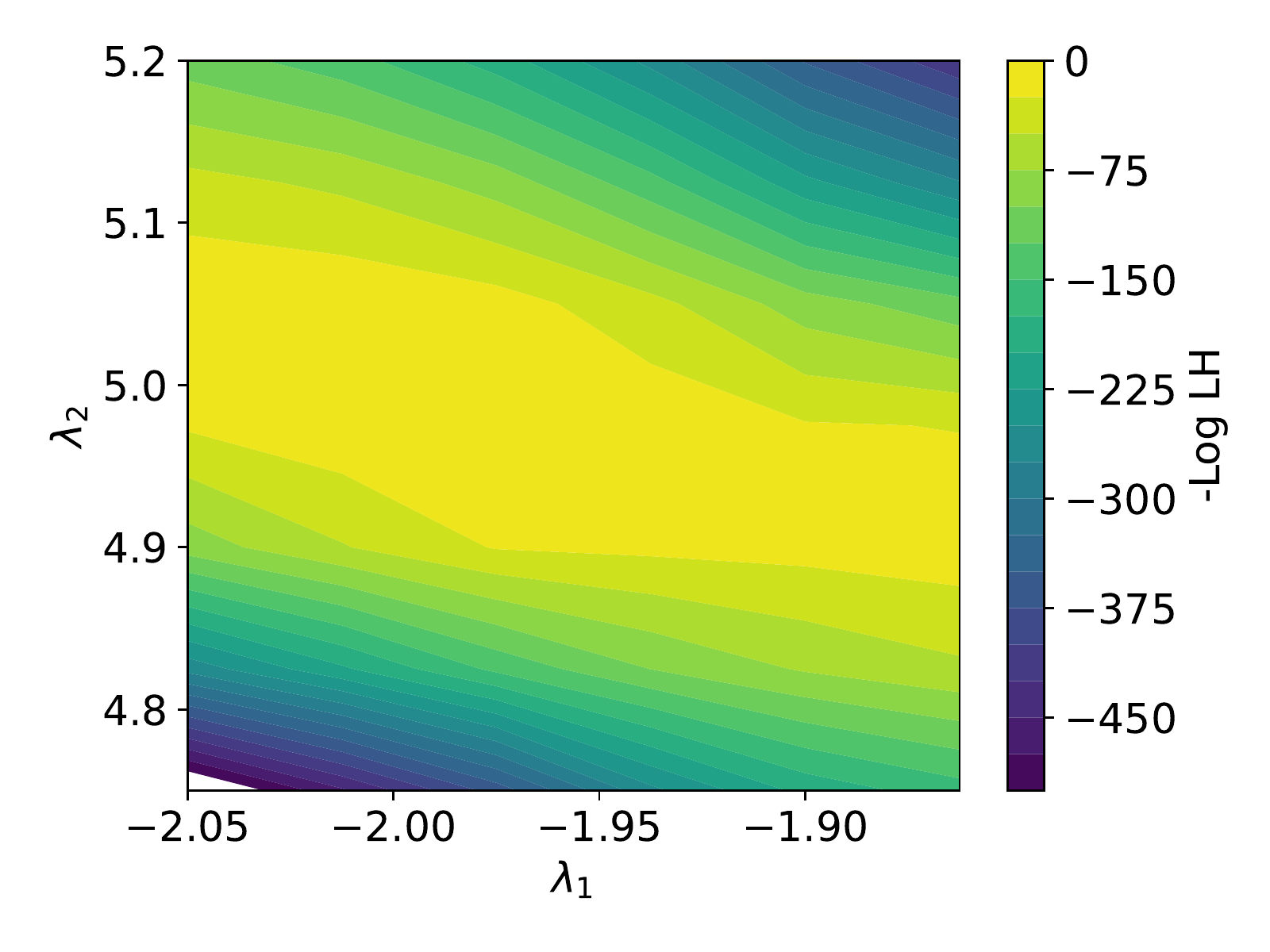}
    \caption{ 
    Scans of the likelihood for two example sets of stellar spectra $s$ (left, right) versus EOS parameters $\lambda_1$ and $\lambda_2$. 
    Top demonstrates the ideal nuisance parameter (NP) conditions where the NPs are fixed to their true values. For the same simulated observed spectra, the center shows a more realistic ``tight'' scenario, and the bottom shows a ``loose'' scenario in which the NPs are not well constrained by priors. In the ``loose'' and ``tight'' scenarios, dependence on the nuisance parameters has been integrated out as described in the text.}
    
    \label{fig:lheos_loose}
\end{figure}

The two networks that model the missing functions $f$ and $h$ allow for an approximate evaluation of the likelihood, Eq.~\ref{eq:eoslhood} as a function of the EOS parameters. Figure~\ref{fig:lheos_loose} shows examples of two individual sets of simulated stars under the three nuisance parameter scenarios.

\begin{figure}[hbt!]
    \centering
    \includegraphics[width=0.45\textwidth]{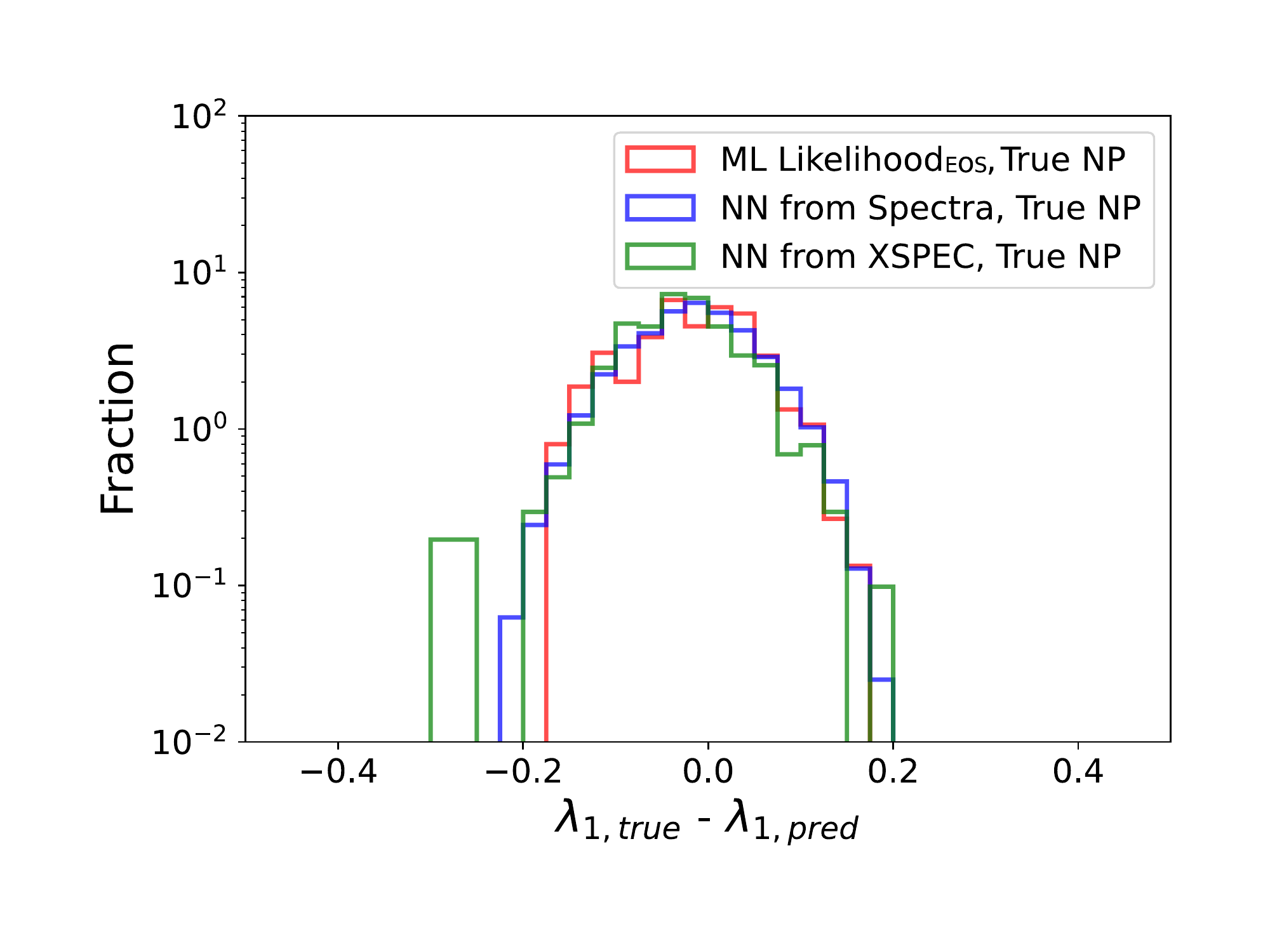}    \includegraphics[width=0.45\textwidth]{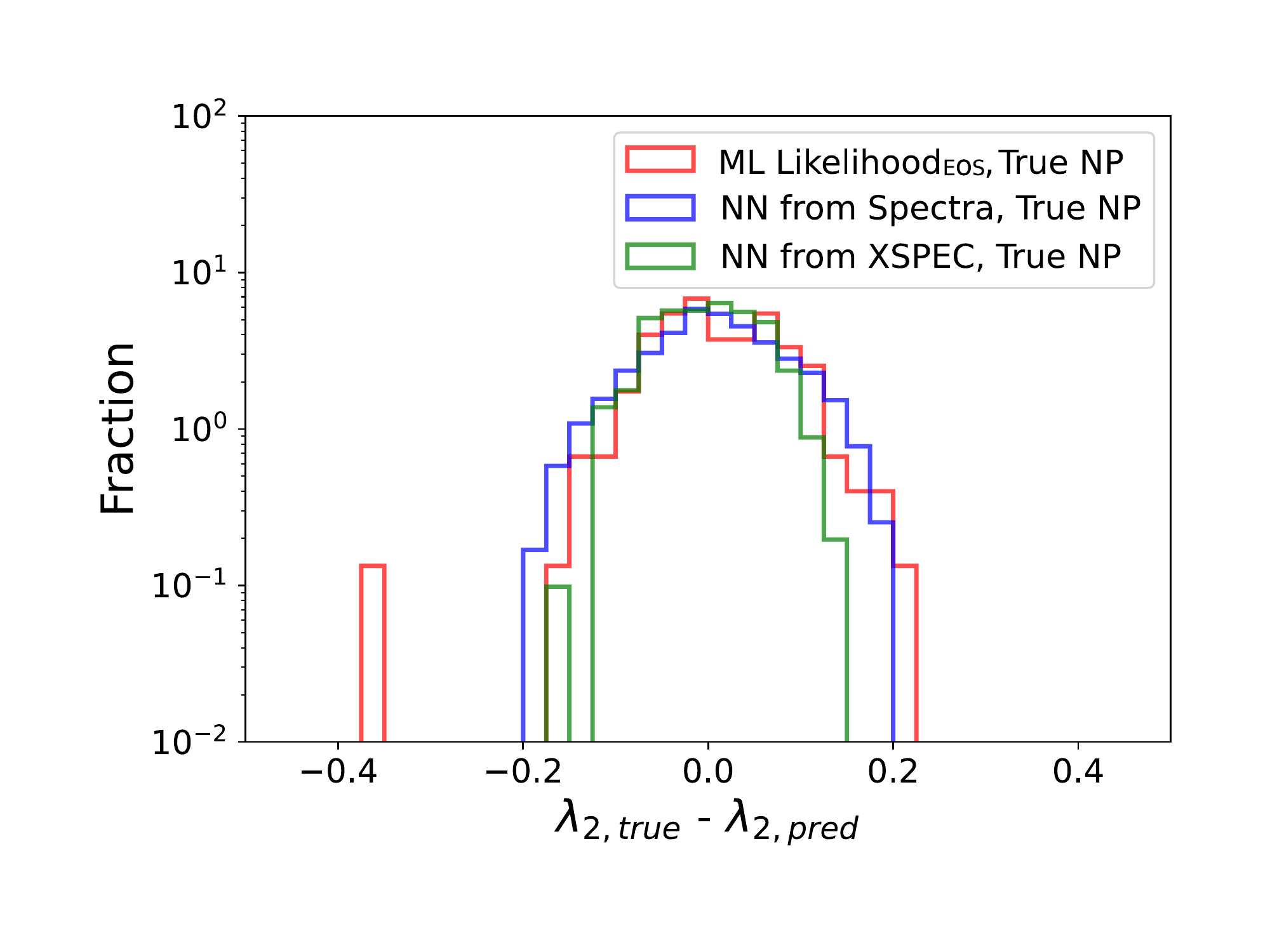}
    \includegraphics[width=0.45\textwidth]{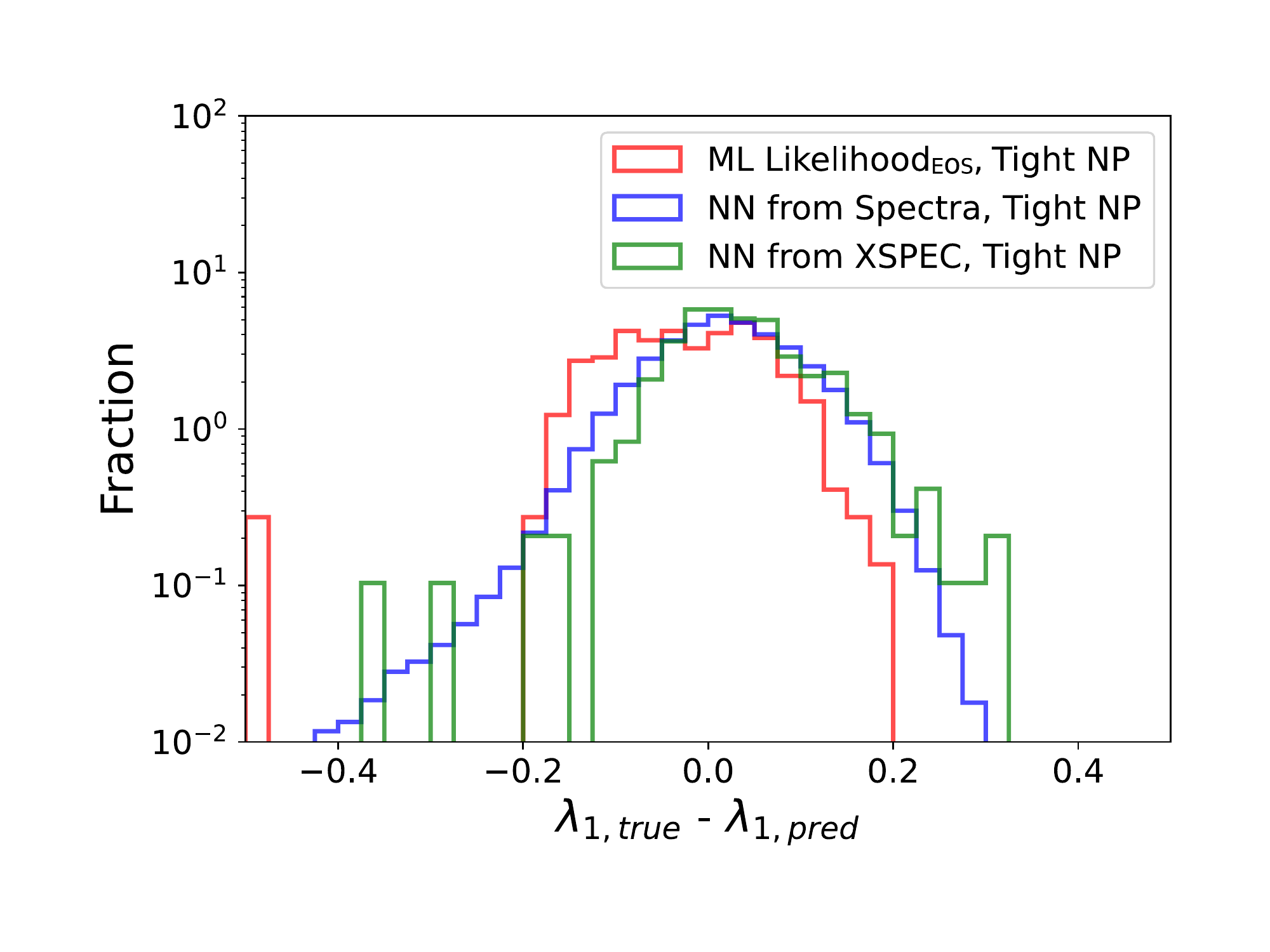}
    \includegraphics[width=0.45\textwidth]{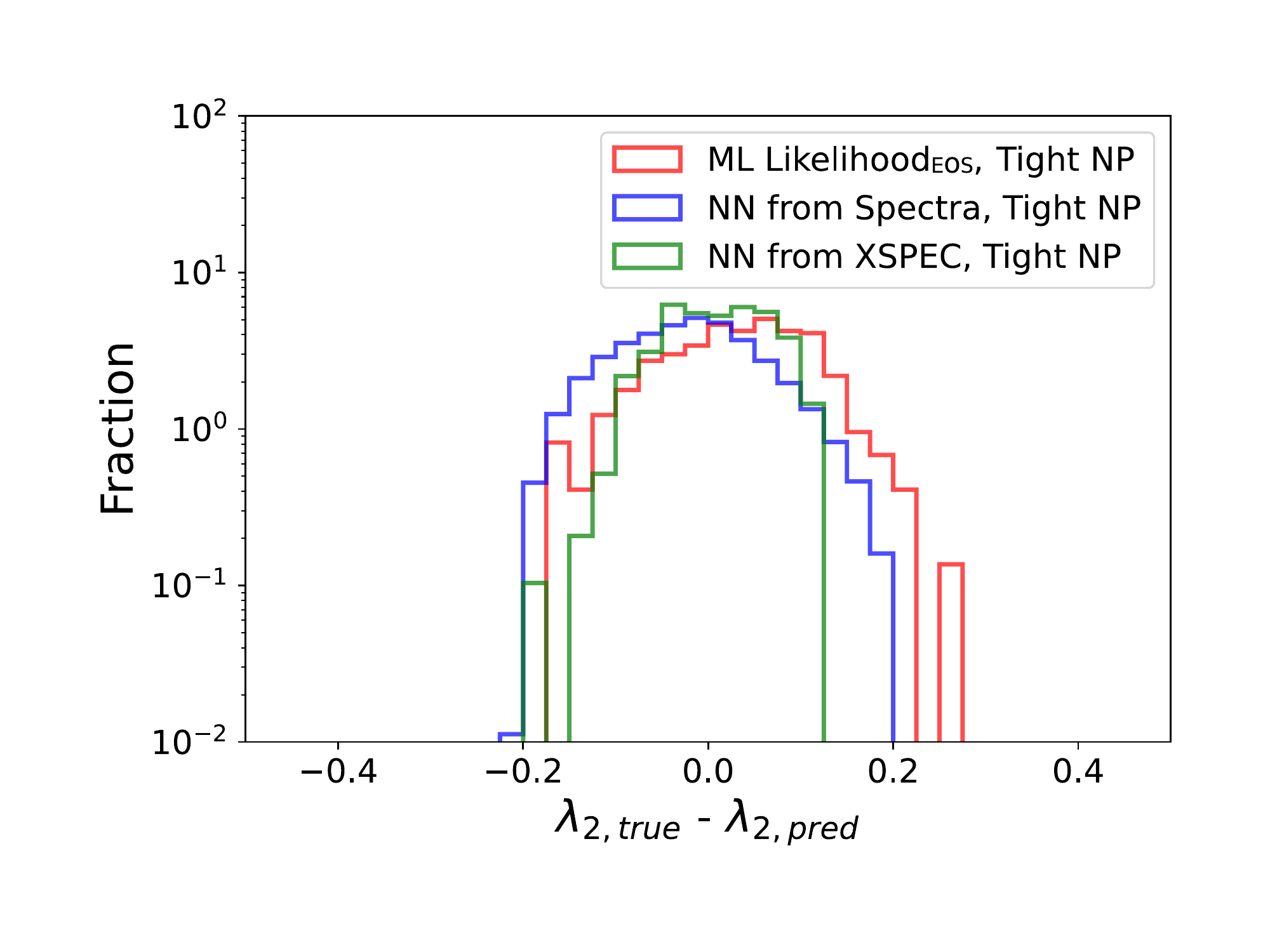}
    \includegraphics[width=0.45\textwidth]{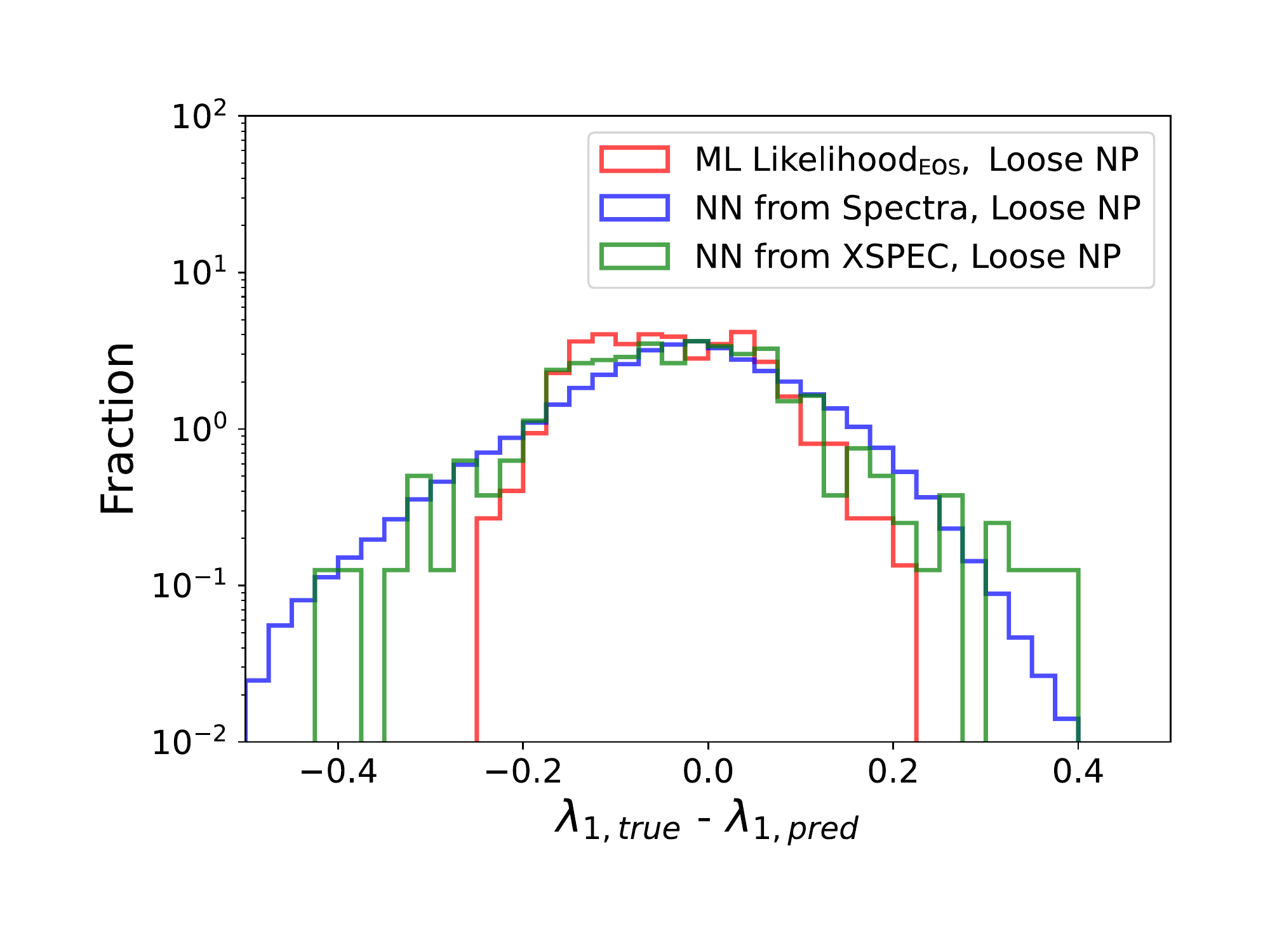}
    \includegraphics[width=0.45\textwidth]{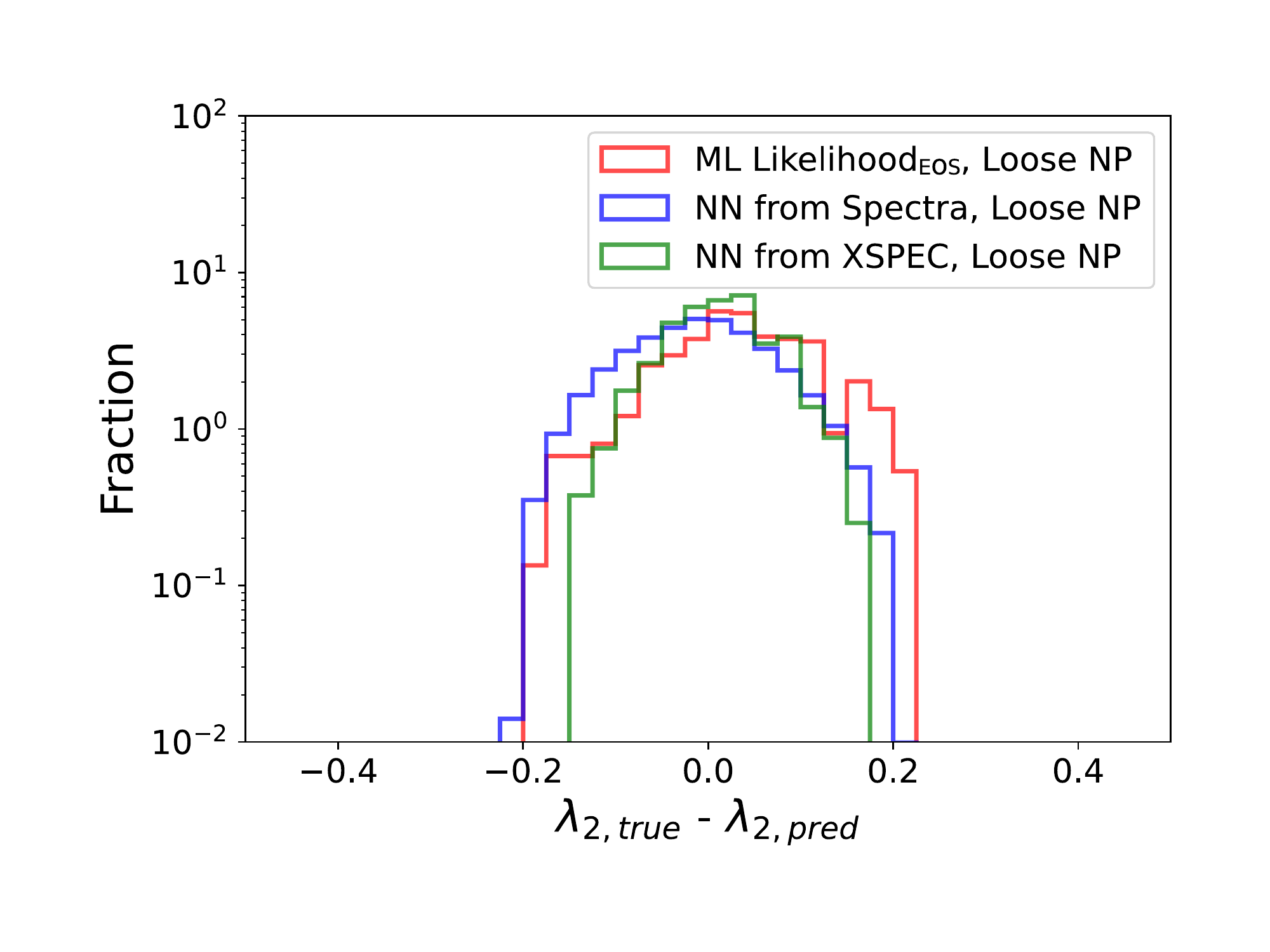}
    \caption{ Performance of our estimation of neutron star EOS parameters $\lambda_1$ (left) and $\lambda_2$ (right)
     using an approximate likelihood that incorporates a neural network, \surrEOS, in  comparison to the performance of a  spectra-to-EOS regression network and network which regresses EOS parameters from $M,R$ values estimated by \xspec, both from Ref.~\cite{delaney-eos}.  Shown are the residual distributions, the difference between the true and predicted values, under three  scenarios of nuisance parameter uncertainties. See Table~\ref{tab:eos} for quantitative analysis.    In the ``true" case, the NPs are fixed to their true values; in the ``tight" and ``loose" cases, they are drawn from narrow or wide priors, respectively; see text for details}
    \label{fig:res_forward_eos}
\end{figure}

To estimate the EOS parameters from a fixed set of stellar spectra, the likelihood is maximized via a course scan over EOS parameter space followed by the use of an optimization algorithm for a more refined location of the optimal EOS parameters. Each evaluation of the likelihood involves nested loops over the stars, an integral over possible masses, and a loop over the spectral bins. Performance of \surrEOS~ and comparison with benchmarks are shown in Fig.~\ref{fig:res_forward_eos} and Table~\ref{tab:eos}. We note that the data used in evaluations are generated via \xspec, not from the models $f$ and $h$, allowing for a test of the fidelity of the machine-learned models. Experiments in which simulated spectra are generated using the models $f$ and $h$ show equivalent performance, indicating that any bias due to mis-estimation by $f$ or $h$ is negligible in this context.

\begin{table}[]
    \caption{ Performance of our estimation of neutron star EOS parameters $\lambda_1$ (left) and $\lambda_2$ (right)
     using an approximate likelihood that incorporates a neural network, \surrEOS, in comparison to the performance of a  spectra-to-EOS regression network and network which regresses EOS parameters from $M,R$ values estimated by \xspec, both from Ref.~\cite{delaney-eos}.  Shown are the mean ($\mu$) and standard deviation ($\sigma$) of the residual distributions under three scenarios of nuisance parameter uncertainties. See Fig~\ref{fig:res_forward_eos} for distributions.    In the ``true" case, the NPs are fixed to their true values; in the ``tight" and ``loose" cases, they are drawn from narrow or wide priors, respectively; see text for details}
    \label{tab:eos}
    \centering
    \begin{tabular}{lcrrrrrr}
             \hline \hline
         & &\multicolumn{2}{c}{$\lambda_1$} & \ \ &\multicolumn{2}{c}{$\lambda_2$} & Combined\\
         \cline{3-4}\cline{6-7}\\
         Method & Nuis. Params. &  $\mu$ & $\sigma$ & &$\mu$ & $\sigma$ & $\sigma$\\
                  \hline 
        \surrEOS         & True  & -0.02 & 0.066  && 0.01  & 0.070   & 0.096 \\
        NN(Spectra)          & True  & -0.02 & 0.066  && 0.01  & 0.075   & 0.099  \\
        NN($M,R$ via \xspec~) & True  & -0.03 & 0.065  && 0.01  & 0.055   & 0.085 \\ \hline
        \surrEOS         & Tight & -0.02 & 0.078  && 0.03  & 0.081   & 0.112 \\
        NN(Spectra)          & Tight & 0.02  & 0.085  && -0.02 & 0.077   & 0.115  \\
        NN($M,R$ via \xspec~) & Tight & -0.03 & 0.081  && 0.01  & 0.056   & 0.098 \\ \hline
        \surrEOS         & Loose & -0.04 & 0.089  && 0.03  & 0.081   & 0.120  \\
        NN(Spectra)          & Loose & -0.03 & 0.131  && -0.01 & 0.078   & 0.152 \\
        NN($M,R$ via \xspec~) & Loose & -0.03 & 0.123  && 0.01  & 0.058   & 0.136  \\
         \hline \hline
    \end{tabular}
\end{table}

\section{Discussion} \label{sec:disc}

The results above demonstrate that machine-learning-derived likelihoods are useful statistical tools, allowing for traditional inference such as parameter estimation for quantities of interest (eg stellar $M$ and $R$) as well as profiling over nuisance parameters (eg stellar distances and temperatures).  

In the case of $M$,$R$-estimation for an individual star, the performance of the \surrMR~ method matches the performance of \xspec~ when the nuisance parameters are known. This is an important validation of the technique, as the simulated samples are generated by \xspec~ and so its internal likelihood estimation represents something of an upper bound on possible performance. Though \xspec~ can provide point estimates and other analysis, \mbox{\surrMR} in this case is valuable as a building block for further analysis, as \xspec~ does not provide an efficient interface to its internal calculations. For example, in the cases where nuisance parameters weaken the inference, \surrMR~ is able to improve on \xspec's performance by marginalizing over the stellar nuisance parameters. Given access to the full likelihood, one could also choose to profile over the nuisance parameters. In addition, while \xspec's inference is linked to a particular theoretical model, \surrMR~ can be trained on a variety or mixture of models, providing a smooth interpolation between otherwise distinct conceptual approaches~\cite{Ghosh:2022zdz}.  

The $M,R$-likelihood estimation is a building block toward the estimation of EOS parameters for sets of stars. In this case, as well, the likelihood provides for reliable inference of the EOS parameters, as demonstrated by the performance of \surrEOS. The residuals in this case again are narrower than the pure regression approach, nearly matching the performance of \xspec~ in the true case, and exceeding its unmarginalized estimates in the realistic case where nuisance parameter uncertainty is important.

Our method uses machine learning to enable what is typically termed a {\it forward} process, in that it aids the calculation of the likelihood of experimental data from the parameters, rather than {\it backward} inference of the parameters from the data. In this sense, it can be considered a fast and flexible simulation tool. The neural networks developed for this work effectively enable end-to-end, fast, and convenient simulation of neutron star spectra for a range of EOS parameters and nuisance parameters, including the intermediate step of generating plausible neutron star properties (mass and radius) for a given set of EOS parameters. In the case where the true likelihood exists but is not made conveniently accessible, our approach provides a powerful and flexible new interface, even without speed enhancements. In the more general case, such as for EOS inference, the approach additionally allows for more rapid generation of simulated stellar masses and radii for specific EOS parameters without solving the complex sets of equations underlying the physical model.

Once released, this framework will serve as a convenient simulation tool that can be used by the larger machine learning research community to generate neutron star samples without the expert knowledge required to run \xspec~ or solve the TOV equations, and spur further innovation, such as in simulator-based inference techniques that can handle a large number of nuisance parameters, or to setup ML challenges where participants are given access to the simulator.
%(such as Refs~\cite{Amrouche:2019wmx,Amrouche:2021tio,https://doi.org/10.48550/arxiv.2206.14642}
\section{Conclusion}
\label{sec:conc}

In this work, we demonstrate an alternative approach to deducing neutron star mass, radius, and EOS from simulated stellar X-ray spectra using machine learning. The alternative approach employs a  technique novel to neutron star astrophysics, machine-learning-derived likelihoods, in which intractable elements of the calculation are replaced with machine-learned functions. This allows for an approximate likelihood calculation with increased interpretability. Our forward neural network model allows us to predict the X-ray spectra given a value of nuisance parameters, the equation of state parameters, and the neutron star mass. Forward modeling of this kind demonstrates how these networks are interpretable, testable, and often with the assumptions being explicit rather than implicit when compared to inverse models~\cite{Ozel15,Steiner18ct}. The \surrEOS~ model outperforms our previous best-performing regression model, demonstrating the power of such a technique for point estimation. Note that it additionally provides access to the full likelihood, which can inform uncertainty estimation. 

While the studies shown in this work demonstrate the power of the method to provide a tractable likelihood for stellar spectra generated from a specific model, the natural ability of neural networks to interpolate allows for the treatment of inherent model uncertainties, by training on samples with mixed or varying models~\cite{Ghosh:2021roe}.

The machine learning-driven calculations lend themselves to various future directions. In terms of training data, EOS models with different parametrization can be tested as well as using a different response matrix (e.g. from NuStar, NICER, or XMM-Newton as described in \cite{ludlam2018detection}) in \xspec~ to generate simulated X-ray spectra. In addition, this technique can be rapidly applied to the analysis of  X-ray spectra from recent observations. The networks developed for this work will become a convenient tool for fast simulation of neutron star spectra which can be run without the expert knowledge required to run \xspec~ or to solve the TOV equations.

\section{Acknowledgements}

DW and AG are supported by The Department of Energy Office of Science grant DE-SC0009920, and AG is also supported under contract DE-AC02-05CH11231. LL was supported by NSF Grant No. 2012857 to the University of California at San Diego. AWS was supported by NSF AST 22-06322, PHY 21-16686, and the Department of Energy Office of Nuclear Physics. DF and FW are supported by the National Science Foundation (USA) under Grant No. PHY-2012152. The authors are grateful to C. O. Heinke and W. C. G. Ho for their assistance in understanding and navigating the \xspec~ software, and to Kyle Cranmer for conceptual contributions.

\clearpage

\iffalse
\section{Integration and Optimization}

This section describes the numerical integration and optimization methods used.

\subsection{Integration}

Integration is performed by a Monte Carlo method.  In order to ensure convergence, four independent sets of 100 samples are made, leading to four estimates of the value of the integral.  Additional sets of 100 samples are performed until the uncertainty on the mean over the sets is reduced to less than 3\% relative. The mean value over all sets is then used as the estimate.

\subsection{Optimization}

Optimization is performed in two stages. First, the rough location of the maximum is located, followed by the use of a maximization algorithm for fine-tuning.

The first stage performs a simple grid scan of the parameter space, dividing the space into $N$ tiles centered on the full parameter range. The process is repeated, but now centered over the expectation value calculated from the previous round, and with a search space the width of three grid tiles, again divided into $N$ smaller tiles. The process is repeated $M$ times.  In the results shown here, $N=4$ and $M=3$.

The second stage runs an algorithm to find a minimum. The {\tt scipy.minimize} function with Nelder-Mead and conjugate-gradient algorithms.  In addition, a method which performs two one-dimensional parabolic fits through the center of the region was explored. The results shown in the paper use the Nelder-Mead algorithm.

The two-stage method performed more reliably than running the algorithm over the full parameter space.
\fi

\appendix
\section{Hyperparameter Optimization} \label{appendix:hypo}

This section describes the hyperparameter optimization methods for model $f$, which generates simulated X-ray spectra from mass, radius, and the nuisance parameters $\nu$.

The network has two input branches, the first to process the mass and radius values and the second to process $\nu$. The output of the two branches is combined together and processed through the remaining layers to predict the X-ray spectra. In both the input branches as well as the secondary portion of the network, we trained models with random combinations of the following parameters:

\begin{enumerate}
    \item Number of layers: one to fifteen,
    \item Number of hidden nodes: 64 to 2048,
    \item Dropout: 0 to 1,
    \item Learning rate: 0.0001 to 0.005,
    \item Activation functions,
    \item Skip connections,
    \item Loss functions: mean squared error (MSE), mean absolute percentage error (MAPE), and the Huber loss function,
    \item Data scalers for both the input and the output: zero-mean, log, min-max, and standard scaler. The scalers preprocess the data before it enters the network, meaning scaled input values (mass, radius, and $\nu$) with a scaled output (spectra) are used for training. After training, the scaling procedure is inverted for the predicted spectra to produce meaningful values. 
\end{enumerate}

The hyperparameters chosen were the combination that had the best performance on the validation data set, described in Section \ref{sec:mllh}.

\bibliography{ns,baldi}

\begin{thebibliography}{58}
\expandafter\ifx\csname natexlab\endcsname\relax\def\natexlab#1{#1}\fi
\expandafter\ifx\csname bibnamefont\endcsname\relax
  \def\bibnamefont#1{#1}\fi
\expandafter\ifx\csname bibfnamefont\endcsname\relax
  \def\bibfnamefont#1{#1}\fi
\expandafter\ifx\csname citenamefont\endcsname\relax
  \def\citenamefont#1{#1}\fi
\expandafter\ifx\csname url\endcsname\relax
  \def\url#1{\texttt{#1}}\fi
\expandafter\ifx\csname urlprefix\endcsname\relax\def\urlprefix{URL }\fi
\providecommand{\bibinfo}[2]{#2}
\providecommand{\eprint}[2][]{\url{#2}}

\bibitem[{\citenamefont{Tolos and Fabbietti}(2020)}]{Tolos:2020aln}
\bibinfo{author}{\bibfnamefont{L.}~\bibnamefont{Tolos}} \bibnamefont{and}
  \bibinfo{author}{\bibfnamefont{L.}~\bibnamefont{Fabbietti}},
  \bibinfo{journal}{Prog. Part. Nucl. Phys.} \textbf{\bibinfo{volume}{112}},
  \bibinfo{pages}{103770} (\bibinfo{year}{2020}), \eprint{2002.09223}.

\bibitem[{\citenamefont{Li et~al.}(2018)\citenamefont{Li, Sedrakian, and
  Weber}}]{li2018competition}
\bibinfo{author}{\bibfnamefont{J.~J.} \bibnamefont{Li}},
  \bibinfo{author}{\bibfnamefont{A.}~\bibnamefont{Sedrakian}},
  \bibnamefont{and} \bibinfo{author}{\bibfnamefont{F.}~\bibnamefont{Weber}},
  \bibinfo{journal}{Physics Letters B} \textbf{\bibinfo{volume}{783}},
  \bibinfo{pages}{234} (\bibinfo{year}{2018}).

\bibitem[{\citenamefont{Spinella and Weber}(2019)}]{spinella2019hyperonic}
\bibinfo{author}{\bibfnamefont{W.~M.} \bibnamefont{Spinella}} \bibnamefont{and}
  \bibinfo{author}{\bibfnamefont{F.}~\bibnamefont{Weber}},
  \bibinfo{journal}{Astronomische Nachrichten} \textbf{\bibinfo{volume}{340}},
  \bibinfo{pages}{145} (\bibinfo{year}{2019}).

\bibitem[{\citenamefont{Malfatti et~al.}(2020)\citenamefont{Malfatti, Orsaria,
  Ranea-Sandoval, Contrera, and Weber}}]{Malfatti:2020}
\bibinfo{author}{\bibfnamefont{G.}~\bibnamefont{Malfatti}},
  \bibinfo{author}{\bibfnamefont{M.~G.} \bibnamefont{Orsaria}},
  \bibinfo{author}{\bibfnamefont{I.~F.} \bibnamefont{Ranea-Sandoval}},
  \bibinfo{author}{\bibfnamefont{G.~A.} \bibnamefont{Contrera}},
  \bibnamefont{and} \bibinfo{author}{\bibfnamefont{F.}~\bibnamefont{Weber}},
  \bibinfo{journal}{Phys. Rev. D} \textbf{\bibinfo{volume}{102}},
  \bibinfo{pages}{063008} (\bibinfo{year}{2020}),
  \urlprefix\url{https://link.aps.org/doi/10.1103/PhysRevD.102.063008}.

\bibitem[{\citenamefont{Sedrakian et~al.}(2023)\citenamefont{Sedrakian, Li, and
  Weber}}]{Sedrakian:2023PPNP}
\bibinfo{author}{\bibfnamefont{A.}~\bibnamefont{Sedrakian}},
  \bibinfo{author}{\bibfnamefont{J.~J.} \bibnamefont{Li}}, \bibnamefont{and}
  \bibinfo{author}{\bibfnamefont{F.}~\bibnamefont{Weber}},
  \bibinfo{journal}{Progress in Particle and Nuclear Physics} p.
  \bibinfo{pages}{104041} (\bibinfo{year}{2023}), ISSN
  \bibinfo{issn}{0146-6410},
  \urlprefix\url{https://www.sciencedirect.com/science/article/pii/S0146641023000224}.

\bibitem[{\citenamefont{{Alcock} et~al.}(1986)\citenamefont{{Alcock}, {Farhi},
  and {Olinto}}}]{Alcock:1986}
\bibinfo{author}{\bibfnamefont{C.}~\bibnamefont{{Alcock}}},
  \bibinfo{author}{\bibfnamefont{E.}~\bibnamefont{{Farhi}}}, \bibnamefont{and}
  \bibinfo{author}{\bibfnamefont{A.}~\bibnamefont{{Olinto}}},
  \bibinfo{journal}{\apj} \textbf{\bibinfo{volume}{310}}, \bibinfo{pages}{261}
  (\bibinfo{year}{1986}).

\bibitem[{\citenamefont{Orsaria et~al.}(2014)\citenamefont{Orsaria, Rodrigues,
  Weber, and Contrera}}]{orsaria2014quark}
\bibinfo{author}{\bibfnamefont{M.}~\bibnamefont{Orsaria}},
  \bibinfo{author}{\bibfnamefont{H.}~\bibnamefont{Rodrigues}},
  \bibinfo{author}{\bibfnamefont{F.}~\bibnamefont{Weber}}, \bibnamefont{and}
  \bibinfo{author}{\bibfnamefont{G.}~\bibnamefont{Contrera}},
  \bibinfo{journal}{Physical Review C} \textbf{\bibinfo{volume}{89}},
  \bibinfo{pages}{015806} (\bibinfo{year}{2014}).

\bibitem[{\citenamefont{Alford}(2001)}]{Alford:2001}
\bibinfo{author}{\bibfnamefont{M.}~\bibnamefont{Alford}},
  \bibinfo{journal}{Annual Review of Nuclear and Particle Science}
  \textbf{\bibinfo{volume}{51}}, \bibinfo{pages}{131} (\bibinfo{year}{2001}),
  \eprint{https://doi.org/10.1146/annurev.nucl.51.101701.132449},
  \urlprefix\url{https://doi.org/10.1146/annurev.nucl.51.101701.132449}.

\bibitem[{\citenamefont{Alford et~al.}(2008)\citenamefont{Alford, Schmitt,
  Rajagopal, and Sch\"afer}}]{Alford:2007xm}
\bibinfo{author}{\bibfnamefont{M.~G.} \bibnamefont{Alford}},
  \bibinfo{author}{\bibfnamefont{A.}~\bibnamefont{Schmitt}},
  \bibinfo{author}{\bibfnamefont{K.}~\bibnamefont{Rajagopal}},
  \bibnamefont{and}
  \bibinfo{author}{\bibfnamefont{T.}~\bibnamefont{Sch\"afer}},
  \bibinfo{journal}{Rev. Mod. Phys.} \textbf{\bibinfo{volume}{80}},
  \bibinfo{pages}{1455} (\bibinfo{year}{2008}), \eprint{0709.4635}.

\bibitem[{\citenamefont{Zdunik and Haensel}(2013)}]{zdunik2013maximum}
\bibinfo{author}{\bibfnamefont{J.}~\bibnamefont{Zdunik}} \bibnamefont{and}
  \bibinfo{author}{\bibfnamefont{P.}~\bibnamefont{Haensel}},
  \bibinfo{journal}{Astronomy \& Astrophysics} \textbf{\bibinfo{volume}{551}},
  \bibinfo{pages}{A61} (\bibinfo{year}{2013}).

\bibitem[{\citenamefont{Baym}(1973)}]{baym1973pion}
\bibinfo{author}{\bibfnamefont{G.}~\bibnamefont{Baym}},
  \bibinfo{journal}{Physical Review Letters} \textbf{\bibinfo{volume}{30}},
  \bibinfo{pages}{1340} (\bibinfo{year}{1973}).

\bibitem[{\citenamefont{Kaplan and Nelson}(1986)}]{KAPLAN}
\bibinfo{author}{\bibfnamefont{D.}~\bibnamefont{Kaplan}} \bibnamefont{and}
  \bibinfo{author}{\bibfnamefont{A.}~\bibnamefont{Nelson}},
  \bibinfo{journal}{Physics Letters B} \textbf{\bibinfo{volume}{175}},
  \bibinfo{pages}{57} (\bibinfo{year}{1986}), ISSN \bibinfo{issn}{0370-2693},
  \urlprefix\url{https://www.sciencedirect.com/science/article/pii/037026938690331X}.

\bibitem[{\citenamefont{Glendenning and
  Schaffner-Bielich}(1999)}]{Glendenning:1999}
\bibinfo{author}{\bibfnamefont{N.~K.} \bibnamefont{Glendenning}}
  \bibnamefont{and}
  \bibinfo{author}{\bibfnamefont{J.}~\bibnamefont{Schaffner-Bielich}},
  \bibinfo{journal}{Phys. Rev. C} \textbf{\bibinfo{volume}{60}},
  \bibinfo{pages}{025803} (\bibinfo{year}{1999}),
  \urlprefix\url{https://link.aps.org/doi/10.1103/PhysRevC.60.025803}.

\bibitem[{\citenamefont{Ellis et~al.}(1995)\citenamefont{Ellis, Knorren, and
  Prakash}}]{ellis1995kaon}
\bibinfo{author}{\bibfnamefont{P.~J.} \bibnamefont{Ellis}},
  \bibinfo{author}{\bibfnamefont{R.}~\bibnamefont{Knorren}}, \bibnamefont{and}
  \bibinfo{author}{\bibfnamefont{M.}~\bibnamefont{Prakash}},
  \bibinfo{journal}{Physics Letters B} \textbf{\bibinfo{volume}{349}},
  \bibinfo{pages}{11} (\bibinfo{year}{1995}).

\bibitem[{\citenamefont{Ramos et~al.}(2001)\citenamefont{Ramos,
  Schaffner-Bielich, and Wambach}}]{Ramos:2001}
\bibinfo{author}{\bibfnamefont{A.}~\bibnamefont{Ramos}},
  \bibinfo{author}{\bibfnamefont{J.}~\bibnamefont{Schaffner-Bielich}},
  \bibnamefont{and} \bibinfo{author}{\bibfnamefont{J.}~\bibnamefont{Wambach}},
  \emph{\bibinfo{title}{Kaon Condensation in Neutron Stars}}
  (\bibinfo{publisher}{Springer Berlin Heidelberg}, \bibinfo{address}{Berlin,
  Heidelberg}, \bibinfo{year}{2001}), pp. \bibinfo{pages}{175--202}, ISBN
  \bibinfo{isbn}{978-3-540-44578-4},
  \urlprefix\url{https://doi.org/10.1007/3-540-44578-1_6}.

\bibitem[{\citenamefont{Steiner et~al.}(2013)\citenamefont{Steiner, Hempel, and
  Fischer}}]{steiner2013core}
\bibinfo{author}{\bibfnamefont{A.~W.} \bibnamefont{Steiner}},
  \bibinfo{author}{\bibfnamefont{M.}~\bibnamefont{Hempel}}, \bibnamefont{and}
  \bibinfo{author}{\bibfnamefont{T.}~\bibnamefont{Fischer}},
  \bibinfo{journal}{The Astrophysical Journal} \textbf{\bibinfo{volume}{774}},
  \bibinfo{pages}{17} (\bibinfo{year}{2013}).

\bibitem[{\citenamefont{Hotokezaka et~al.}(2011)\citenamefont{Hotokezaka,
  Kyutoku, Okawa, Shibata, and Kiuchi}}]{hotokezaka2011binary}
\bibinfo{author}{\bibfnamefont{K.}~\bibnamefont{Hotokezaka}},
  \bibinfo{author}{\bibfnamefont{K.}~\bibnamefont{Kyutoku}},
  \bibinfo{author}{\bibfnamefont{H.}~\bibnamefont{Okawa}},
  \bibinfo{author}{\bibfnamefont{M.}~\bibnamefont{Shibata}}, \bibnamefont{and}
  \bibinfo{author}{\bibfnamefont{K.}~\bibnamefont{Kiuchi}},
  \bibinfo{journal}{Physical Review D} \textbf{\bibinfo{volume}{83}},
  \bibinfo{pages}{124008} (\bibinfo{year}{2011}).

\bibitem[{\citenamefont{{Rutledge} et~al.}(1999)\citenamefont{{Rutledge},
  {Bildsten}, {Brown}, {Pavlov}, and {Zavlin}}}]{Rutledge}
\bibinfo{author}{\bibfnamefont{R.~E.} \bibnamefont{{Rutledge}}},
  \bibinfo{author}{\bibfnamefont{L.}~\bibnamefont{{Bildsten}}},
  \bibinfo{author}{\bibfnamefont{E.~F.} \bibnamefont{{Brown}}},
  \bibinfo{author}{\bibfnamefont{G.~G.} \bibnamefont{{Pavlov}}},
  \bibnamefont{and} \bibinfo{author}{\bibfnamefont{V.~E.}
  \bibnamefont{{Zavlin}}}, \bibinfo{journal}{\apj}
  \textbf{\bibinfo{volume}{514}}, \bibinfo{pages}{945} (\bibinfo{year}{1999}),
  \eprint{astro-ph/9810288}.

\bibitem[{\citenamefont{{Heinke} et~al.}(2006)\citenamefont{{Heinke},
  {Rybicki}, {Narayan}, and {Grindlay}}}]{Heinke06}
\bibinfo{author}{\bibfnamefont{C.~O.} \bibnamefont{{Heinke}}},
  \bibinfo{author}{\bibfnamefont{G.~B.} \bibnamefont{{Rybicki}}},
  \bibinfo{author}{\bibfnamefont{R.}~\bibnamefont{{Narayan}}},
  \bibnamefont{and} \bibinfo{author}{\bibfnamefont{J.~E.}
  \bibnamefont{{Grindlay}}}, \bibinfo{journal}{\apj}
  \textbf{\bibinfo{volume}{644}}, \bibinfo{pages}{1090} (\bibinfo{year}{2006}),
  \eprint{astro-ph/0506563}.

\bibitem[{\citenamefont{Lattimer and Prakash}(2001)}]{Lattimer01}
\bibinfo{author}{\bibfnamefont{J.~M.} \bibnamefont{Lattimer}} \bibnamefont{and}
  \bibinfo{author}{\bibfnamefont{M.}~\bibnamefont{Prakash}},
  \bibinfo{journal}{Astrophys. J.} \textbf{\bibinfo{volume}{550}},
  \bibinfo{pages}{426} (\bibinfo{year}{2001}),
  \urlprefix\url{https://doi.org/10.1086/319702}.

\bibitem[{\citenamefont{Lindblom}(2010)}]{PhysRevD.82.103011}
\bibinfo{author}{\bibfnamefont{L.}~\bibnamefont{Lindblom}},
  \bibinfo{journal}{Phys. Rev. D} \textbf{\bibinfo{volume}{82}},
  \bibinfo{pages}{103011} (\bibinfo{year}{2010}),
  \urlprefix\url{https://link.aps.org/doi/10.1103/PhysRevD.82.103011}.

\bibitem[{\citenamefont{Steiner et~al.}(2010)\citenamefont{Steiner, Lattimer,
  and {Brown}}}]{Steiner10te}
\bibinfo{author}{\bibfnamefont{A.~W.} \bibnamefont{Steiner}},
  \bibinfo{author}{\bibfnamefont{J.~M.} \bibnamefont{Lattimer}},
  \bibnamefont{and} \bibinfo{author}{\bibfnamefont{E.~F.}
  \bibnamefont{{Brown}}}, \bibinfo{journal}{Astrophys. J.}
  \textbf{\bibinfo{volume}{722}}, \bibinfo{pages}{33} (\bibinfo{year}{2010}),
  \eprint{1005.0811},
  \urlprefix\url{http://doi.org/10.1088/0004-637X/722/1/33}.

\bibitem[{\citenamefont{Lindblom and Indik}(2014)}]{Lindblom:2013xkra}
\bibinfo{author}{\bibfnamefont{L.}~\bibnamefont{Lindblom}} \bibnamefont{and}
  \bibinfo{author}{\bibfnamefont{N.~M.} \bibnamefont{Indik}},
  \bibinfo{journal}{Phys. Rev. D} \textbf{\bibinfo{volume}{89}},
  \bibinfo{pages}{064003} (\bibinfo{year}{2014}), \bibinfo{note}{[Erratum:
  Phys.Rev.D 93, 129903 (2016)]}, \eprint{1310.0803}.

\bibitem[{\citenamefont{Lindblom and Indik}(2013)}]{issp}
\bibinfo{author}{\bibfnamefont{L.}~\bibnamefont{Lindblom}} \bibnamefont{and}
  \bibinfo{author}{\bibfnamefont{N.}~\bibnamefont{Indik}},
  \bibinfo{journal}{Physical Review D} \textbf{\bibinfo{volume}{86}}
  (\bibinfo{year}{2013}).

\bibitem[{\citenamefont{Baldi}(2021)}]{baldi2021deep}
\bibinfo{author}{\bibfnamefont{P.}~\bibnamefont{Baldi}},
  \emph{\bibinfo{title}{Deep Learning in Science}}
  (\bibinfo{publisher}{Cambridge University Press},
  \bibinfo{address}{Cambridge, UK}, \bibinfo{year}{2021}).

\bibitem[{\citenamefont{Baldi et~al.}(2014)\citenamefont{Baldi, Sadowski, and
  Whiteson}}]{baldi2014searching}
\bibinfo{author}{\bibfnamefont{P.}~\bibnamefont{Baldi}},
  \bibinfo{author}{\bibfnamefont{P.}~\bibnamefont{Sadowski}}, \bibnamefont{and}
  \bibinfo{author}{\bibfnamefont{D.}~\bibnamefont{Whiteson}},
  \bibinfo{journal}{Nature Communications} \textbf{\bibinfo{volume}{5}}
  (\bibinfo{year}{2014}).

\bibitem[{\citenamefont{Ghosh et~al.}(2021)\citenamefont{Ghosh, Nachman, and
  Whiteson}}]{Ghosh:2021roe}
\bibinfo{author}{\bibfnamefont{A.}~\bibnamefont{Ghosh}},
  \bibinfo{author}{\bibfnamefont{B.}~\bibnamefont{Nachman}}, \bibnamefont{and}
  \bibinfo{author}{\bibfnamefont{D.}~\bibnamefont{Whiteson}},
  \bibinfo{journal}{Phys. Rev. D} \textbf{\bibinfo{volume}{104}},
  \bibinfo{pages}{056026} (\bibinfo{year}{2021}), \eprint{2105.08742}.

\bibitem[{\citenamefont{Baldi et~al.}(2016{\natexlab{a}})\citenamefont{Baldi,
  Cranmer, Faucett, Sadowski, and Whiteson}}]{Baldi:2016fzo}
\bibinfo{author}{\bibfnamefont{P.}~\bibnamefont{Baldi}},
  \bibinfo{author}{\bibfnamefont{K.}~\bibnamefont{Cranmer}},
  \bibinfo{author}{\bibfnamefont{T.}~\bibnamefont{Faucett}},
  \bibinfo{author}{\bibfnamefont{P.}~\bibnamefont{Sadowski}}, \bibnamefont{and}
  \bibinfo{author}{\bibfnamefont{D.}~\bibnamefont{Whiteson}},
  \bibinfo{journal}{Eur. Phys. J. C} \textbf{\bibinfo{volume}{76}},
  \bibinfo{pages}{235} (\bibinfo{year}{2016}{\natexlab{a}}),
  \eprint{1601.07913}.

\bibitem[{\citenamefont{Guest et~al.}(2016)\citenamefont{Guest, Collado, Baldi,
  Hsu, Urban, and Whiteson}}]{Guest:2016iqz}
\bibinfo{author}{\bibfnamefont{D.}~\bibnamefont{Guest}},
  \bibinfo{author}{\bibfnamefont{J.}~\bibnamefont{Collado}},
  \bibinfo{author}{\bibfnamefont{P.}~\bibnamefont{Baldi}},
  \bibinfo{author}{\bibfnamefont{S.-C.} \bibnamefont{Hsu}},
  \bibinfo{author}{\bibfnamefont{G.}~\bibnamefont{Urban}}, \bibnamefont{and}
  \bibinfo{author}{\bibfnamefont{D.}~\bibnamefont{Whiteson}},
  \bibinfo{journal}{Phys. Rev. D} \textbf{\bibinfo{volume}{94}},
  \bibinfo{pages}{112002} (\bibinfo{year}{2016}), \eprint{1607.08633}.

\bibitem[{\citenamefont{Cranmer et~al.}(2020)\citenamefont{Cranmer, Brehmer,
  and Louppe}}]{Cranmer_2020}
\bibinfo{author}{\bibfnamefont{K.}~\bibnamefont{Cranmer}},
  \bibinfo{author}{\bibfnamefont{J.}~\bibnamefont{Brehmer}}, \bibnamefont{and}
  \bibinfo{author}{\bibfnamefont{G.}~\bibnamefont{Louppe}},
  \bibinfo{journal}{Proceedings of the National Academy of Sciences}
  \textbf{\bibinfo{volume}{117}}, \bibinfo{pages}{30055}
  (\bibinfo{year}{2020}),
  \urlprefix\url{https://doi.org/10.1073%2Fpnas.1912789117}.

\bibitem[{\citenamefont{Fujimoto et~al.}(2020)\citenamefont{Fujimoto,
  Fukushima, and Murase}}]{Fujimoto:2019hxv}
\bibinfo{author}{\bibfnamefont{Y.}~\bibnamefont{Fujimoto}},
  \bibinfo{author}{\bibfnamefont{K.}~\bibnamefont{Fukushima}},
  \bibnamefont{and} \bibinfo{author}{\bibfnamefont{K.}~\bibnamefont{Murase}},
  \bibinfo{journal}{Phys. Rev. D} \textbf{\bibinfo{volume}{101}},
  \bibinfo{pages}{054016} (\bibinfo{year}{2020}), \eprint{1903.03400}.

\bibitem[{\citenamefont{Fujimoto et~al.}(2021)\citenamefont{Fujimoto,
  Fukushima, and Murase}}]{Fujimoto:2021zas}
\bibinfo{author}{\bibfnamefont{Y.}~\bibnamefont{Fujimoto}},
  \bibinfo{author}{\bibfnamefont{K.}~\bibnamefont{Fukushima}},
  \bibnamefont{and} \bibinfo{author}{\bibfnamefont{K.}~\bibnamefont{Murase}},
  \bibinfo{journal}{JHEP} \textbf{\bibinfo{volume}{03}}, \bibinfo{pages}{273}
  (\bibinfo{year}{2021}), \eprint{2101.08156}.

\bibitem[{\citenamefont{Morawski and Bejger}(2020)}]{Morawski:2020izm}
\bibinfo{author}{\bibfnamefont{F.}~\bibnamefont{Morawski}} \bibnamefont{and}
  \bibinfo{author}{\bibfnamefont{M.}~\bibnamefont{Bejger}},
  \bibinfo{journal}{Astron. Astrophys.} \textbf{\bibinfo{volume}{642}},
  \bibinfo{pages}{A78} (\bibinfo{year}{2020}), \eprint{2006.07194}.

\bibitem[{\citenamefont{Soma et~al.}(2022)\citenamefont{Soma, Wang, Shi,
  St{\"o}cker, and Zhou}}]{soma2022neural}
\bibinfo{author}{\bibfnamefont{S.}~\bibnamefont{Soma}},
  \bibinfo{author}{\bibfnamefont{L.}~\bibnamefont{Wang}},
  \bibinfo{author}{\bibfnamefont{S.}~\bibnamefont{Shi}},
  \bibinfo{author}{\bibfnamefont{H.}~\bibnamefont{St{\"o}cker}},
  \bibnamefont{and} \bibinfo{author}{\bibfnamefont{K.}~\bibnamefont{Zhou}},
  \bibinfo{journal}{Journal of Cosmology and Astroparticle Physics}
  \textbf{\bibinfo{volume}{2022}}, \bibinfo{pages}{071} (\bibinfo{year}{2022}).

\bibitem[{\citenamefont{Ferreira and Provid\^encia}(2019)}]{Ferreira:2019bny}
\bibinfo{author}{\bibfnamefont{M.}~\bibnamefont{Ferreira}} \bibnamefont{and}
  \bibinfo{author}{\bibfnamefont{C.}~\bibnamefont{Provid\^encia}}
  (\bibinfo{year}{2019}), \eprint{1910.05554}.

\bibitem[{\citenamefont{Farrell et~al.}(2022)}]{delaney-eos}
\bibinfo{author}{\bibfnamefont{D.}~\bibnamefont{Farrell}} \bibnamefont{et~al.}
  (\bibinfo{year}{2022}).

\bibitem[{\citenamefont{Papamakarios}(2019)}]{nle}
\bibinfo{author}{\bibfnamefont{G.}~\bibnamefont{Papamakarios}},
  \emph{\bibinfo{title}{Neural density estimation and likelihood-free
  inference}} (\bibinfo{year}{2019}),
  \urlprefix\url{https://arxiv.org/abs/1910.13233}.

\bibitem[{\citenamefont{Romani et~al.}(2022)\citenamefont{Romani, Kandel,
  Filippenko, Brink, and Zheng}}]{romani2022psr}
\bibinfo{author}{\bibfnamefont{R.~W.} \bibnamefont{Romani}},
  \bibinfo{author}{\bibfnamefont{D.}~\bibnamefont{Kandel}},
  \bibinfo{author}{\bibfnamefont{A.~V.} \bibnamefont{Filippenko}},
  \bibinfo{author}{\bibfnamefont{T.~G.} \bibnamefont{Brink}}, \bibnamefont{and}
  \bibinfo{author}{\bibfnamefont{W.}~\bibnamefont{Zheng}},
  \bibinfo{journal}{The Astrophysical Journal Letters}
  \textbf{\bibinfo{volume}{934}}, \bibinfo{pages}{L18} (\bibinfo{year}{2022}).

\bibitem[{\citenamefont{Hebeler et~al.}(2013)\citenamefont{Hebeler, Lattimer,
  Pethick, and Schwenk}}]{hebeler2013equation}
\bibinfo{author}{\bibfnamefont{K.}~\bibnamefont{Hebeler}},
  \bibinfo{author}{\bibfnamefont{J.}~\bibnamefont{Lattimer}},
  \bibinfo{author}{\bibfnamefont{C.~J.} \bibnamefont{Pethick}},
  \bibnamefont{and} \bibinfo{author}{\bibfnamefont{A.}~\bibnamefont{Schwenk}},
  \bibinfo{journal}{The Astrophysical Journal} \textbf{\bibinfo{volume}{773}},
  \bibinfo{pages}{11} (\bibinfo{year}{2013}).

\bibitem[{\citenamefont{Steiner et~al.}(2018)\citenamefont{Steiner, Heinke,
  Bogdanov, Li, Ho, Bahramian, and Han}}]{Steiner18ct}
\bibinfo{author}{\bibfnamefont{A.~W.} \bibnamefont{Steiner}},
  \bibinfo{author}{\bibfnamefont{C.~O.} \bibnamefont{Heinke}},
  \bibinfo{author}{\bibfnamefont{S.}~\bibnamefont{Bogdanov}},
  \bibinfo{author}{\bibfnamefont{C.}~\bibnamefont{Li}},
  \bibinfo{author}{\bibfnamefont{W.~C.~G.} \bibnamefont{Ho}},
  \bibinfo{author}{\bibfnamefont{A.}~\bibnamefont{Bahramian}},
  \bibnamefont{and} \bibinfo{author}{\bibfnamefont{S.}~\bibnamefont{Han}},
  \bibinfo{journal}{Mon. Not. Roy. Astron. Soc.}
  \textbf{\bibinfo{volume}{476}}, \bibinfo{pages}{421} (\bibinfo{year}{2018}),
  \eprint{1709.05013}, \urlprefix\url{https://doi.org/10.1093/mnras/sty215}.

\bibitem[{\citenamefont{Heinke et~al.}(2003)\citenamefont{Heinke, Grindlay,
  Lugger, Cohn, Edmonds, Lloyd, and Cool}}]{Heinke_2003}
\bibinfo{author}{\bibfnamefont{C.~O.} \bibnamefont{Heinke}},
  \bibinfo{author}{\bibfnamefont{J.~E.} \bibnamefont{Grindlay}},
  \bibinfo{author}{\bibfnamefont{P.~M.} \bibnamefont{Lugger}},
  \bibinfo{author}{\bibfnamefont{H.~N.} \bibnamefont{Cohn}},
  \bibinfo{author}{\bibfnamefont{P.~D.} \bibnamefont{Edmonds}},
  \bibinfo{author}{\bibfnamefont{D.~A.} \bibnamefont{Lloyd}}, \bibnamefont{and}
  \bibinfo{author}{\bibfnamefont{A.~M.} \bibnamefont{Cool}},
  \bibinfo{journal}{The Astrophysical Journal} \textbf{\bibinfo{volume}{598}},
  \bibinfo{pages}{501} (\bibinfo{year}{2003}),
  \urlprefix\url{https://doi.org/10.1086/378885}.

\bibitem[{\citenamefont{{Bogdanov} et~al.}(2016)\citenamefont{{Bogdanov},
  {Heinke}, {{\"O}zel}, and {G{\"u}ver}}}]{2016ApJ...831..184B}
\bibinfo{author}{\bibfnamefont{S.}~\bibnamefont{{Bogdanov}}},
  \bibinfo{author}{\bibfnamefont{C.~O.} \bibnamefont{{Heinke}}},
  \bibinfo{author}{\bibfnamefont{F.}~\bibnamefont{{{\"O}zel}}},
  \bibnamefont{and}
  \bibinfo{author}{\bibfnamefont{T.}~\bibnamefont{{G{\"u}ver}}},
  \bibinfo{journal}{\apj} \textbf{\bibinfo{volume}{831}}, \bibinfo{eid}{184}
  (\bibinfo{year}{2016}), \eprint{1603.01630}.

\bibitem[{\citenamefont{Campana et~al.}(1998)\citenamefont{Campana, Colpi,
  Mereghetti, Stella, and Tavani}}]{campana1998neutron}
\bibinfo{author}{\bibfnamefont{S.}~\bibnamefont{Campana}},
  \bibinfo{author}{\bibfnamefont{M.}~\bibnamefont{Colpi}},
  \bibinfo{author}{\bibfnamefont{S.}~\bibnamefont{Mereghetti}},
  \bibinfo{author}{\bibfnamefont{L.}~\bibnamefont{Stella}}, \bibnamefont{and}
  \bibinfo{author}{\bibfnamefont{M.}~\bibnamefont{Tavani}},
  \bibinfo{journal}{The Astronomy and Astrophysics Review}
  \textbf{\bibinfo{volume}{8}}, \bibinfo{pages}{279} (\bibinfo{year}{1998}).

\bibitem[{\citenamefont{Potekhin}(2014)}]{potekhin2014atmospheres}
\bibinfo{author}{\bibfnamefont{A.~Y.} \bibnamefont{Potekhin}},
  \bibinfo{journal}{Physics-Uspekhi} \textbf{\bibinfo{volume}{57}},
  \bibinfo{pages}{735} (\bibinfo{year}{2014}).

\bibitem[{\citenamefont{Slane et~al.}(2002)\citenamefont{Slane, Helfand, and
  Murray}}]{cooling}
\bibinfo{author}{\bibfnamefont{P.}~\bibnamefont{Slane}},
  \bibinfo{author}{\bibfnamefont{D.}~\bibnamefont{Helfand}}, \bibnamefont{and}
  \bibinfo{author}{\bibfnamefont{S.}~\bibnamefont{Murray}},
  \bibinfo{journal}{Astrophysical Journal - ASTROPHYS J}
  \textbf{\bibinfo{volume}{571}} (\bibinfo{year}{2002}).

\bibitem[{\citenamefont{Wijnands et~al.}(2017)\citenamefont{Wijnands, Degenaar,
  and Page}}]{wijnands2017cooling}
\bibinfo{author}{\bibfnamefont{R.}~\bibnamefont{Wijnands}},
  \bibinfo{author}{\bibfnamefont{N.}~\bibnamefont{Degenaar}}, \bibnamefont{and}
  \bibinfo{author}{\bibfnamefont{D.}~\bibnamefont{Page}},
  \bibinfo{journal}{Journal of Astrophysics and Astronomy}
  \textbf{\bibinfo{volume}{38}}, \bibinfo{pages}{1} (\bibinfo{year}{2017}).

\bibitem[{\citenamefont{Sun et~al.}(2019)\citenamefont{Sun, Li, Zhang, Zhang,
  Bauer, Xue, and Yuan}}]{mag}
\bibinfo{author}{\bibfnamefont{H.}~\bibnamefont{Sun}},
  \bibinfo{author}{\bibfnamefont{Y.}~\bibnamefont{Li}},
  \bibinfo{author}{\bibfnamefont{B.-B.} \bibnamefont{Zhang}},
  \bibinfo{author}{\bibfnamefont{B.}~\bibnamefont{Zhang}},
  \bibinfo{author}{\bibfnamefont{F.}~\bibnamefont{Bauer}},
  \bibinfo{author}{\bibfnamefont{Y.}~\bibnamefont{Xue}}, \bibnamefont{and}
  \bibinfo{author}{\bibfnamefont{W.}~\bibnamefont{Yuan}}, \bibinfo{journal}{The
  Astrophysical Journal} \textbf{\bibinfo{volume}{886}}, \bibinfo{pages}{129}
  (\bibinfo{year}{2019}).

\bibitem[{\citenamefont{{Arnaud}}(1996)}]{xspec}
\bibinfo{author}{\bibfnamefont{K.~A.} \bibnamefont{{Arnaud}}}, in
  \emph{\bibinfo{booktitle}{Astronomical Data Analysis Software and Systems
  V}}, edited by \bibinfo{editor}{\bibfnamefont{G.~H.} \bibnamefont{{Jacoby}}}
  \bibnamefont{and} \bibinfo{editor}{\bibfnamefont{J.}~\bibnamefont{{Barnes}}}
  (\bibinfo{year}{1996}), vol. \bibinfo{volume}{101} of
  \emph{\bibinfo{series}{Astronomical Society of the Pacific Conference
  Series}}, p.~\bibinfo{pages}{17}.

\bibitem[{\citenamefont{Baldi et~al.}(2016{\natexlab{b}})\citenamefont{Baldi,
  Bauer, Eng, Sadowski, and Whiteson}}]{baldi2016jet}
\bibinfo{author}{\bibfnamefont{P.}~\bibnamefont{Baldi}},
  \bibinfo{author}{\bibfnamefont{K.}~\bibnamefont{Bauer}},
  \bibinfo{author}{\bibfnamefont{C.}~\bibnamefont{Eng}},
  \bibinfo{author}{\bibfnamefont{P.}~\bibnamefont{Sadowski}}, \bibnamefont{and}
  \bibinfo{author}{\bibfnamefont{D.}~\bibnamefont{Whiteson}},
  \bibinfo{journal}{Physical Review D} \textbf{\bibinfo{volume}{93}},
  \bibinfo{pages}{094034} (\bibinfo{year}{2016}{\natexlab{b}}).

\bibitem[{\citenamefont{Fenton et~al.}(2022)\citenamefont{Fenton, Shmakov, Ho,
  Hsu, Whiteson, and Baldi}}]{fenton2022permutationless}
\bibinfo{author}{\bibfnamefont{M.~J.} \bibnamefont{Fenton}},
  \bibinfo{author}{\bibfnamefont{A.}~\bibnamefont{Shmakov}},
  \bibinfo{author}{\bibfnamefont{T.-W.} \bibnamefont{Ho}},
  \bibinfo{author}{\bibfnamefont{S.-C.} \bibnamefont{Hsu}},
  \bibinfo{author}{\bibfnamefont{D.}~\bibnamefont{Whiteson}}, \bibnamefont{and}
  \bibinfo{author}{\bibfnamefont{P.}~\bibnamefont{Baldi}},
  \bibinfo{journal}{Physical Review D} \textbf{\bibinfo{volume}{105}},
  \bibinfo{pages}{112008} (\bibinfo{year}{2022}), \bibinfo{note}{also
  arXiv:2010.09206}.

\bibitem[{\citenamefont{Typel et~al.}(2010)\citenamefont{Typel, R\"opke,
  Kl\"ahn, Blaschke, and Wolter}}]{Typel}
\bibinfo{author}{\bibfnamefont{S.}~\bibnamefont{Typel}},
  \bibinfo{author}{\bibfnamefont{G.}~\bibnamefont{R\"opke}},
  \bibinfo{author}{\bibfnamefont{T.}~\bibnamefont{Kl\"ahn}},
  \bibinfo{author}{\bibfnamefont{D.}~\bibnamefont{Blaschke}}, \bibnamefont{and}
  \bibinfo{author}{\bibfnamefont{H.~H.} \bibnamefont{Wolter}},
  \bibinfo{journal}{Phys. Rev. C} \textbf{\bibinfo{volume}{81}},
  \bibinfo{pages}{015803} (\bibinfo{year}{2010}),
  \urlprefix\url{https://link.aps.org/doi/10.1103/PhysRevC.81.015803}.

\bibitem[{\citenamefont{Oertel et~al.}(2017)\citenamefont{Oertel, Hempel,
  Kl\"ahn, and Typel}}]{Oertel}
\bibinfo{author}{\bibfnamefont{M.}~\bibnamefont{Oertel}},
  \bibinfo{author}{\bibfnamefont{M.}~\bibnamefont{Hempel}},
  \bibinfo{author}{\bibfnamefont{T.}~\bibnamefont{Kl\"ahn}}, \bibnamefont{and}
  \bibinfo{author}{\bibfnamefont{S.}~\bibnamefont{Typel}},
  \bibinfo{journal}{Rev. Mod. Phys.} \textbf{\bibinfo{volume}{89}},
  \bibinfo{pages}{015007} (\bibinfo{year}{2017}),
  \urlprefix\url{https://link.aps.org/doi/10.1103/RevModPhys.89.015007}.

\bibitem[{\citenamefont{Lindblom}(2018)}]{PhysRevD.97.123019}
\bibinfo{author}{\bibfnamefont{L.}~\bibnamefont{Lindblom}},
  \bibinfo{journal}{Phys. Rev. D} \textbf{\bibinfo{volume}{97}},
  \bibinfo{pages}{123019} (\bibinfo{year}{2018}),
  \urlprefix\url{https://link.aps.org/doi/10.1103/PhysRevD.97.123019}.

\bibitem[{\citenamefont{Heinke et~al.}(2006)\citenamefont{Heinke, Rybicki,
  Narayan, and Grindlay}}]{Heinke_2006}
\bibinfo{author}{\bibfnamefont{C.~O.} \bibnamefont{Heinke}},
  \bibinfo{author}{\bibfnamefont{G.~B.} \bibnamefont{Rybicki}},
  \bibinfo{author}{\bibfnamefont{R.}~\bibnamefont{Narayan}}, \bibnamefont{and}
  \bibinfo{author}{\bibfnamefont{J.~E.} \bibnamefont{Grindlay}},
  \bibinfo{journal}{The Astrophysical Journal} \textbf{\bibinfo{volume}{644}},
  \bibinfo{pages}{1090} (\bibinfo{year}{2006}),
  \urlprefix\url{https://doi.org/10.1086/503701}.

\bibitem[{\citenamefont{Lattimer and Steiner}(2014)}]{Lattimer14ns}
\bibinfo{author}{\bibfnamefont{J.~M.} \bibnamefont{Lattimer}} \bibnamefont{and}
  \bibinfo{author}{\bibfnamefont{A.~W.} \bibnamefont{Steiner}},
  \bibinfo{journal}{Astrophys. J.} \textbf{\bibinfo{volume}{784}},
  \bibinfo{pages}{123} (\bibinfo{year}{2014}), \eprint{1305.3242},
  \urlprefix\url{http://doi.org/10.1088/0004-637X/784/2/123}.

\bibitem[{\citenamefont{Ghosh et~al.}(2022)\citenamefont{Ghosh, Ju, Nachman,
  and Siodmok}}]{Ghosh:2022zdz}
\bibinfo{author}{\bibfnamefont{A.}~\bibnamefont{Ghosh}},
  \bibinfo{author}{\bibfnamefont{X.}~\bibnamefont{Ju}},
  \bibinfo{author}{\bibfnamefont{B.}~\bibnamefont{Nachman}}, \bibnamefont{and}
  \bibinfo{author}{\bibfnamefont{A.}~\bibnamefont{Siodmok}},
  \bibinfo{journal}{Phys. Rev. D} \textbf{\bibinfo{volume}{106}},
  \bibinfo{pages}{096020} (\bibinfo{year}{2022}), \eprint{2203.12660}.

\bibitem[{\citenamefont{{{\"O}zel} and {Psaltis}}(2015)}]{Ozel15}
\bibinfo{author}{\bibfnamefont{F.}~\bibnamefont{{{\"O}zel}}} \bibnamefont{and}
  \bibinfo{author}{\bibfnamefont{D.}~\bibnamefont{{Psaltis}}},
  \bibinfo{journal}{\apj} \textbf{\bibinfo{volume}{810}}, \bibinfo{eid}{135}
  (\bibinfo{year}{2015}), \eprint{1505.05156}.

\bibitem[{\citenamefont{Ludlam et~al.}(2018)\citenamefont{Ludlam, Miller,
  Arzoumanian, Bult, Cackett, Chakrabarty, Dauser, Enoto, Fabian, Garc{\'\i}a
  et~al.}}]{ludlam2018detection}
\bibinfo{author}{\bibfnamefont{R.}~\bibnamefont{Ludlam}},
  \bibinfo{author}{\bibfnamefont{J.}~\bibnamefont{Miller}},
  \bibinfo{author}{\bibfnamefont{Z.}~\bibnamefont{Arzoumanian}},
  \bibinfo{author}{\bibfnamefont{P.}~\bibnamefont{Bult}},
  \bibinfo{author}{\bibfnamefont{E.}~\bibnamefont{Cackett}},
  \bibinfo{author}{\bibfnamefont{D.}~\bibnamefont{Chakrabarty}},
  \bibinfo{author}{\bibfnamefont{T.}~\bibnamefont{Dauser}},
  \bibinfo{author}{\bibfnamefont{T.}~\bibnamefont{Enoto}},
  \bibinfo{author}{\bibfnamefont{A.}~\bibnamefont{Fabian}},
  \bibinfo{author}{\bibfnamefont{J.}~\bibnamefont{Garc{\'\i}a}},
  \bibnamefont{et~al.}, \bibinfo{journal}{The Astrophysical Journal Letters}
  \textbf{\bibinfo{volume}{858}}, \bibinfo{pages}{L5} (\bibinfo{year}{2018}).

\end{thebibliography}

\end{document}